\documentclass[useAMS, usenatbib]{mn2e}

\usepackage{rotating}
\usepackage{stmaryrd}
\usepackage{aas_macros}
\usepackage{times}
\usepackage{graphicx}
\usepackage{subfigure}
\usepackage{amsmath}
\usepackage{amsfonts}

\begin{document}

\graphicspath{{figures/}} 

\def\HI{\hbox{H$\,\rm \scriptstyle I\ $}}
\def\HII{\hbox{H$\,\rm \scriptstyle II\ $}} 
\def\HeI{\hbox{He$\,\rm \scriptstyle I\ $}}
\def\HeII{\hbox{He$\,\rm \scriptstyle II\ $}}
\def\HeIII{\hbox{He$\,\rm \scriptstyle III\ $}} 

\title[The intergalactic medium thermal history at $z=1.7$--$3.2$]{The
  intergalactic medium thermal history at redshift
  \boldmath{$z=1.7$}--\boldmath{$3.2$} from the Ly-\boldmath{$\alpha$}
  forest: a comparison of measurements using wavelets and the flux
  distribution}

\author[A.~Garzilli, J.S.~Bolton, T.-S.~Kim, S.~Leach and M.~Viel]{A.~Garzilli$^{1}$\thanks{E-mail: garzilli@sissa.it},  
  J.S.~Bolton$^{2}$, T.-S.~Kim$^{3}$, S.~Leach$^{1}$,  M.~Viel$^{4,5}$\\
  $^1$  SISSA, Astrophysics Sector, via Bonomea 265, 34136 Trieste, Italy \\
  $^2$ School of Physics, University of Melbourne, Parkville, VIC
  3010, Australia\\
  $^3$ Department of Astronomy, University of Wisconsin-Madison, 475 N. Charter St., Madison, WI53706, USA \\
  $^4$ INAF-Osservatorio Astronomico di Trieste, via G. B. Tiepolo 11,
  I-34131 Trieste, Italy\\
  $^5$ INFN/National Institute for Nuclear Physics, via Valerio 2,
  I-34127 Trieste, Italy\\
} \date{Accepted ---. Received ---; in original form \today}

\maketitle

\begin{abstract}
  We investigate the thermal history of the intergalactic medium (IGM)
  in the redshift interval $z=1.7$--$3.2$ by studying the small-scale
  fluctuations in the Lyman-$\alpha$ forest transmitted flux.  We
  apply a wavelet filtering technique to eighteen high resolution
  quasar spectra obtained with the Ultraviolet and Visual Echelle
  Spectrograph (UVES), and compare these data to synthetic spectra
  drawn from a suite of hydrodynamical simulations in which the IGM
  thermal state and cosmological parameters are varied.  From the
  wavelet analysis we obtain estimates of the IGM thermal state that
  are in good agreement with other recent, independent wavelet-based
  measurements.  We also perform a reanalysis of the same data set
  using the Lyman-$\alpha$ forest flux probability distribution
  function (PDF), which has previously been used to measure the IGM
  temperature-density relation.  This provides an important
  consistency test for measurements of the IGM thermal state, as it
  enables a direct comparison of the constraints obtained using these
  two different methodologies.  We find the constraints obtained from
  wavelets and the flux PDF are formally consistent with each other,
  although in agreement with previous studies, the flux PDF constraints
  favour an isothermal or inverted IGM temperature-density relation.
  We also perform a joint analysis by combining our wavelet and flux
  PDF measurements, constraining the IGM thermal state at $z=2.1$ to
  have a temperature at mean density of $T_0/[10^3$ K$]=17.3\pm 1.9$
  and a power-law temperature-density relation exponent $\gamma=1.1
  \pm 0.1$ (1$\sigma$).  Our results are consistent with previous
  observations that indicate there may be additional sources of
  heating in the IGM at $z<4$.

\end{abstract}


\section{Introduction}

The intergalactic medium (IGM) is the largest reservoir of baryonic
matter in the early Universe, and so gaining an understanding of its
physical state and chemical composition is an important goal of modern
cosmology.  In the current picture for the evolution of the baryons,
there are two reionisation events which turned the neutral gas in the
IGM into an ionised medium. The first reionisation event is thought to
be caused by hydrogen (and neutral helium) ionising radiation
produced by early galaxies. The precise redshift of this reionisation
event is not well constrained but it is thought to initiate at a
redshift no later than $z=11$ (\citealt{larson2010}) and end by
$z\simeq 6$, which is when the Universe becomes transparent to redshifted
Lyman-$\alpha$ photons from quasars~\citep{becker2001,fan06}. The
second reionisation event is expected to instead be driven by quasars
at lower redshifts, which produce a hard ionising spectrum that can
reionise singly ionised helium by $z\sim
3$~\citep{madau1999,FurlanettoOh2008,mcquinn2009}.  Photo-heating
during both of these reionisation events leaves a `footprint' on the
thermal state of the IGM; determining the redshift evolution of the
IGM temperature can therefore help pin down the details of these
reionisation eras (e.g. \citealt{theuns2002,hui2003,raskutti2012}).

In the simplest picture, the competition between photo-heating and
cooling due to the adiabatic expansion of the Universe results in a
power-law temperature-density relation following reionisation,
$T=T_{0}\Delta^{\gamma-1}$ for $\Delta = {\rho}/{\langle \rho \rangle}
< 10$.  The density dependence of the recombination rate means that
higher density gas recombines faster, yielding more neutral atoms per
unit time for photo-heating. These regions thus cool less rapidly than
lower density gas, resulting in a temperature-density relation which
evolves from isothermal ($\gamma \sim 1$) following reionisation
toward a power-law with $\gamma\sim 1.6$ (\citealt{hui1997}).  In
principle, if the redshift dependence of the temperature-density
relation can be measured then the timing of reionisation can thus be
inferred.  Indeed, by studying the thermal widths of absorption lines
in the Lyman-$\alpha$ forest, \cite{schaye2000} observed an increase
in the temperature at mean density and a flattening of the
temperature-density relation at $z\sim 3$, which may indicate the
epoch of \HeII reionisation occurred around this time.

However, this is a somewhat simplified picture; according to the
numerical simulations presented by \cite{mcquinn2009}, \HeII
reionisation is inhomogeneous and long-range heating by hard photons
will induce large-scale fluctuations of the order of 50 comoving Mpc
in the IGM temperature and \HeII ionisation state.  Another question
mark hangs over the slope of the temperature-density relation
describing the IGM thermal state. Observational work from
\cite{2007ApJ...662...72B}, \cite{bolton2008} and \cite{viel2009} has
suggested that the IGM may obey an `inverted' ($\gamma<1$)
temperature-density relation in which, somewhat counter-intuitively,
less dense gas is hotter than denser gas.  Although it appears
difficult to produce this result by \HeII photo-heating by quasars
(\citealt{mcquinn2009,bolton2009}), it has recently been suggested by
\cite{2011arXiv1106.5504C} and \cite{puchwein2011} that it could be a
consequence of volumetric heating by TeV emission from blazars.

Given these uncertainties, it is important to investigate the
observational constraints in more detail.  One way to achieve this is
to directly compare different methods used to measure the IGM thermal
state.  This allows one to establish whether these different
approaches are consistent, or whether there are systematic
uncertainties which impact differently upon the competing
approaches.  The methodologies used in the literature thus far to
measure the IGM thermal state from the Lyman-$\alpha$ forest can be
broadly divided in two classes. The common feature in both approaches
is that they measure the IGM temperature via the impact of Jeans
(pressure) smoothing and thermal Doppler broadening on the
Lyman-$\alpha$ forest. The first class consists of methods that fit
Voigt profiles to each absorption line in the Lyman-$\alpha$ forest.
Examples of this class are the earlier work of \cite{schaye2000},
\cite{ricotti2000} and \cite{mcdonald2000} and more recently
\cite{bolton2010}.  The second class consists of methods in which the
transmitted flux is analysed with a global statistical approach,
without decomposing the spectra into separate features. Power spectra
studies belong to this class \citep{zaldarriaga2001,viel2009}, as well
as methods that examine other statistical properties of the forest,
such as the flux probability distribution function (PDF)
\citep{bolton2008,Calura2012}, the wavelet analysis method applied in
\cite{lidz2010} and the `curvature' statistic used by
\cite{becker2010}.

The main aim of this work is to compare two of these competing
techniques, the flux PDF and wavelets, by applying them to the
metal-cleaned Lyman-$\alpha$ forest spectra presented by
\cite{kim2007}.  These data have previously been used in studies of
the flux PDF which have found the IGM temperature-density relation may
be isothermal or inverted at $2<z<3$ (\citealt{bolton2008,viel2009}).
In order to interpret the results we have also utilised and extended
the suite of hydrodynamical simulations used in the analysis
of~\cite{becker2010}.  This comparison is of particular interest
because, as pointed out by \cite{lidz2010}, there appears to be some
tension between the IGM thermal parameters inferred from wavelets and
the flux PDF, particularly with respect to measurements of the slope
of the temperature-density relation.  This paper is therefore
organised as follows: in Section 2 we review the observations and
numerical simulations used; in Section 3 our implementation of the
wavelet analysis is presented; in Section 4 we describe our
interpolation and parameter determination methodology and in Section 5
we present our results along with a comparison to previous studies.
We conclude in Section 6.  An appendix at the end of the paper lists
the simulations we use in this study, along with several tests for
numerical convergence and the sensitivity of our results to parameter
assumptions.

\section{Observations and numerical simulations}

In this section we briefly review our observational dataset, the
numerical simulations used for our theoretical interpretation and the
procedure used to generate synthetic spectra.

\subsection{Spectra, metal removal and continuum placement}\label{sec:spectra}

Our analysis uses the eighteen metal-cleaned quasar spectra from
\cite{kim2007} obtained with the Ultraviolet and Visual Echelle
Spectrograph (UVES) on the VLT. The spectra are sampled with pixels of
width $0.05\mbox\,{\rm \AA}$ and have a signal to noise ratio per pixel of
order $30 \mbox{--}50$.  In order to avoid the proximity effect, the
region 4000 $\rm km\,s^{-1}$ bluewards of the Lyman-$\alpha$ emission
line has been excluded. These spectra contain Lyman limit systems with
column densities in the interval $10^{17.2} {\rm cm}^{-2}\leq N_{\rm
  HI}\leq 10^{19}{\rm cm}^{-2}$, but no damped Lyman-$\alpha$ systems,
defined by a column density $N_{\rm HI}\geq 10^{20.3}{\rm cm}^{-2}$.

In order to correct the \HI absorption for metal contamination, metal
absorption lines in the spectra were identified and fitted with Voigt
profiles.  These were substituted by a continuum level or by a
Lyman-$\alpha$ only absorption profile generated from the fitted
Lyman-$\alpha$ parameters (see \citealt{kim2007} for details).  Note
this approach differs from the metal removal procedure used in
\cite{lidz2010}, where only narrow absorption lines (with $b < 7\,{\rm
  km /s}$) were identified as metals and excised from the spectra.

The continuum level in the spectra was determined by locally
connecting regions that are thought to be absorption-free. This is an
iterative procedure, which starts with connecting non absorbed regions
and is subsequently updated during the process of Voigt profile
fitting the \HI and metal lines.  Note that in our analysis we neglect
the possibility of having an extended and slowly varying continuum
absorption.  This means that spectral regions which are considered to
be absorption free could actually suffer from absorption by a broad
\HI density fluctuation, and the measured optical depth would then be
underestimated.  We shall discuss the effect of continuum placement on
our results further in Section~\ref{sec:constraints}.

\begin{table}
\centering
\begin{tabular}{lcccc}
\hline
\noalign{\smallskip}
QSO & $z_{\mathrm{em}}$ & $z_{\mathrm{Ly\alpha}}$ & $\lambda_{\mathrm{Ly\alpha}}(\rm{\mbox{\AA}})$
& S/N \\
\noalign{\smallskip}
\hline
\noalign{\smallskip}
Q0055--269          & 3.655 & 2.936$-$3.205 & 4785$-$5112 & 80$-$50   \\ 
PKS2126--158        & 3.279 & 2.815$-$3.205 & 4638$-$5112 & 50$-$200 \\ 
Q0420--388          & 3.116 & 2.480$-$3.038 & 4231$-$4909 & 100$-$140\\
HE0940--1050        & 3.078 & 2.452$-$3.006 & 4197$-$4870 & 50$-$130 \\
HE2347--4342        & 2.874 & 2.336$-$2.819 & 4055$-$4643 & 100$-$160 \\
Q0002--422          & 2.767 & 2.209$-$2.705 & 3901$-$4504 & 60$-$70 \\
PKS0329--255        & 2.704 & 2.138$-$2.651 & 3815$-$4439 & 30$-$55\\
Q0453--423          & 2.658 & 2.359$-$2.588 & 4084$-$4362 & 90$-$100\\
HE1347--2457        & 2.609 & 2.048$-$2.553 & 3705$-$4319 & 85$-$100 \\
Q0329--385          & 2.434 & 1.902$-$2.377 & 3528$-$4105 & 50$-$55 \\ 
HE2217--2818        & 2.413 & 1.886$-$2.365 & 3509$-$4091 & 65$-$120 \\ 
Q0109--3518         & 2.405 & 1.905$-$2.348 & 3532$-$4070 & 60$-$80 \\
HE1122--1648        & 2.404 & 1.891$-$2.358 & 3514$-$4082 & 70$-$170\\
J2233--606          & 2.250 & 1.756$-$2.197 & 3335$-$3886 & 30$-$50 \\
PKS0237--23         & 2.223 & 1.765$-$2.179 & 3361$-$3865 & 75$-$110\\ 
PKS1448--232        & 2.219 & 1.719$-$2.175 & 3306$-$3860 & 30$-$90 \\
Q0122--380          & 2.193 & 1.700$-$2.141 & 3282$-$3819 & 30$-$80 \\
Q1101--264          & 2.141 & 1.880$-$2.097 & 3503$-$3765 & 80$-$110 \\
\noalign{\smallskip}
\hline
\end{tabular}
\caption{Properties of the quasar spectra from
  \citet{kim2007}. $z_{\mathrm{em}}$ is the approximate redshift of
  the quasar measured from the Lyman-$\alpha$ emission line;
  $z_{\mathrm{Ly\alpha}}$ and $\lambda_{\mathrm{Ly\alpha}}$ are the
  redshift and wavelength intervals associated with the Lyman-$\alpha$
  absorption; S/N is the signal to noise ratio per 0.05$\mbox{\AA}$
  pixel.}
\label{tab:real_spectra}
\end{table}

\subsection{Numerical simulations}\label{sec:sims}

The simulations we use are based on the suite of models used
by~\citet{becker2010}, which we have extended by further varying the
cosmological and IGM thermal parameters assumed.  We make use of {\sc
  Gadget-3} code which is a parallel smoothed particle hydrodynamics
code~\citep{springel2005}.  Our simulations are performed in a
periodic box of $10$ comoving Mpc$/h$ in linear size. We describe the
evolution of both the dark matter and the gas, using $256^3$
particles for simulations in which the cosmological parameters are
varied, or $2 \times 512^3$ particles for simulations in which the IGM
thermal state parameters are varied.  A summary of the simulations is
given in Tab~\ref{tab:simulations} in the appendix.

The simulations all start at $z=99$ with initial conditions generated
using the \citet{eisenstein1999} transfer function. Star formation is
incorporated using a simplified prescription in which all gas
particles with $\Delta>10^3$ and temperature $T<10^5$ K are converted
into collisionless stars. Since we are not interested in the details
of star formation, we thereby avoid the small dynamical times that
would arise due to these overdense regions. As the bulk of the
Lyman-$\alpha$ forest absorption corresponds to densities $\Delta <
10$, this prescription has an impact at below the percent level on the
final computation of flux PDF and flux power~\citep{viel2004} and so
we expect this will not affect our work. To check numerical
convergence of our simulations we have also performed a series of
simulations with varying gas particle and box size.  We conclude from
these tests, demonstrated in Figure~\ref{fig:resolution} in the
appendix, that our study of the statistics of small-scale structure of
the Lyman-$\alpha$ forest demand the relatively high mass resolution
afforded by a large number of particles, $2\times512^3$, in a
10$h^{-1\,}$Mpc volume.  With greater computational resources, we
could improve the accuracy of our simulations by increasing the box
size from 10 $h^{-1\,}$Mpc, while maintaining high mass
resolution. However we believe that the possible improvements are
small compared to the error budget we have assumed in
Section~\ref{sec:errorbar}.
  
A spatially uniform ultraviolet (UV) background applied in the
optically thin limit determines the photo-heating and photo-ionisation
of the gas in the simulations. In order to generate different thermal
histories, we have rescaled all of the \HI, \HeI and \HeII
photo-heating rates using the~\citet{haardt2001} UV background model
for galaxy and QSO emission. The photo-heating rates, $\epsilon_{\rm
  i}$, were changed by rescaling their values in a density dependent
fashion, $\epsilon^{\rm new}_{\rm i} = \zeta \Delta^{\xi}
\epsilon_{\rm i}^{\rm HM01}$.  A list of the values used for the
scaling coefficients $\zeta$ and $\xi$, along with the corresponding
values for $T_0$ and $\gamma$ and the other main simulations
parameters are given in Table~\ref{tab:simulations} in the appendix.
The resulting thermal histories are self-consistent in the sense that
the gas pressure, and hence the Jeans smoothing, is compatible with
the gas temperature. Computing simulated spectra with this treatment
requires a large computational effort, because each thermal history
requires its own simulation. By comparison, the \citet{lidz2010}
reference simulation was run with a fixed chosen ionisation state, and
the thermal state was then superimposed by applying a
temperature-density relation in a post-processing step, thereby
neglecting a full treatment of Jeans smoothing in this approximation.
However, note that both our approach and the \citet{lidz2010}
technique neglect radiative transfer effects
(e.g. \citealt{tittley2007,mcquinn2009}) because of the computational
effort that solving the cosmological radiative transfer equation would
imply.

\begin{figure}
  \centering
  \includegraphics[width=0.25\textwidth, angle=270]{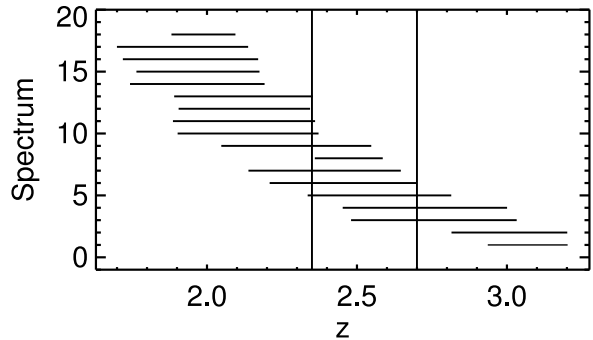}
  \caption{Redshift range of the observed spectra listed in Table
    1. The two vertical lines indicate the edges of the redshift bins
    with effective redshifts of $\left<z\right> = [2.07,2.52,2.93]$, and
    which are roughly centred on the redshift of our simulation
    outputs. The percentage of spectra in each redshift bin are 57, 25
    and 18 per cent, respectively.}\label{fig:real_spectra}
\end{figure}

We have selected three simulation snapshots at $z=[2.17, 2.55, 2.98]$
from the~\citet{becker2010} models, which cover the redshift range of
our data, and the values of $T_0$ and $\gamma$ have been determined by
fitting the temperature-density relation at each redshift for each
simulation.  These snapshots then determine the way in which we split
the data into three redshift ranges, divided at $z=[2.35, 2.70]$, as
shown in Figure~\ref{fig:real_spectra}. These three redshift bins
contain 57, 25 and 18 percent of the data, respectively, and the
effective (average) redshift of the data in each bin is
$\left<z\right>=[2.07,2.52,2.93]$. Hence-forward, these effective
redshifts will be the nominal values used when quoting our results.

In order to assess the effect of astrophysical and cosmological
uncertainties on the IGM physics we vary the simulation parameters on
a grid, one at a time. There are two sets of simulations: in the first
set we vary only the IGM thermal state parameters around a reference
model with $[\Omega_{\rm m},\Omega_{\rm b}, h,\sigma_8, n_s] =
[0.26, 0.0444, 0.72, 0.8, 0.96]$ and with a reference IGM thermal state
$[T_0/{\rm 10^3K}, \gamma] = [15, 1.6]$. The range of thermal state
parameters covered by our simulations extend from around
$T_{0}=4\,600-31\,000$\,K and $\gamma=0.7-1.6$. In the second set we
vary only the cosmological parameters around $[\Omega_{\rm m},
h,\sigma_8, n_s] = [0.26, 0.72, 0.85, 0.95]\pm
[0.04,0.08,0.05,0.05]$, with a reference IGM thermal state $[T_0/{\rm
10^3 K}, \gamma] = [20, 1.06]$ and fixed $\Omega_{\rm
b}=0.0444$. The difference in cosmological parameters for these two
simulations sets owes to imperfect planning, though this has not been
a great source of concern or bias since, as we demonstrate in the
appendix of this paper, the uncertainties in the cosmological
parameters are not the limiting factor in our predictions for
Lyman-$\alpha$ spectra or in our interpretation of the data.

\subsection{Synthetic spectra}\label{sec:syn_spec}

Our synthetic spectra are obtained using the following procedure
described in~\citet{theuns98} and~\citet{bolton2008}, and briefly
reviewed here. At each redshift slice, approximately $10^3$ randomly
chosen lines of sight are selected. We convolve the resulting \HI
density with a Voigt profile using the approximation introduced
in~\cite{garcia2006} -- this approximation is sufficiently accurate
for the range of column densities considered here. We resample the
spectra into velocity intervals of $4.4 \, {\rm km\,s}^{-1}$ and we
account for instrumental resolution by convolving the spectra with a
Gaussian with full width at half maximum of $7 \, {\rm
  km\,s}^{-1}$. We add noise to the spectra with the same level of the
observed data.

\begin{figure}
  \centering
  \includegraphics[totalheight=0.34\textheight, angle=270]{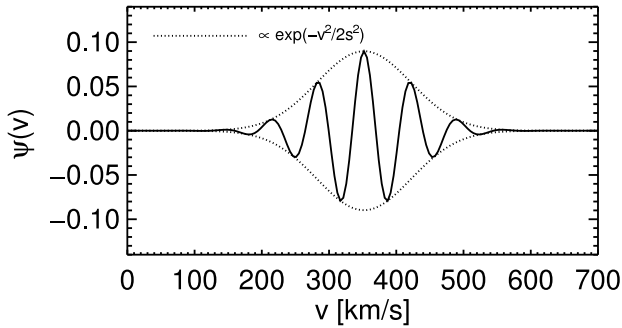}
  \caption{Real part of the Morlet wavelet,
    Equation~(\ref{eqn:morlet}), with smoothing scale $s = 70\, {\rm
    km\,s}^{-1}$. The width of the envelope depends on the scale probed by
    the oscillations.}\label{fig:morlet}
\end{figure}

In order to leave the \HI effective optical depth, $\tau_{\rm eff}$, as a
free parameter we adjust the mean transmitted flux as $\langle F
\rangle \rightarrow \langle F' \rangle = \langle \exp{[-\tau A]}
\rangle$, where $A$ is chosen such that the global mean normalized
flux $\langle F' \rangle $ satisfies $\tau_{\rm eff} = - \ln ( \langle
F' \rangle )$.  As we will discuss, $\tau_{\rm eff}$ is one of the
main uncertainties limiting the determination of the IGM thermal state
in our wavelet analysis.

\section{Wavelet analysis of spectra}
\label{sec:wavelet_pdf}

The wavelet decomposition of the Lyman-$\alpha$ forest was first
proposed by \cite{theuns2000}, \cite{meiksin2000}, \cite{theuns2002} and
\cite{zaldarriaga2002}.  The idea is to filter the spectra in such a
way as to construct observables that are sensitive to the thermal
state of the IGM. The wavelet decomposition was therefore suggested
because it might be used to detect temporal or even spatial variations
of physical properties of the IGM
(e.g. \citealt{2011MNRAS.415..977M}).

The physical motivation for this type of filtering relates to the
effects of Doppler broadening and Jeans smoothing on small scale
structure in the Lyman-$\alpha$ forest.  The thermal Doppler effect
arises due to the velocity dispersion of the hydrogen atoms, which
causes broadening of the absorption lines ( for a recent analysis of
thermal broadening and Jeans smoothing, see
\citet{peeples2010_1,peeples2010_2}).  The velocity distribution is
described by the Maxwell distribution
\begin{eqnarray}
P(v) dv = \sqrt{\frac{m_{\rm p}}{2 \pi k_{\rm B}
T}}\exp\left[{-\frac{m_{\rm p} v^2}{2 k_{\rm B} T}}\right]dv,
\end{eqnarray}
from which it can be seen that the velocity dispersion is proportional
to $\sqrt{T}$. This results in a broadening of the absorption spectra
by a factor $\sqrt{\frac{2 k_{\rm B} T}{m_{\rm p}}}\approx 13 \, {\rm
  km\,s^{-1}}$ at $T=10^4\, {\rm K}$. The other important effect is
Jeans smoothing. Because the ideal equation of state holds, an
increase in the temperature of the gas corresponds to an increase in
pressure which then also smoothes the small-scale structures (see
  for example \cite{pawlik2009}). The Jeans smoothing depends on the
full thermal history of the IGM, because pressure forces alter the
dynamical state of the gas~\citep{hui1997}.

\subsection{The wavelet amplitude PDF}

Following~\cite{lidz2010}, we have implemented the `Morlet wavelet' as
a probe of this thermal smoothing, which we briefly review here. The
Morlet wavelet is a Gaussian in the complex plane, defined in velocity
space by
\begin{eqnarray}\label{eqn:morlet} 
\psi_k (v) = {\cal A}\exp\left[-i k v/ 2 \pi\right] \exp\left[- v^2/2 s^2\right], 
\end{eqnarray}
where the `smoothing scale' $s = 2 \pi / k$ is chosen so that the
wavelet changes its global width depending on the probed scale $k$.
Our normalization constant ${\cal A}$ is chosen so that the integral
of the squared wavelet function is unity. An example is shown in
Figure~\ref{fig:morlet} for a smoothing scale $s=70\, {\rm
  km\,s}^{-1}$. This follows the choice of~\cite{lidz2010} which taken
as a compromise between maximising the sensitivity to the small-scale
structure and avoiding the possible contamination by metal lines.

The spectra $F(v)$ are first convolved with the Morlet wavelet, to
obtain a filtered spectrum
\begin{eqnarray}\label{eqn:filtered_spec}
f(v)= \int dv' F(v) \psi(v-v').
\end{eqnarray}
The filtered spectrum $f$ is then squared and smoothed in order to
compute the `wavelet amplitude'
\begin{eqnarray}\label{eqn:wavelet_amp}
A(v) = \frac{1}{L} \int dv' \Theta(|v - v'|, L/2) f^2(v'),
\end{eqnarray}
where $\Theta(v, L/2)$ is the top-hat function with width $L = 1\,000
\, {\rm km\,s}^{-1}$. This choice of large-scale smoothing
follows~\cite{lidz2010}, though we have also checked that our results
do not depend strongly on the exact value of $L$.

\begin{figure*}
  \centering
  \includegraphics[width=0.6\textwidth, angle=270]{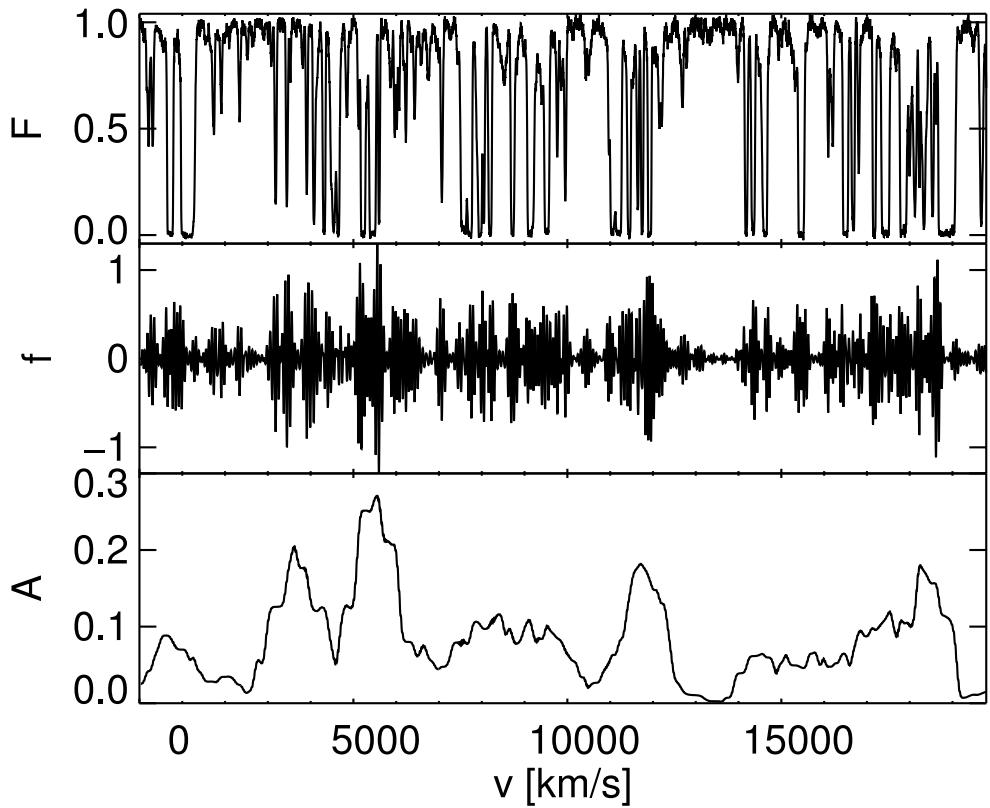}
  \caption{Upper panel: the original QSO spectrum, Q0055--269.  Middle
    panel: the amplitude of the signal,
    Equation~(\ref{eqn:filtered_spec}), obtained by convolving the
    spectrum with a Morlet wavelet with $s=70 \, {\rm km\,s}^{-1}$.
    Lower panel: the same amplitude after squaring and smoothing with
    a $1\,000 \, {\rm km\,s}^{-1}$ wide top-hat filter,
    Equation~(\ref{eqn:wavelet_amp}).
    }\label{fig:wavelet_signal_20}
\end{figure*}

\begin{figure*}
    \centering
    \includegraphics[width=0.45\columnwidth, angle=-90]{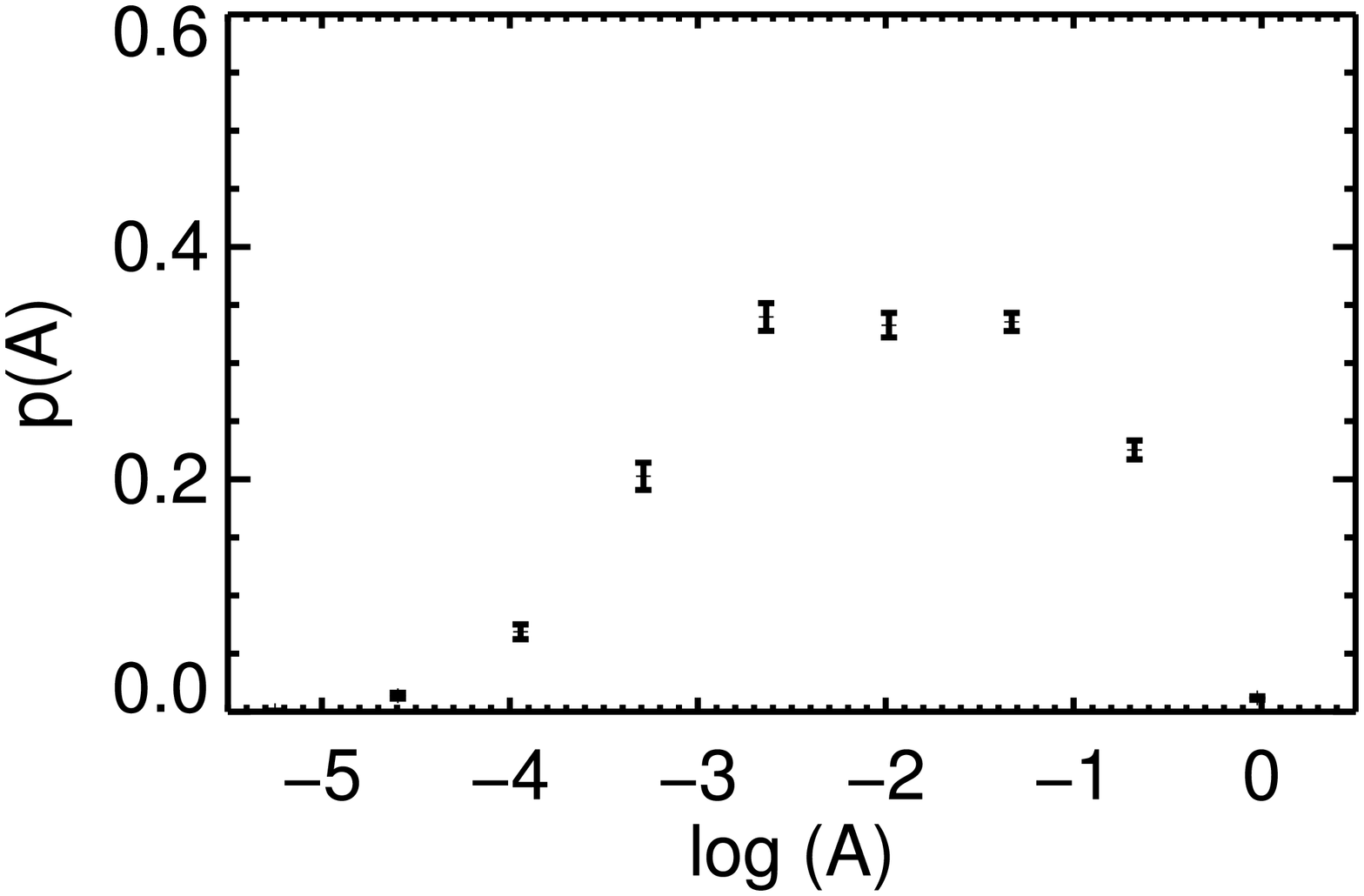}
    \includegraphics[width=0.45\columnwidth, angle=-90]{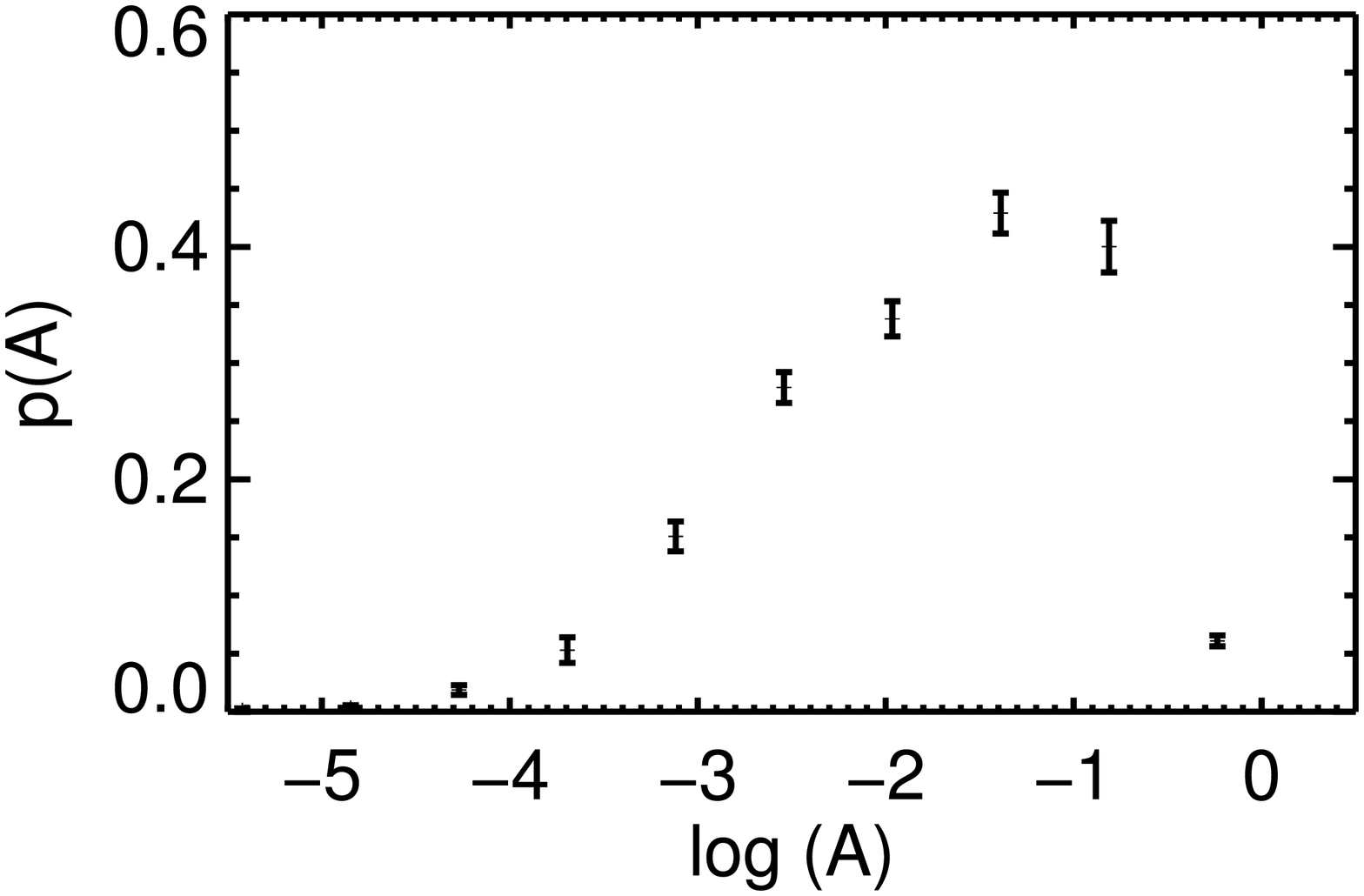}
    \includegraphics[width=0.45\columnwidth, angle=-90]{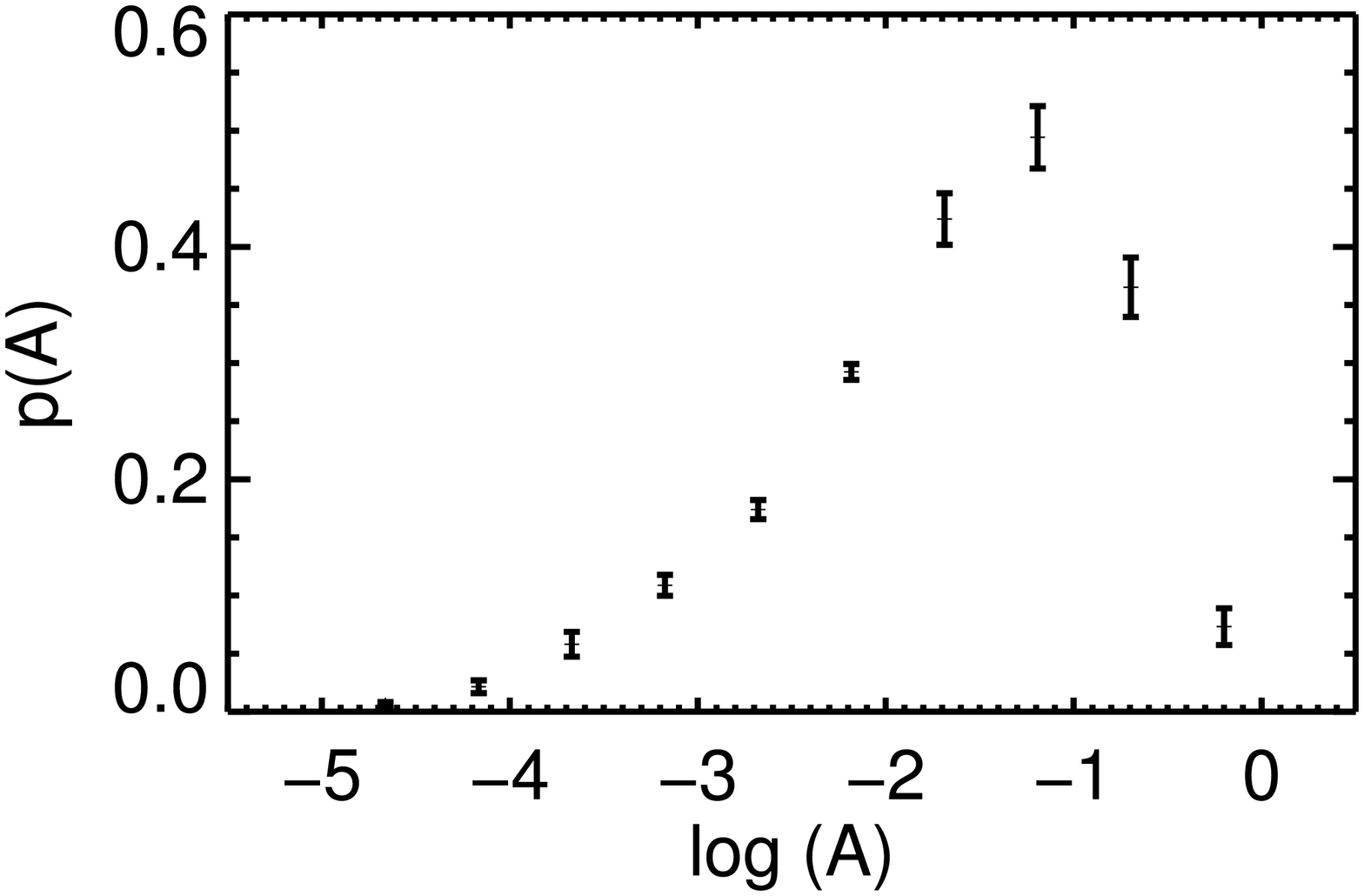}
    \caption{The observed wavelet amplitude PDF,
      Equation~(\ref{eqn:wavelet_amp}), for $s=70 \, {\rm km\,s}^{-1}$
      and $L=1\,000 \, {\rm km\,s}^{-1}$, for redshifts
      $z=[2.1,\,2.5,\,2.9]$ (left to right). The error bars displayed
      here, calculated via jackknife resampling with
      Equation~(\ref{eq:jackknife}), are thought to be underestimated
      and so we have made an allowance for theoretical uncertainty in
      our interpretation of the data, as described in
      Section~\ref{sec:errorbar}.}
    \label{fig:real_wavelet_pdf_0}
\end{figure*}

Figure~\ref{fig:wavelet_signal_20} shows an example of these
processing steps for Q0055-269, our highest redshift spectrum.  It can
be seen that the wavelet amplitude is greater in regions of the
spectrum with absorption lines of width comparable to the sampling
scale. The resulting signal captures some of the inhomogeneity of the
original spectrum.

The final observable is the `wavelet amplitude PDF', $p(A)$, which is
calculated by binning $A$, calculated for each spectra, into a
histogram with ten logarithmically spaced bins, whose minimum and
maximum values are taken from the data. The same processing is applied
to our simulations to obtain the predicted wavelet amplitude PDF. The
resulting wavelet amplitude PDFs computed from our data are presented
in Figure~\ref{fig:real_wavelet_pdf_0}.  These data are the key
observational results of this work, which will be analysed and
interpreted in Section~\ref{sec:results}.

\subsection{Error bar estimates}
\label{sec:errorbar}

Following~\cite{lidz2006} we have attempted to estimate the error bars
of the wavelet amplitude PDF using the `jackknife' method, although as
we describe below, we believe our main uncertainty arises from the
accuracy with which we can predict the wavelet amplitude PDF. The
jackknife method is a resampling method in which our spectra are
divided into $n_{\rm g}=10$ subgroups of equal size, from which a set
of $n_{\rm g}$ wavelet PDFs, $\tilde{p}_k(A)$, are computed by
omitting from the data one subgroup of data at a time.  The covariance
matrix of the wavelet PDF amplitude is then estimated using
\begin{eqnarray}
C_{ij} = \sum^{n_{\rm g}}_{k = 1} [p(A_i) - \tilde{p}_k(A_i)] [p(A_j) -
  \tilde{p}_k(A_j)].
\label{eq:jackknife}
\end{eqnarray}
\cite{lidz2006} tested their covariance values against those estimated
directly from $10\,000$ mock spectra, and found that
Equation~(\ref{eq:jackknife}) holds approximately, but that the error
bars can sometime be underestimated, especially in the tails of the
wavelet amplitude distribution. Partly owing to this observation, and
partly due to our more limited number of mock QSO spectra on which to
perform the jackknife method, we have opted for what should be a
conservative estimate of wavelet PDF uncertainties: the diagonal
elements of Equation~(\ref{eq:jackknife}) are all replaced with $(0.25
\times {\rm max}(p(A_i))^2$, as shown in
Figures~\ref{fig:wavelet_interp} and~\ref{fig:pdf_thermal_1},
and we ignore the off-diagonal elements suggested by the jackknife
resampling. In doing so we are attempting to make an allowance for the
theoretical uncertainty associated with our calculation of the wavelet
amplitude PDF as well as for the level of interpolation errors
suggested by our validation tests described in
Section~\ref{sec:interpolation}.  We note that we implemented the
jackknife method, as defined in Equation~(\ref{eq:jackknife}), (as
well as a bootstrap resampling method) to estimate the covariance and
error bars shown in Figure~\ref{fig:real_wavelet_pdf_0}, but we
concluded that the covariance was being underestimated. Given the
approximation applied in our interpolation scheme, we believe the
current error-bar prescription we apply is sufficiently representative
to constrain the central value and width of the wavelet amplitude PDF,
and that it is on the conservative side.  A full theoretical
understanding of the wavelet amplitude bin-bin covariance remains an
open problem.

\section{Parameter determination methodology}

In this section we describe the core ingredients of our analysis: our
interpolation scheme, parameter sampling method and IGM
parametrization.

\subsection{Interpolation scheme} \label{sec:interpolation}

\begin{figure}
    \centering
    \includegraphics[width=0.8\columnwidth, angle=-90]{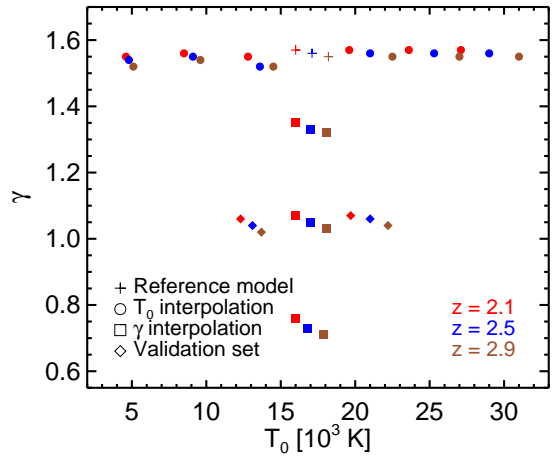}
    \caption{Simulations used in our interpolation scheme: reference
      model (crosses), $T_0$ interpolation (circles), $\gamma$
      interpolation (squares) and the validation set
      (diamonds). Further information about the simulations may be
      found in Table~\ref{tab:simulations}.}
    \label{fig:sims_plot}
\end{figure}

In order to calculate the wavelet PDF for a given location in our
parameter space, we perform an interpolation of the wavelet PDF
calculated over our available simulations.  Our approach is to perform
a cubic-spline interpolation of the wavelet PDF differences as a
function of the cosmological and astrophysical parameters, for which
we have three simulations along each cosmological parameter direction,
three simulations varying $\gamma$ and six simulations varying $T_0$,
as illustrated in Figure~\ref{fig:sims_plot}.

As mentioned in Section~\ref{sec:syn_spec}, the effect of varying
$\tau_{\rm eff}$ is calculated in post-processing. We therefore
calculate the wavelet PDF on a fixed grid of 100 $\tau_{\rm eff}$
values -- checking that an implementation with ten grid points in this
direction would also be acceptable.  We then apply the scheme
\begin{eqnarray}
  p(A)(\vec\theta, \tau_{\rm eff}) &=& 
  p(A)(\vec\theta_0, \tau_{\rm eff}) + \nonumber\\
  &  & \sum_{i=1}^{n} \Delta p(A)(\{\theta_0^1, \ldots, \theta^i, \ldots, \theta_0^n\},\tau_{\rm eff}),
  \label{eq:interp1}
\end{eqnarray}
for the two closest values of $\tau_{\rm eff}$ on our grid and obtain
the final $p(A)( \vec\theta, \tau_{\rm eff})$ via linear interpolation
of these two function evaluations. Here $\vec\theta$ denotes a parameter
vector with $n$ components, and $\Delta p$ is the interpolated PDF
differences relative to the fiducial model, $\vec\theta_0$.

\begin{figure}
    \centering
    \includegraphics[width=0.8\columnwidth, angle=-90]{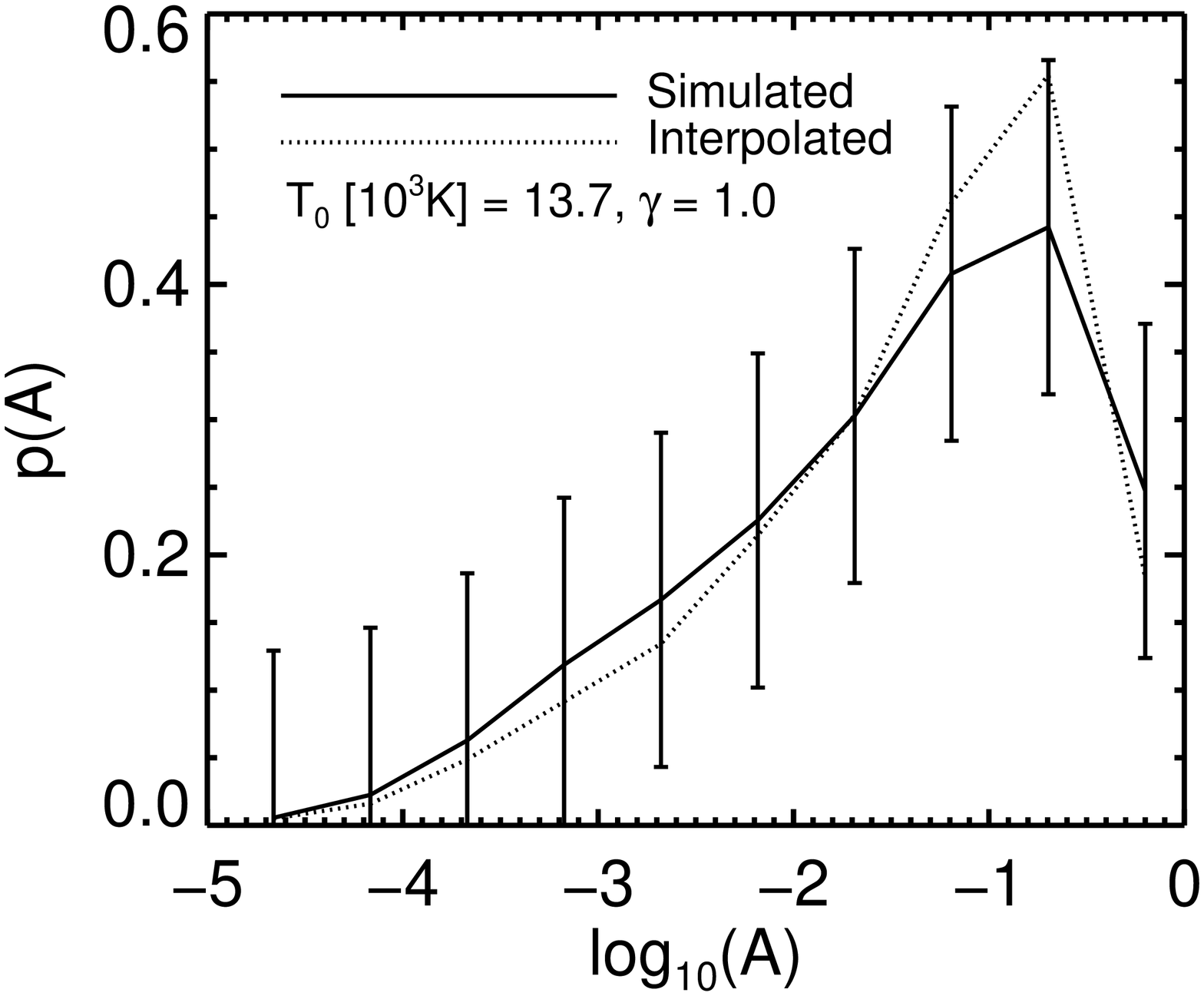}
    \includegraphics[width=0.8\columnwidth, angle=-90]{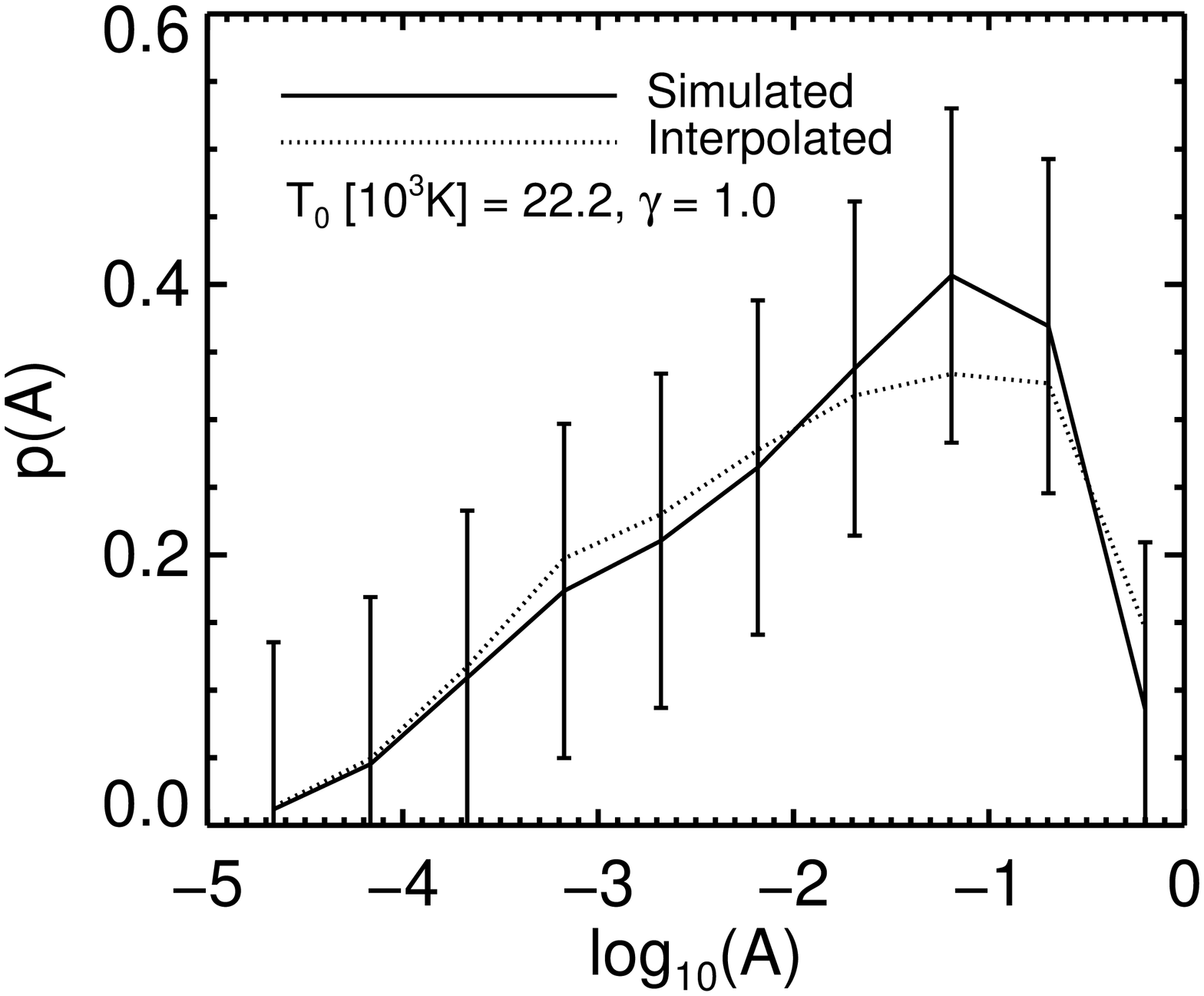}
    \caption{Comparison of the wavelet PDF calculated with our
      interpolation scheme (dotted) to the PDF from a direct
      simulation (solid), for two validation simulations, C10 (upper)
      and E10 (lower). The agreement is satisfactorily within the
      uncertainties that we have assumed.}
    \label{fig:wavelet_interp}
\end{figure}

We have checked the accuracy of this interpolation scheme by comparing
it with the wavelet PDF calculated for a `validation set' of
simulations which are not used in the interpolation procedure. The
wavelet PDF for these simulations, (`D10' and `E10' with parameters
given in Table~\ref{tab:simulations}) were found to be satisfactorily
reproduced, within the uncertainties that we have assumed, as described
in Section~\ref{sec:errorbar}. An illustrative case is shown in
Figure~\ref{fig:wavelet_interp} for the $z=2.9$ bin.

We believe that this interpolation captures the variations in the
wavelet amplitude PDF predicted by our simulations sufficiently
accurately for our purposes and within the generous errors we have assumed.
We therefore expect that our parameter constraints will be
on conservative side. We suggest that, with more computational resources
than currently available to us, the {\sc pico}
\citep{2007ApJ...654....2F} training-set/interpolation scheme could be
applied to accurately and efficiently predict Lyman-$\alpha$ forest
observational quantities.

\subsection{Parameter sampling}

In order to estimate the astrophysical and cosmological parameters and
their uncertainties, we use a sampling-based approach. Bayes' theorem
can be written as ${\cal L}(\mathbf{D}|\theta) \pi(\theta) = Z
P(\theta)$, where ${\cal L}$ is the likelihood of the data
$\mathbf{D}$ given parameters $\theta$, $\pi(\theta)$ are priors on
the parameters, $Z=\int d\theta \pi(\theta){\cal L} $ is the model
likelihood or `evidence' and $P(\theta)$ is the sought-after posterior
distribution of the parameters.

We perform posterior sampling using the `nested sampling' algorithm,
which is technique proposed by~\citet{skilling2004}, principally for
estimation of the evidence $Z$ with posterior sampling as a
by-product. We have chosen this algorithm because it is well-suited
for sampling likelihoods that are multi-modal, strongly degenerate, or
non-Gaussian.  To briefly summarise the nested sampling method, the
multi-dimensional integral $Z$ is remapped onto a particular one
dimensional integral.  A key ingredient for this kind of sampling is
then the ability to draw uniform random samples from the remaining
region of parameter space delimited by an iso-likelihood surface.  We
made use of the publicly available implementation {\sc MultiNest}
\citep{feroz2008,feroz2009}, in which those regions bounded by
iso-likelihood contours are approximated by ellipsoids
\citep{mukherjee2006}. We run {\sc MultiNest} in a configuration with
500-1000 `live points' and with a relatively low sampling efficiency
of around $10^{-3}$. The posterior samples are then analysed with the
{\sc getdist} package~\citep{2002PhRvD..66j3511L} in order to extract
two-dimensional and one-dimensional marginalised constraints.

\subsection{IGM parametrization and priors}\label{sec:param}

The next step in our analysis is to parametrize the redshift evolution
of the quantities we wish to constrain.  We therefore investigate a
model for the possible redshift dependence of the IGM parameters in
which $T_0$, $\gamma$ and $\tau_{\rm eff}$ are allowed to vary as
piecewise constants centred on the redshifts $z=[2.1,2.5,2.9]$, and
which will be referred to as the `redshift bins parametrization'.

We have put wide flat priors on the IGM parameter ranges, $5 <T_0/{\rm
  10^3K}<50$, $0.5 < \gamma < 1.7$, $0.075 < \tau_{\rm eff}[z=2.1] <
  0.2$, $0.13 < \tau_{\rm eff}[z=2.5] < 0.3$, and $0.2 < \tau_{\rm
  eff}[z=2.9] < 0.5$; here the $\tau_{\rm eff}$ prior approximates and
  encompasses the ranges allowed from Figure 13 of~\citet{kim2007}.
  For the cosmological parameters, we have imposed the flat priors
  $0.6 < \sigma_8 < 1.0$, $0.9 < n_s < 1$ and $0.20 < \Omega_{\rm m} <
  0.32$, and $60 <H_0<84$ which is intended to be a conservative range
  encompassing the region favoured by WMAP~\citep{larson2010} and
  $H_0$ constraints~\citep{freedman2001}.

Note that more restrictive power-law parametrizations for $T_0(z)$,
$\gamma(z)$ and $\tau_{\rm eff}(z)$ have been investigated by a number
of authors~\citep{viel2009,bird2010}. Owing to the potentially complex
and uncertain phenomenology suggested by theoretical models of the
IGM, we have opted for the more general redshift bins parametrization.

\subsection{The effect of astrophysical and cosmological parameters on the wavelet PDF}

\begin{figure*}
  \centering
  \includegraphics[width=0.55\columnwidth,angle=270]{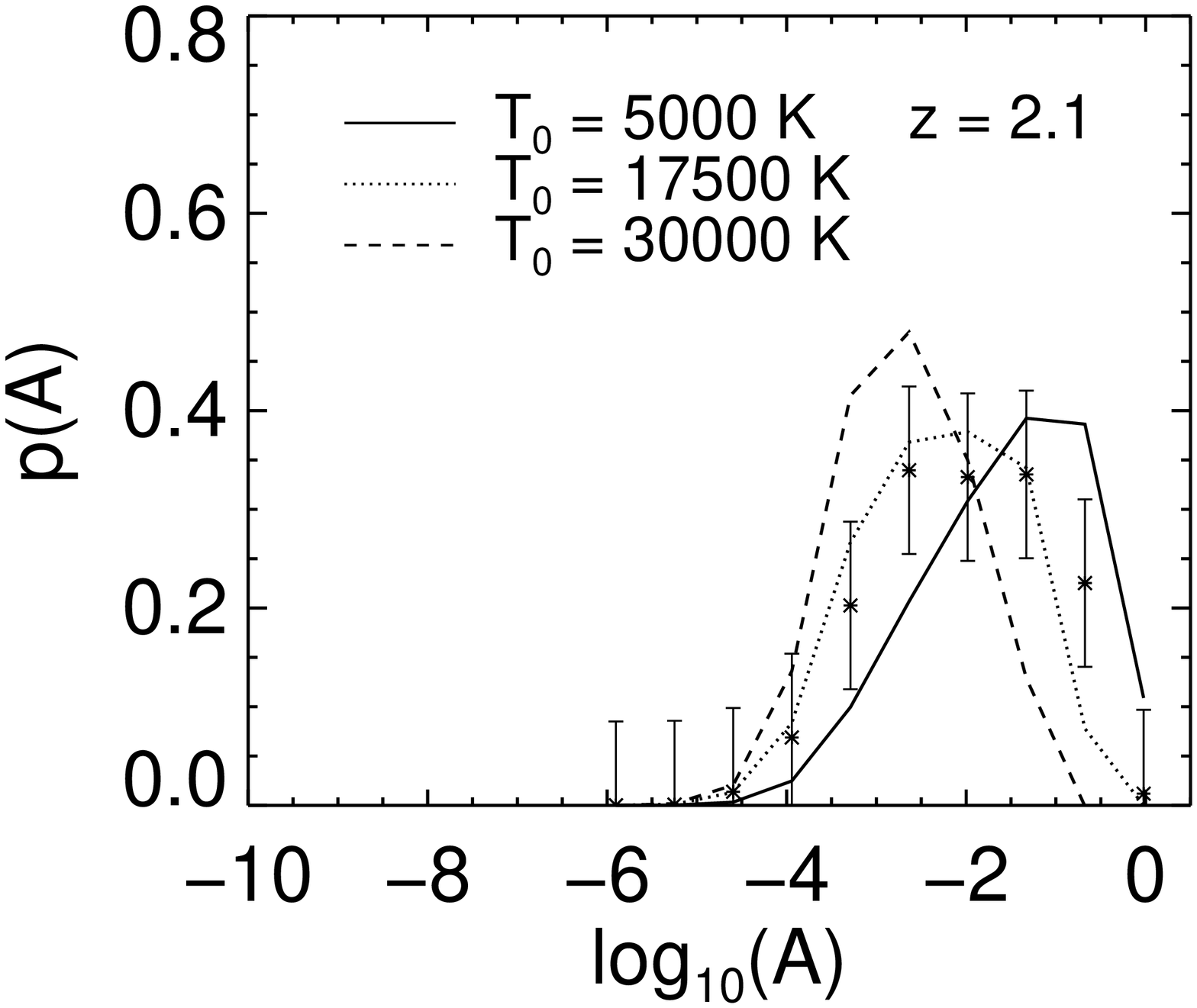}
  \includegraphics[width=0.55\columnwidth,angle=270]{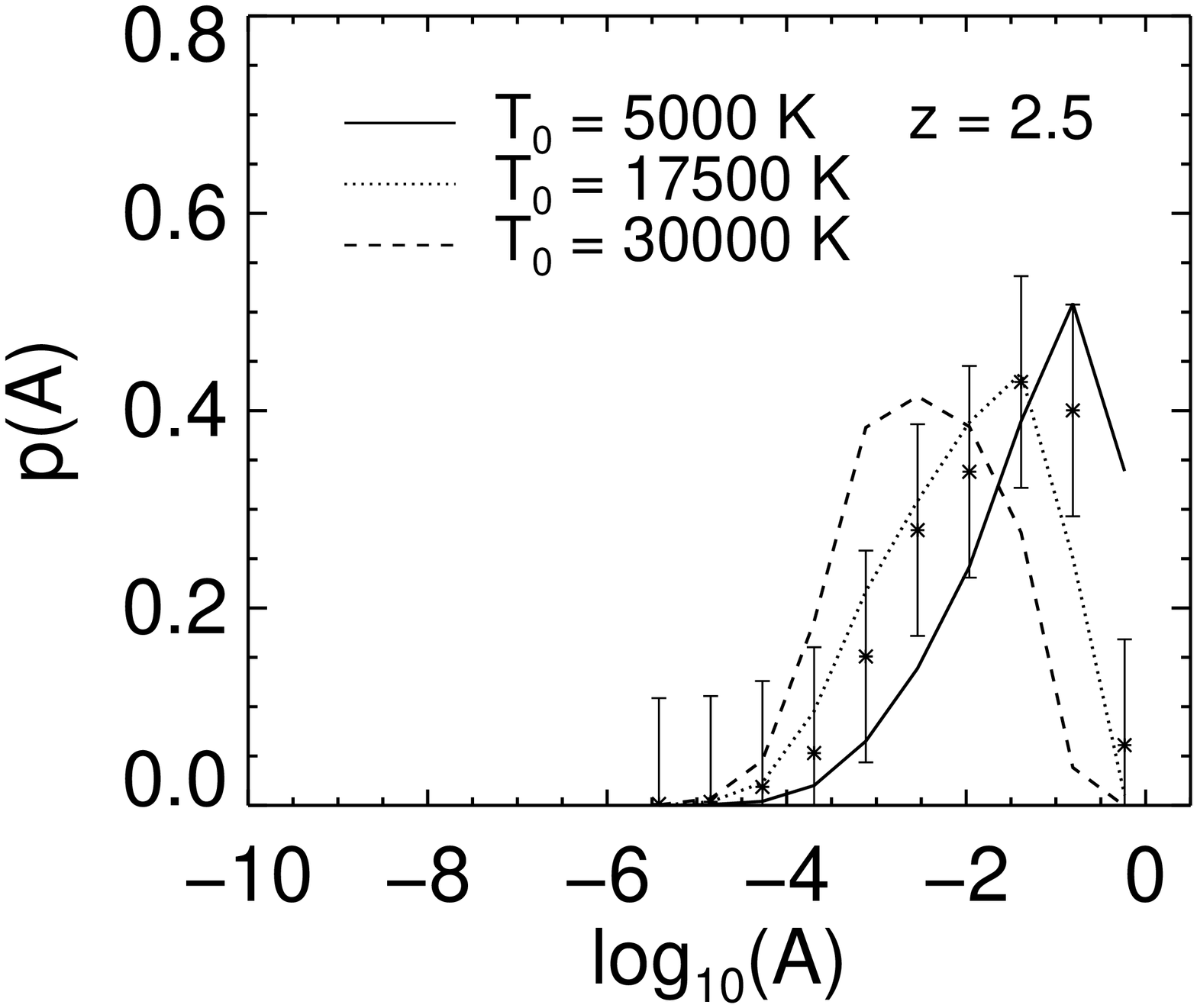}
  \includegraphics[width=0.55\columnwidth,angle=270]{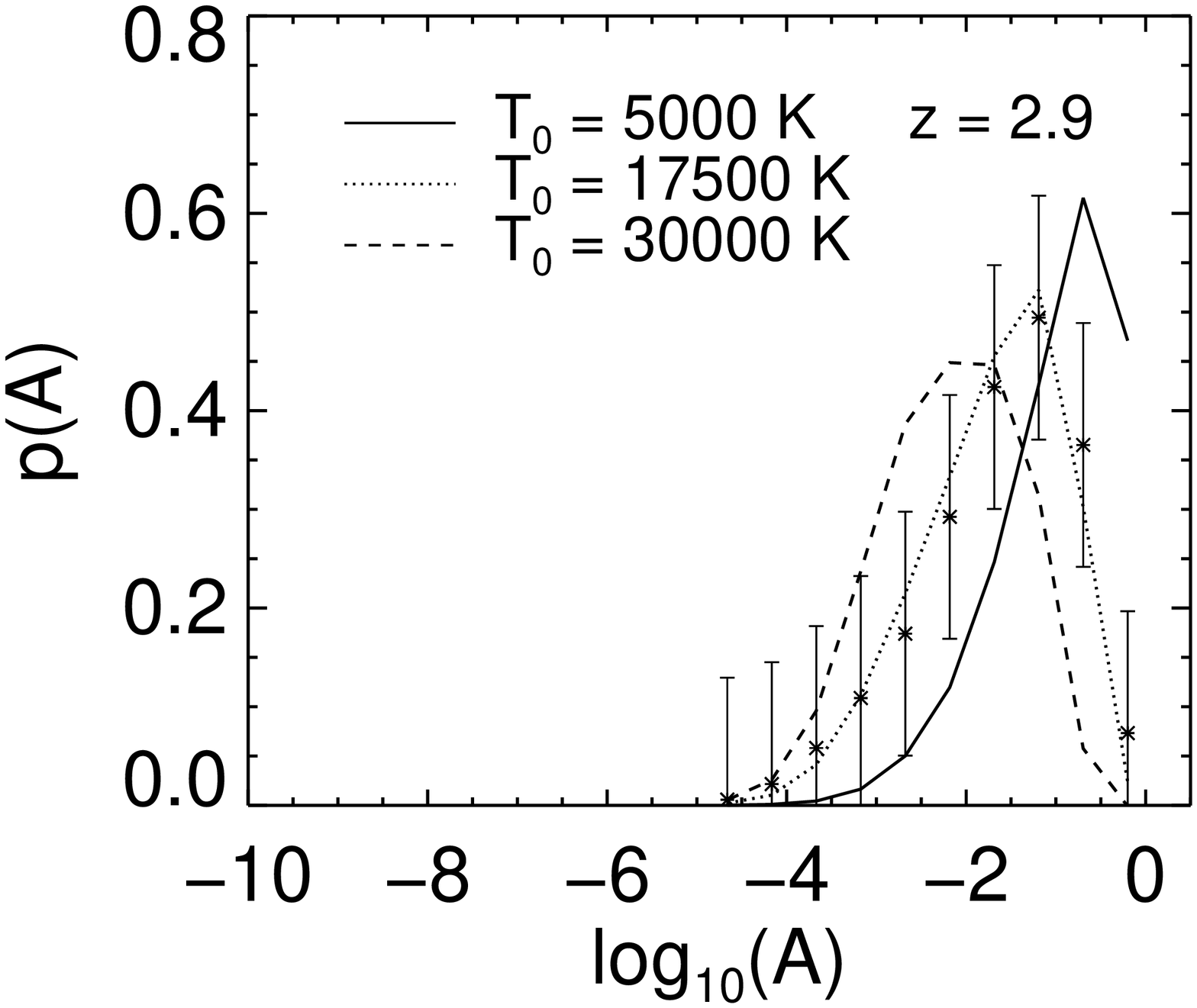}
  \includegraphics[width=0.55\columnwidth,angle=270]{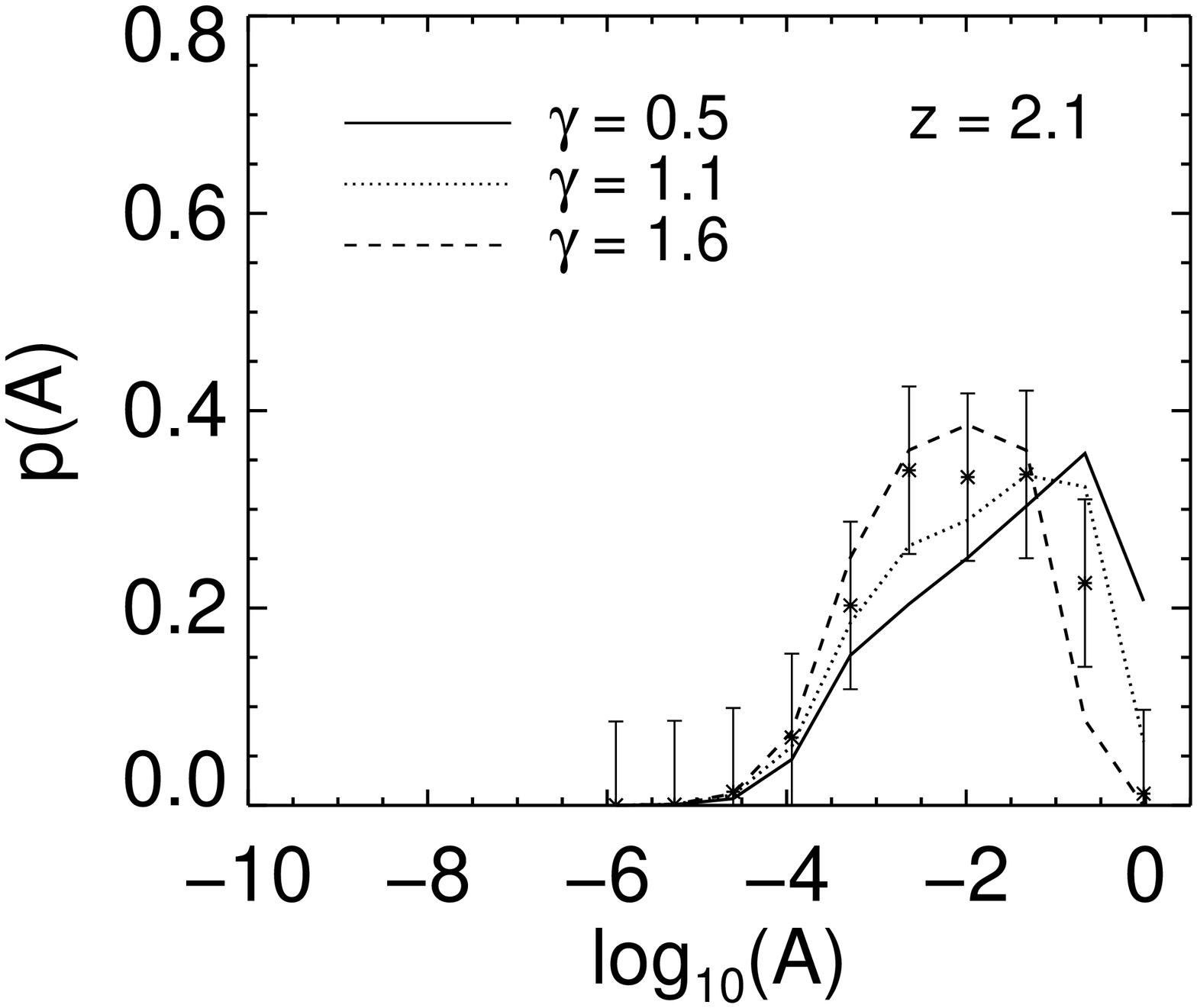}
  \includegraphics[width=0.55\columnwidth,angle=270]{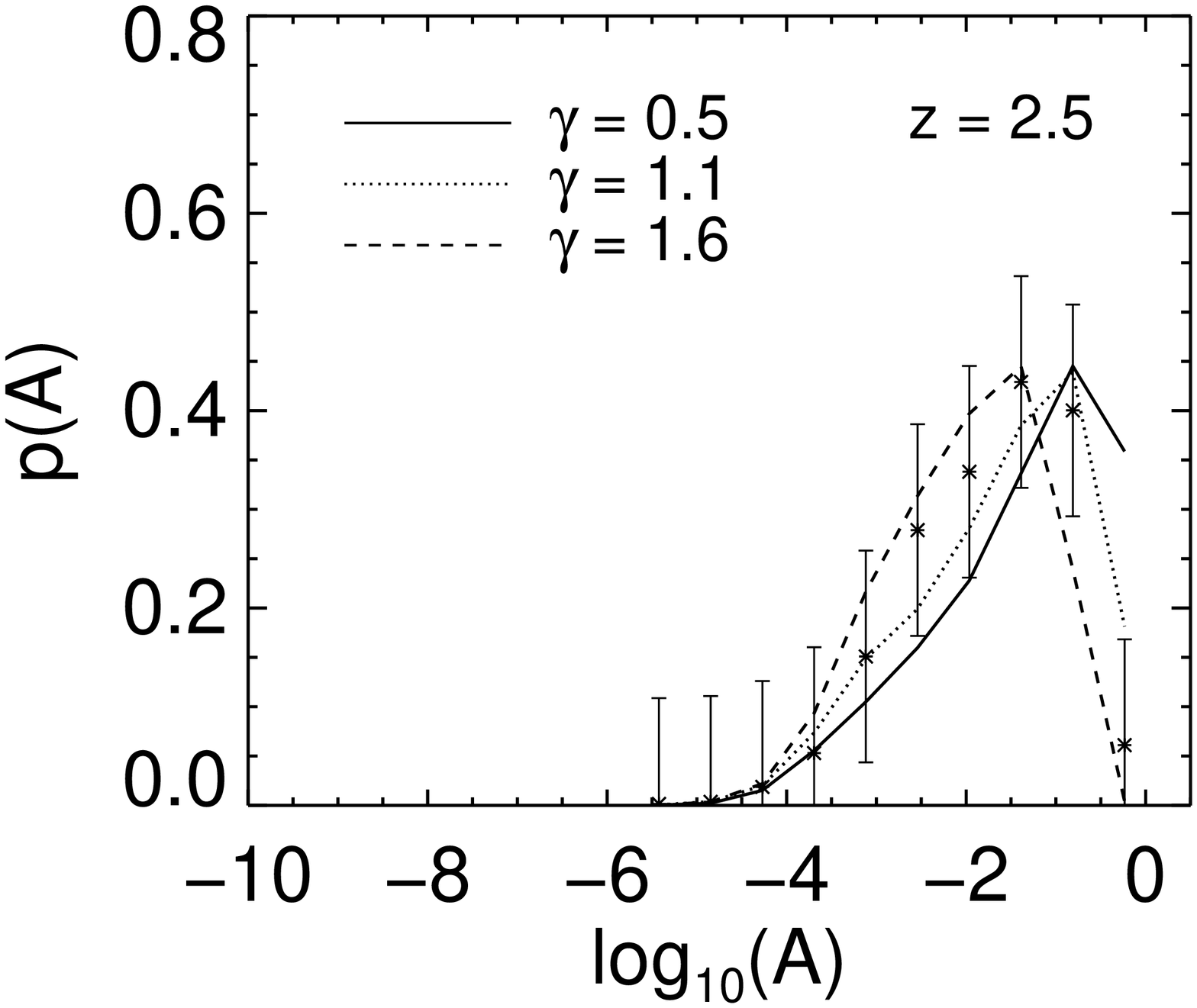}
  \includegraphics[width=0.55\columnwidth,angle=270]{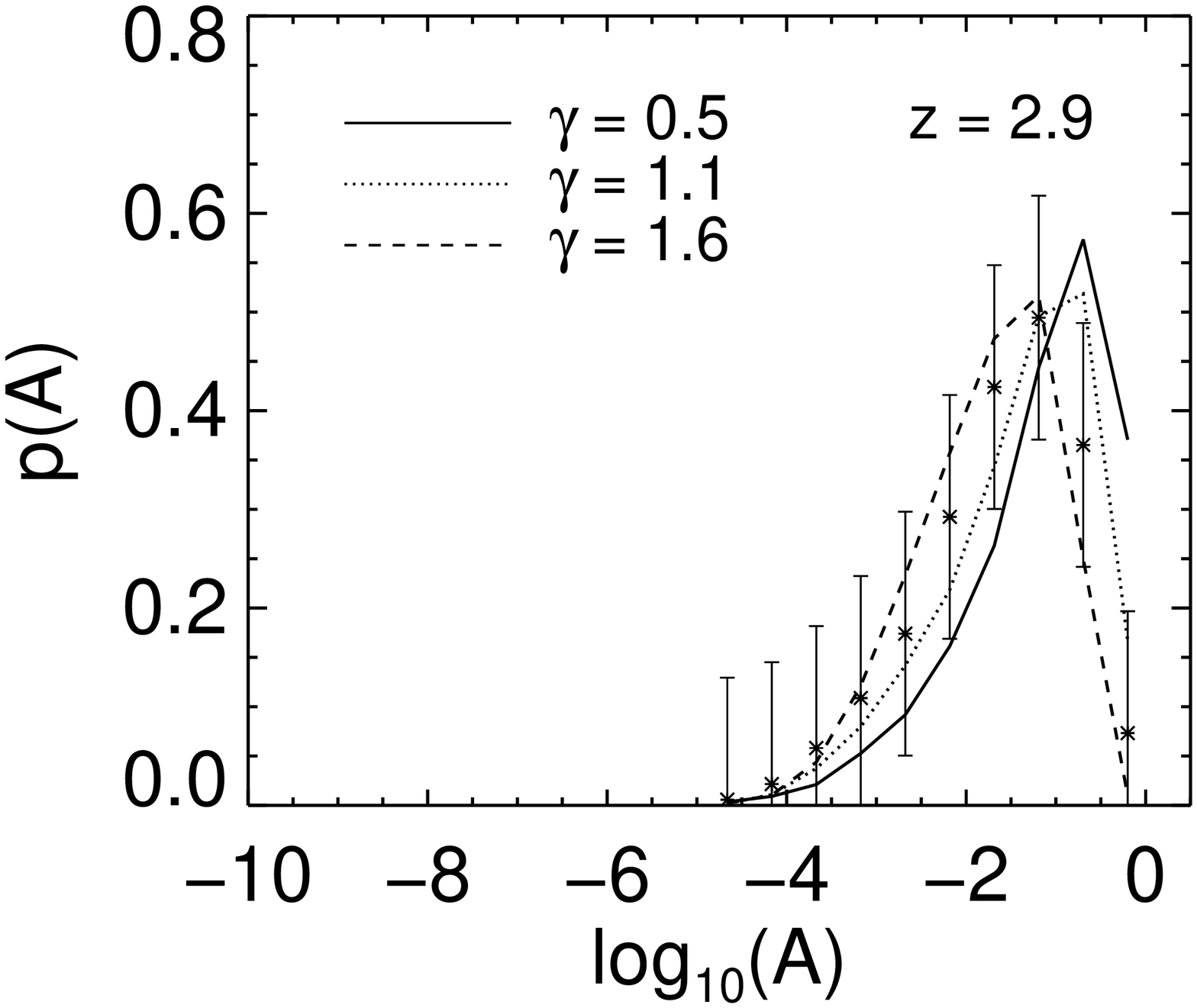}
  \includegraphics[width=0.55\columnwidth,angle=270]{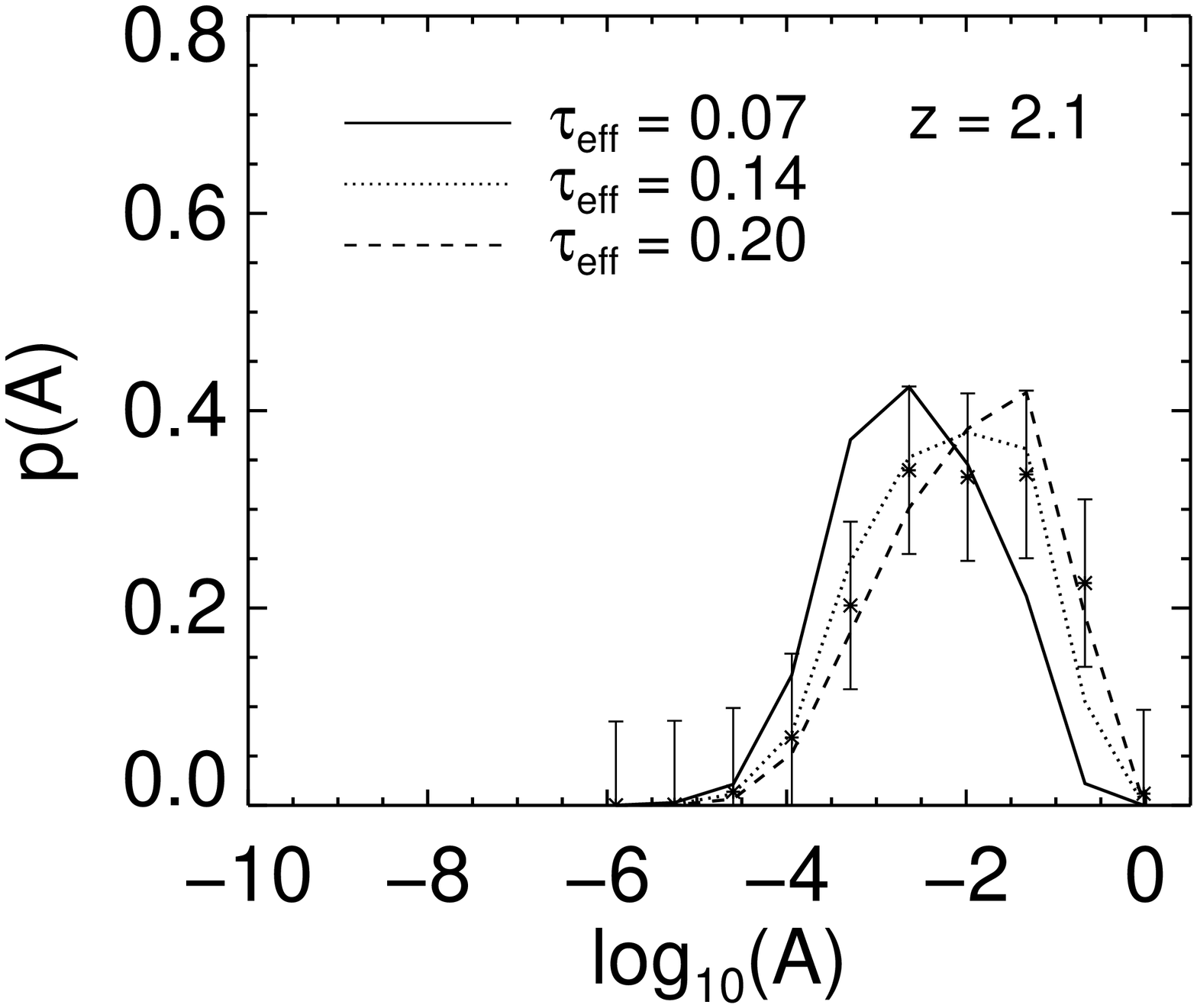}
  \includegraphics[width=0.55\columnwidth,angle=270]{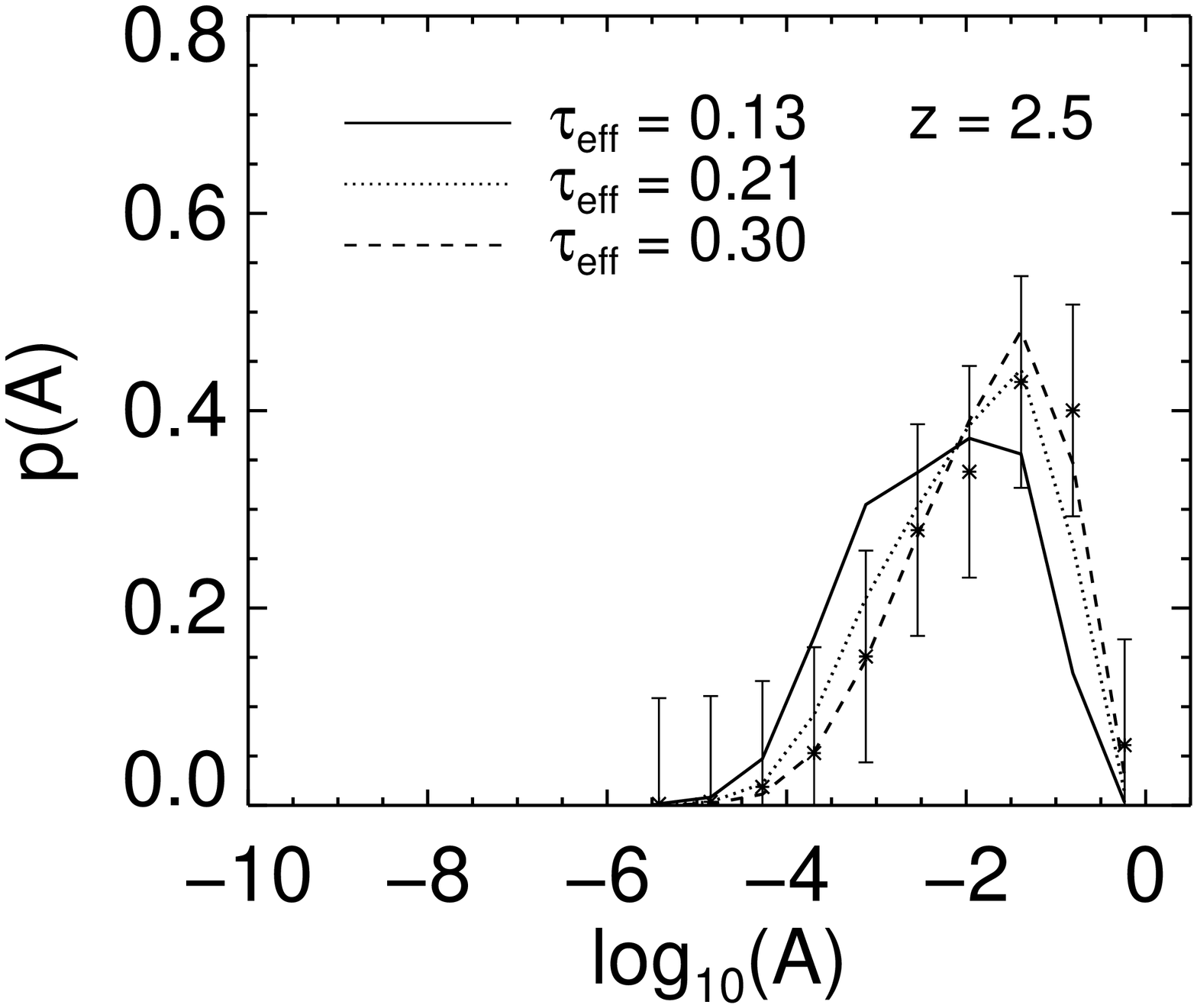}
  \includegraphics[width=0.55\columnwidth,angle=270]{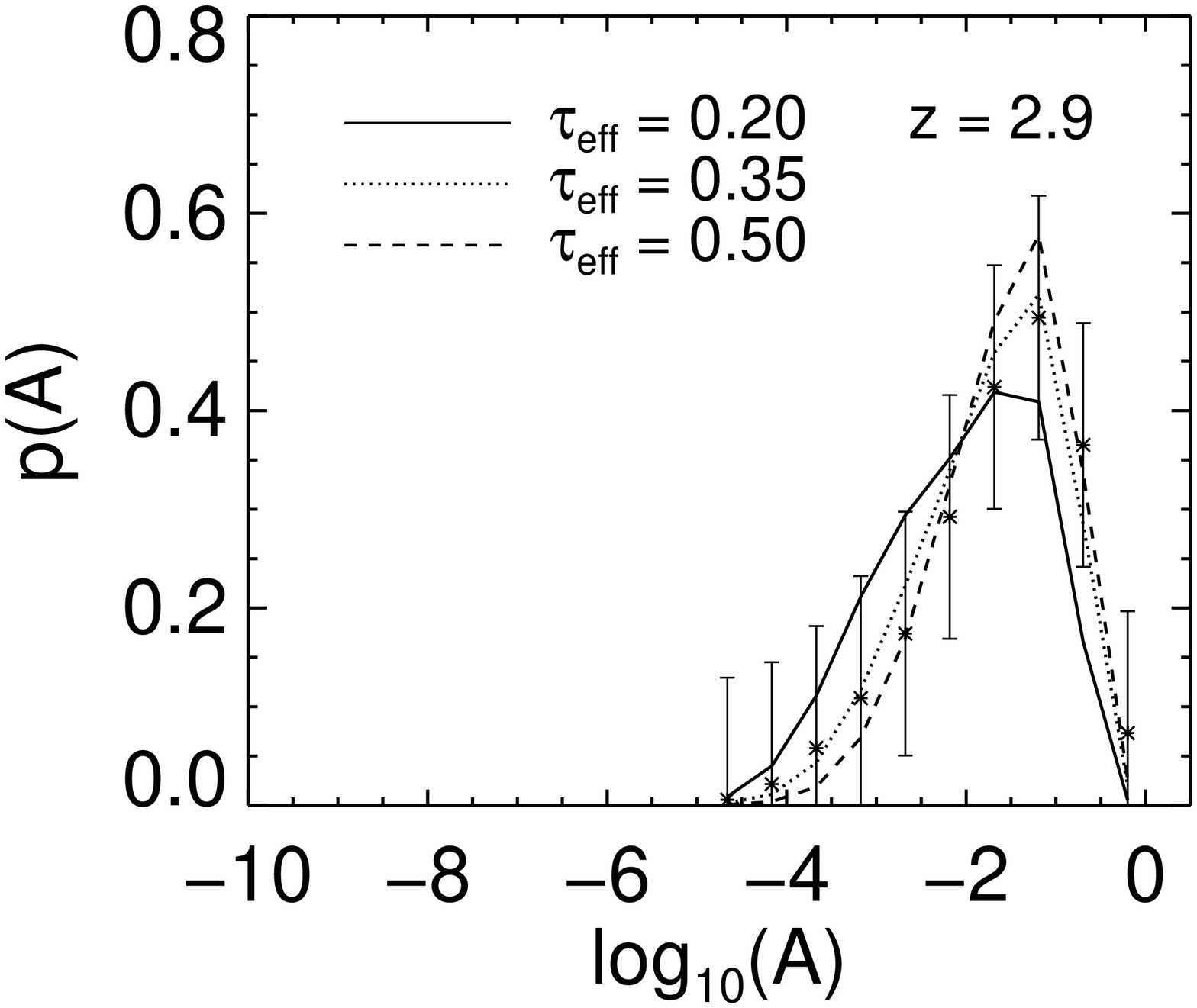}
  \caption{The predicted wavelet amplitude PDF at redshifts $z=[2.1,\,
      2.5,\, 2.9]$ (left to right columns) for our fiducial
    cosmological parameters. Top panels: varying $T_0$. The wavelet
    PDF shifts to lower values for higher temperatures owing to the
    suppression of power at small scales. Middle panels: varying
    $\gamma$. Here $\gamma$ defines the slope of the
    temperature-density relation, and a higher value for $\gamma$
    leads to a higher value for the temperature at fixed $T_0$ in the
    overdense gas predominantly probed by the Lyman-$\alpha$ forest at
    $z<3$. Bottom panels: varying $\tau_{\rm eff}$. Increasing
    $\tau_{\rm eff}$ decreases the mean transmitted flux and shifts
    the characteristic density probed by the Lyman-$\alpha$ forest to
    lower values.  These regions are cooler for our fiducial
    $\gamma=1.6$ than more overdense regions, leading to shift towards
    higher values of the wavelet amplitude.}
  \label{fig:pdf_thermal_1}
\end{figure*}

Having described our IGM parametrization and implemented our
interpolation scheme, we are now in a position to demonstrate how the
wavelet amplitude PDF depends on the astrophysical parameters, $T_0$,
$\gamma$ and $\tau_{\rm eff}$, before proceeding to present our
temperature constraints.  This is illustrated in
Figure~\ref{fig:pdf_thermal_1}, from which we may confirm the
phenomenology for the wavelet amplitude PDF found in~\citet{lidz2010}:
the wavelet amplitude PDF shifts to smaller values for higher
temperatures owing to Doppler broadening and Jeans smoothing which
suppresses small scale power.  Higher values of $\gamma$ also shift
the peak of the wavelet PDF to lower amplitudes at fixed $T_0$ due to the
decreased thermal smoothing associated with absorption from overdense
gas, which dominates the Lyman-$\alpha$ absorption at $z<3$
(e.g. \citealt{becker2010}).  Increasing $\tau_{\rm eff}$ lowers the
characteristic gas density probed by the Lyman-$\alpha$ absorption,
shifting the peak amplitude of the wavelet PDF to higher values due to
the colder underdense gas present for our fiducial $\gamma=1.6$.
Clearly the uncertainty in $\tau_{\rm eff}$ needs to be marginalised
over in order to estimate $T_0$ and $\gamma$, as any physical effect
that affects the power spectrum of the Lyman-$\alpha$ absorption lines
at small scales $k>0.1$ s/km will also substantially affect the
wavelet amplitude PDF.

Finally, the effect of the cosmological parameters on the wavelet
amplitude is found to be weak, except for $\sigma_8$ that has an
slight impact on the wavelet PDF at redshift $z=2.1$ as we demonstrate
in Figure~\ref{fig:pdf_cosmo} of Appendix~\ref{app:cosmo}. The effect
appears not to be as simple as a shift in the wavelet amplitude PDF as
might be expected for a change in the power spectrum, as argued
by~\cite{lidz2010}. Our explicit simulations of the effect of varying
$\sigma_8$ show a change in the width of the wavelet PDF in the lower
redshift bin, where structure formation is more advanced.

\section{Results}
\label{sec:results}

\begin{figure*}
  \centering
  \includegraphics[width=0.65\columnwidth]{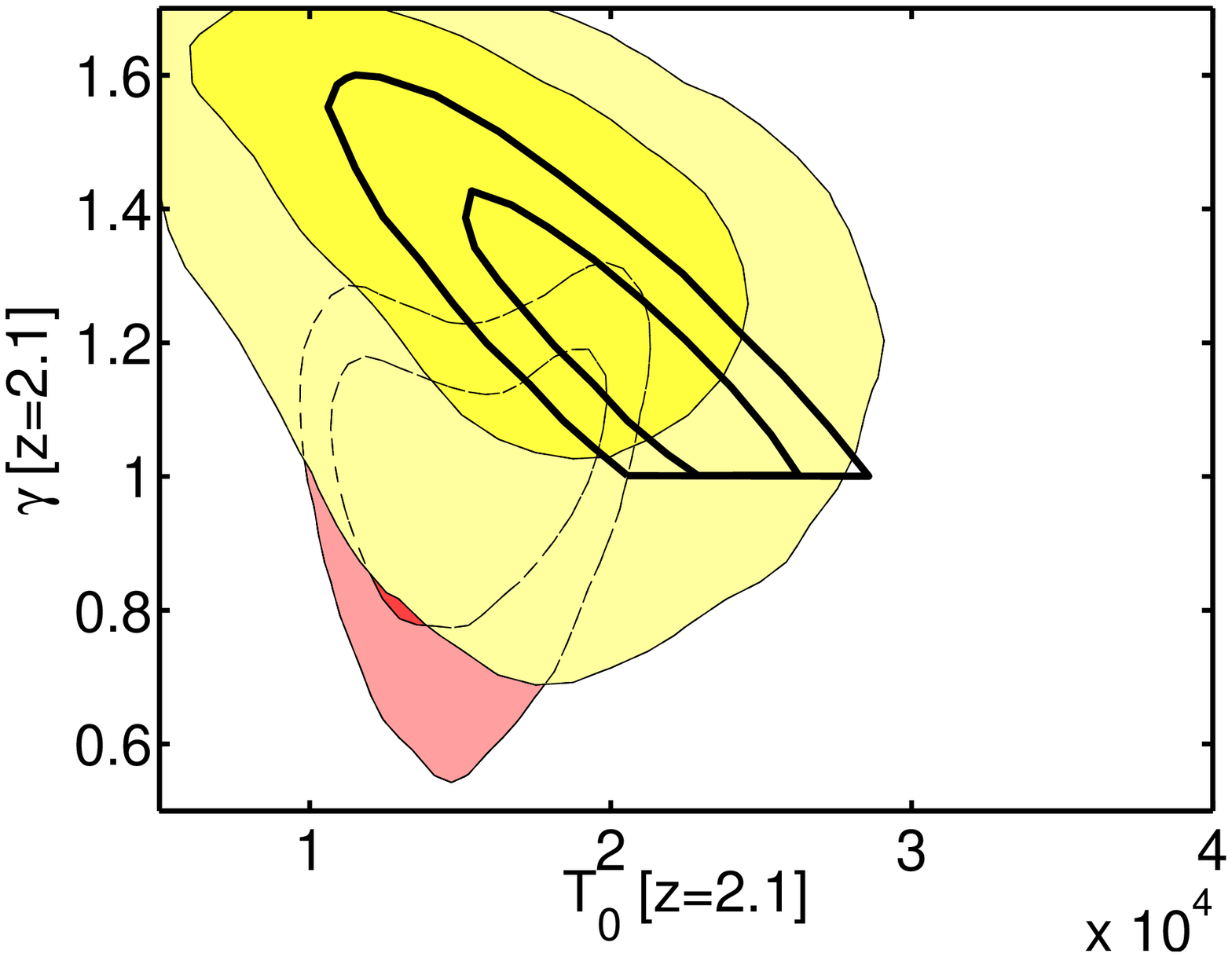}
  \includegraphics[width=0.65\columnwidth]{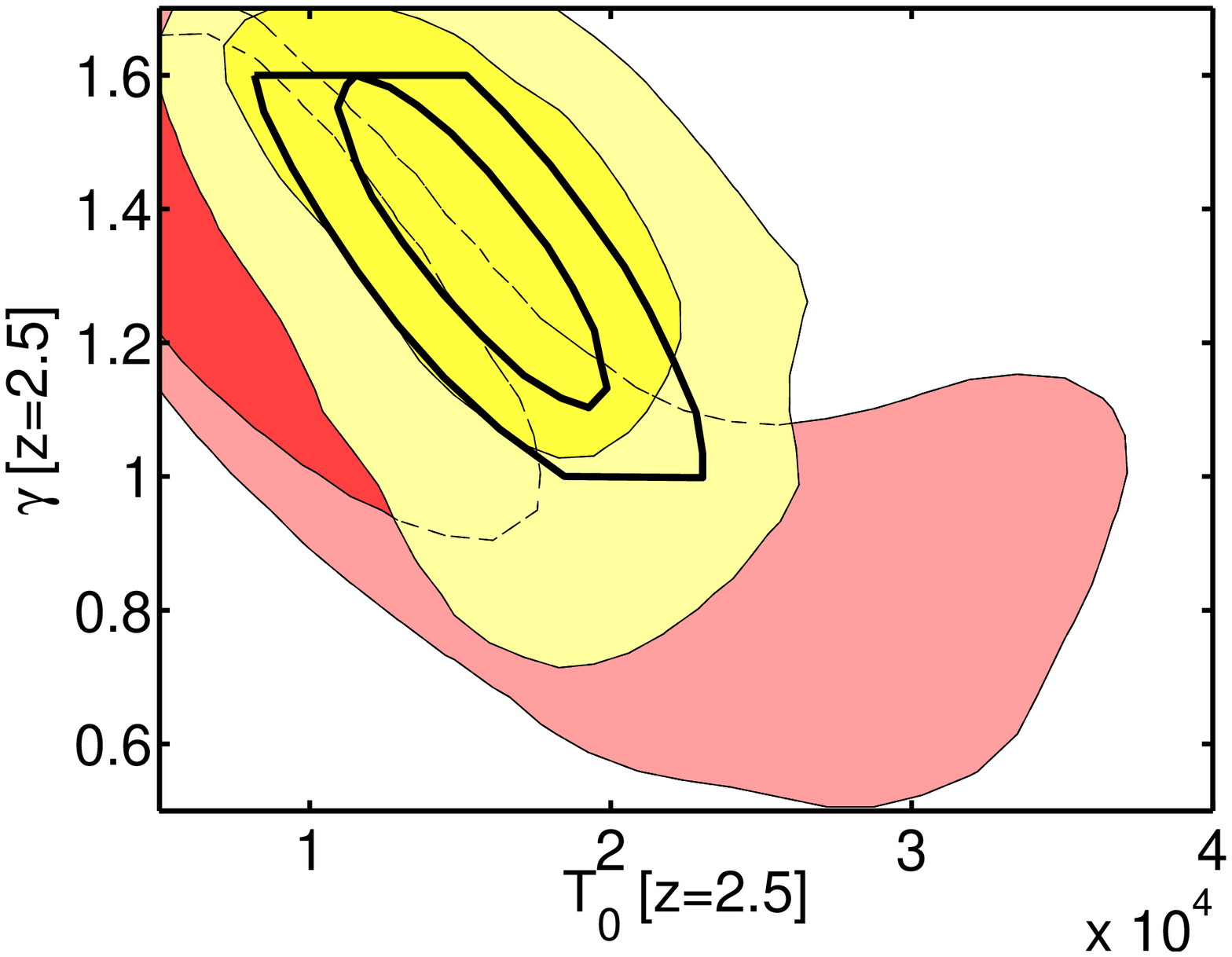}
  \includegraphics[width=0.65\columnwidth]{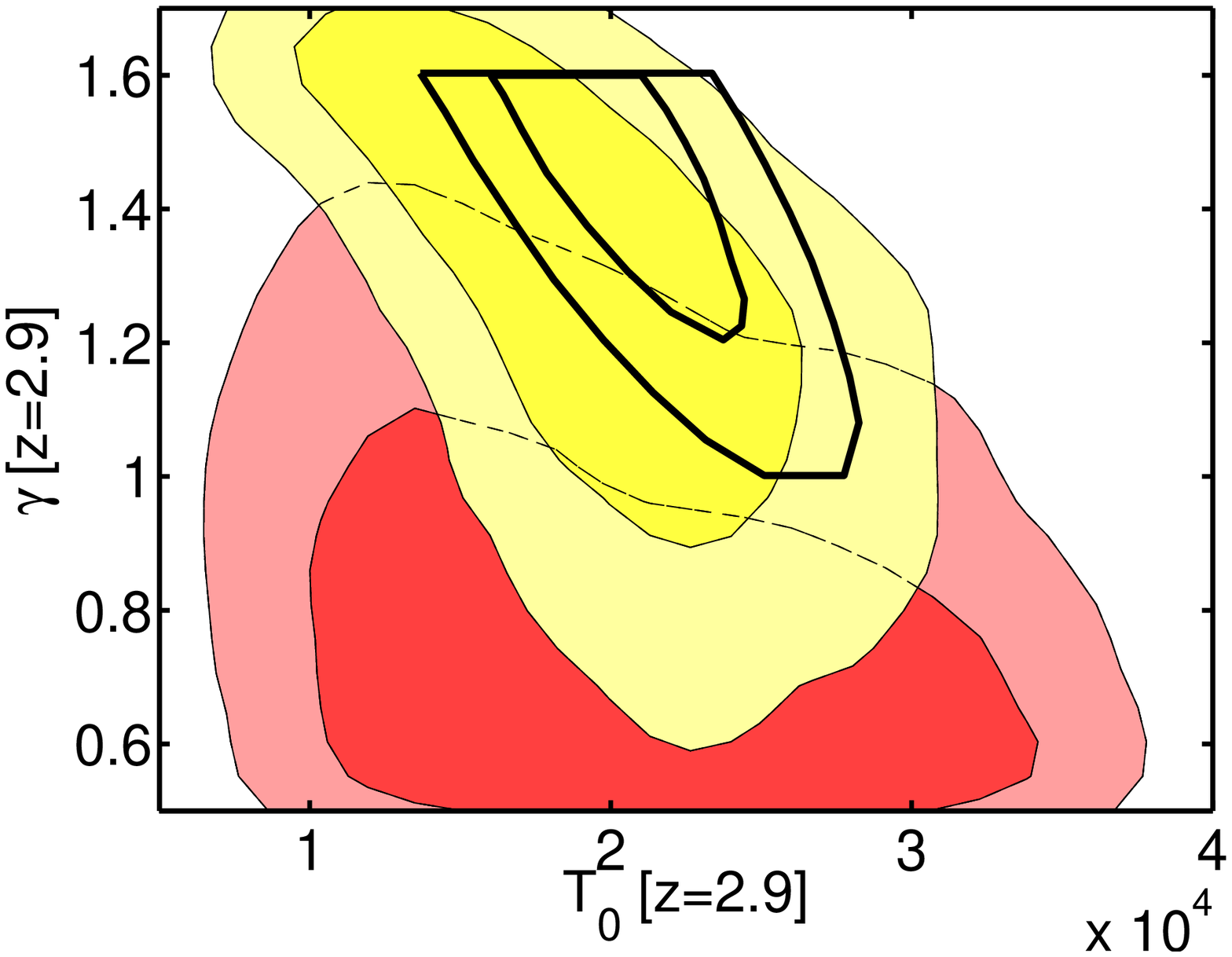}
  \includegraphics[width=0.65\columnwidth]{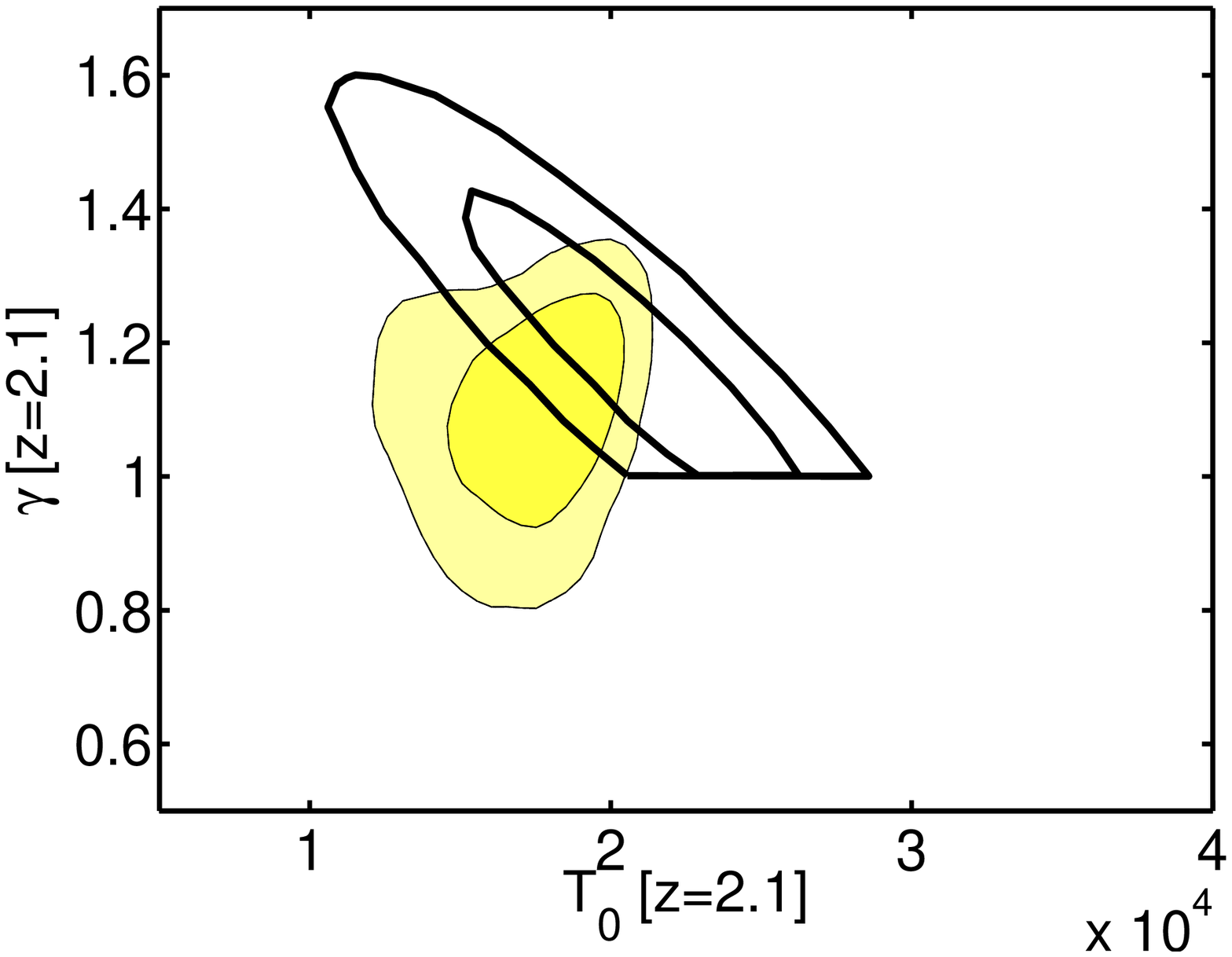}
  \includegraphics[width=0.65\columnwidth]{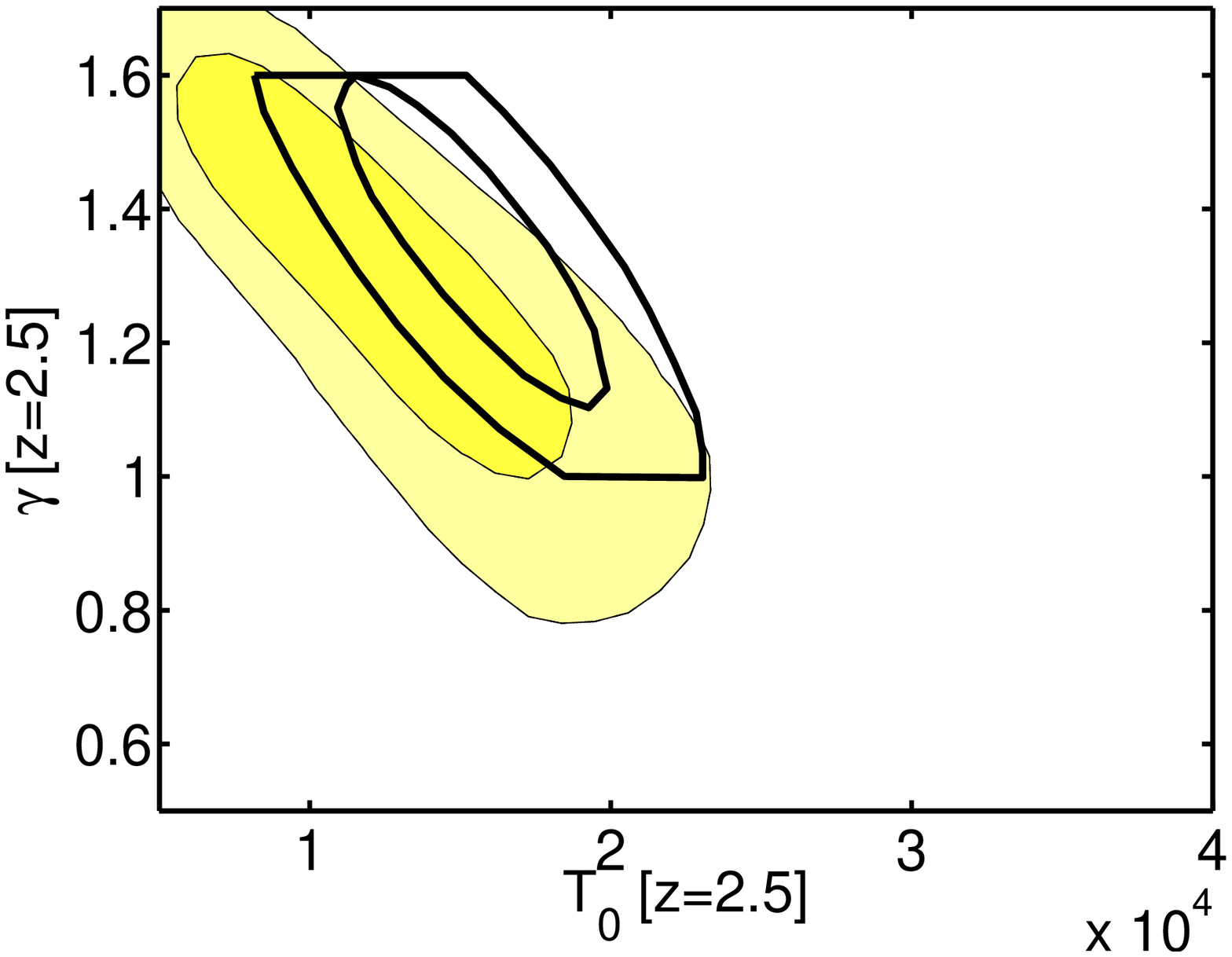}
  \includegraphics[width=0.65\columnwidth]{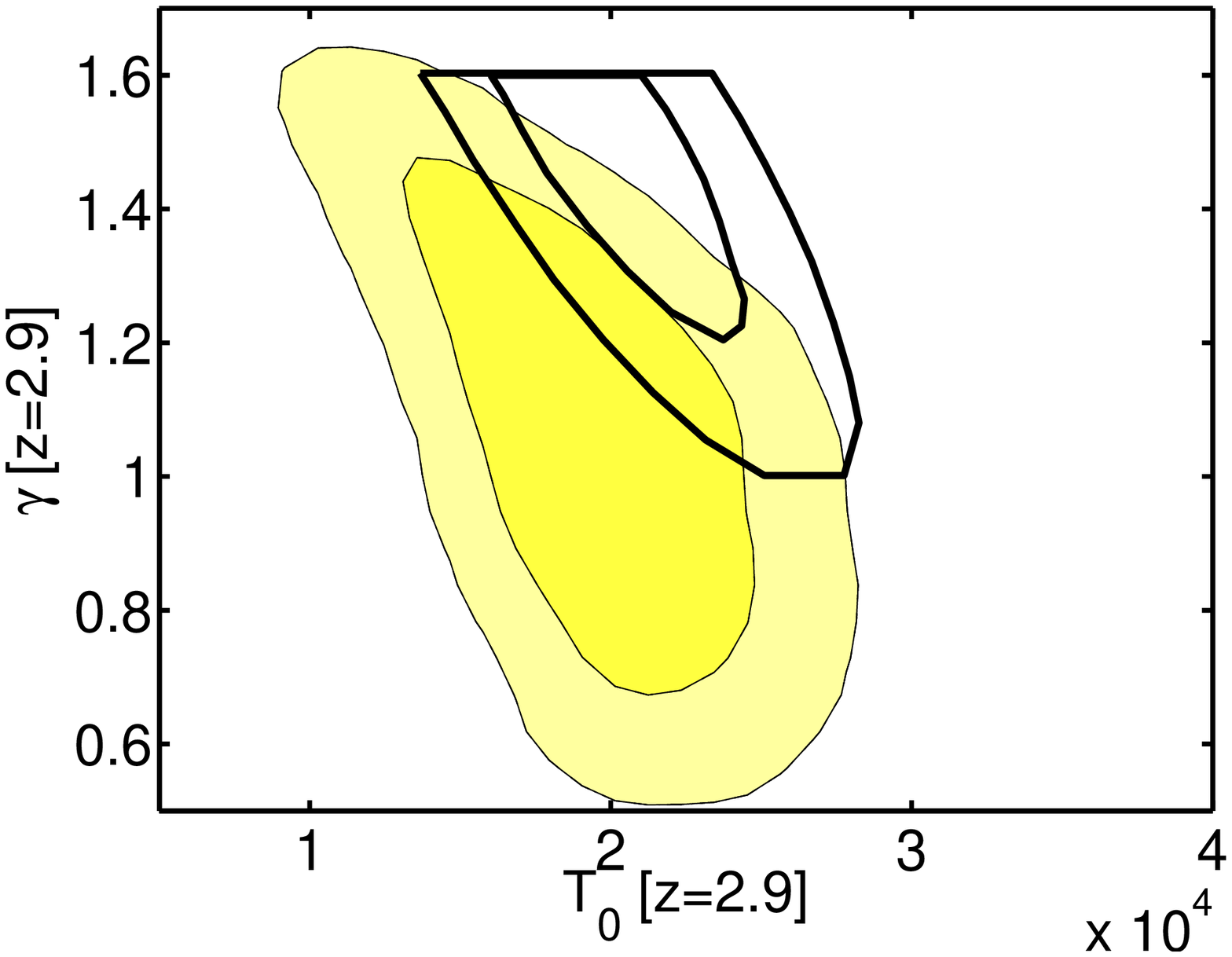}
  \caption{Upper panels: constraints on the IGM thermal parameters $T$
    and $\gamma$ from our implementation of wavelet PDF analysis
    (yellow, light filled contours), compared with our reanalysis of
    the flux PDF from~\citet{viel2009} (red, dark filled contours) and
    the 1$\sigma$ and 2$\sigma$ constraints from~\citet{lidz2010}
    which assume a prior $\gamma=1.0-1.6$ (black contours). Lower panels:
    the joint constraints from the wavelet and flux PDF (filled
    contours) again compared to the \citet{lidz2010} wavelet PDF
    constraints.}
  \label{fig:t0_gamma}
\end{figure*}

We now turn to describing the results of our analysis.  Briefly, to
restate the main aim of this work, we wish to compare the IGM
cosmological and astrophysical constraints derived from our reanalysis of the
Lyman-$\alpha$ flux PDF of~\citet{kim2007} with our new constraints
derived from the wavelet amplitude PDF.  We apply these two
methodologies to the same dataset for the first time, enabling us to
explore any systematic differences between these approaches.  The
computation of the flux PDF likelihood is performed in a similar way
to~\citet{viel2009}, using a second-order Taylor expansion of the
cosmological and astrophysical parameter space.  We have extended
the~\citet{viel2009} analysis of the \cite{kim2007} PDF to use our
redshift bins parametrization, thereby dividing the data set into
three redshift bins as opposed to using a single power-law
parametrization as in~\citet{viel2009}.  Following \cite{viel2009}, we
also fit the simulations to the observed flux PDF in the flux range
$F=0.1$--$0.8$ only, in order to minimise the effect of continuum
uncertainties on our results.

\subsection{Constraints on the IGM thermal history} \label{sec:constraints}

Our main results are shown in Figure~\ref{fig:t0_gamma}, which
displays the 1$\sigma$ and 2$\sigma$ contours obtained from our
analysis of the wavelet PDF (yellow contours) and the flux PDF (red
contours).  Firstly, we note that the values of $T_0$ and $\gamma$
inferred from our analysis of the wavelet and flux PDF are in broad
agreement with each other, and are formally consistent within
1$\sigma$.  In part this is because in this analysis we have, in the
first instance, left $\tau_{\rm eff}$ as free parameter, which
enlarges the parameter space consistent with the wavelet PDF due to
the degeneracy between $\tau_{\rm eff}$ and $T_0$.  It is also
apparent, however, that the flux PDF constraints generally favour a
lower value for the temperature-density relation slope, $\gamma$,
compared to the wavelet PDF (e.g. \citealt{bolton2008,viel2009}).
This implies there may be a systematic difference between the two
methodologies.

Motivated by recent suggestions about possible systematics in the flux
PDF due to continuum
placement~\citep{2011arXiv1103.2780L,2011MNRAS.415..977M} we have
investigated whether the wavelet and flux PDF show any sensitivity to
the continuum level assumed. We performed a check by lowering the
continuum level on the synthetic spectra by 3 per cent
(e.g. \citealt{Tytler04,Faucher08}), and then recalculating both the
flux and wavelet PDF for our $z=2.98$ simulations, for two models with
$T_0[10^3$K$]=18$, $\tau_{\rm eff}=0.350$ and $\gamma=1.6$ and $0.7$
respectively. Figures showing the results from these tests are shown
in Figure~\ref{fig:continuum_effect} in the appendix.  We conclude
that while the flux PDF shows some sensitivity to the continuum level
at high ($F>0.8$) and low ($F<0.1$) fluxes (data that have not been
used in our fit), the overall change in the fit is small compared to
the improvement in the fit from moving (for example) from $\gamma=1.6$
to $\gamma=0.7$.  It therefore appears that continuum errors do not
fully explain the tendency for the flux PDF to favour somewhat lower
values of $\gamma$ at $z=3$.

For comparison, we also show in Figure~\ref{fig:t0_gamma} results from
the wavelet PDF analysis of~\cite{lidz2010} extracted from their
Figure 25.  The three \cite{lidz2010} redshift bins are
$z=[2.1,\,2.6,\,3.0]$ which roughly correspond to our own redshift
bins.  Note that~\citet{lidz2010} imposed a prior $\gamma=1.0-1.6$ and
regarded their constraints as approximate, cautioning against taking
their results too literally. Therefore, in interpreting their results
we will attempt to make an allowance for the caveat they expressed by
examining their 2$\sigma$ constraints. \cite{lidz2010} have analysed
just over double the number of spectra used in this study by using the
40 spectra reduced by~\cite{2008A&A...491..465D}; sixteen of our
eighteen spectra also appear in their sample. Our wavelet PDF error
bar treatment is probably the more conservative, explaining why our
constraints are somewhat looser.  In general, however, we find very
good agreement with the~\cite{lidz2010} constraints, which is
encouraging given the independently reduced observational data set and
different simulation method used by these authors.

\begin{figure}
  \centering
  \includegraphics[width=\columnwidth]{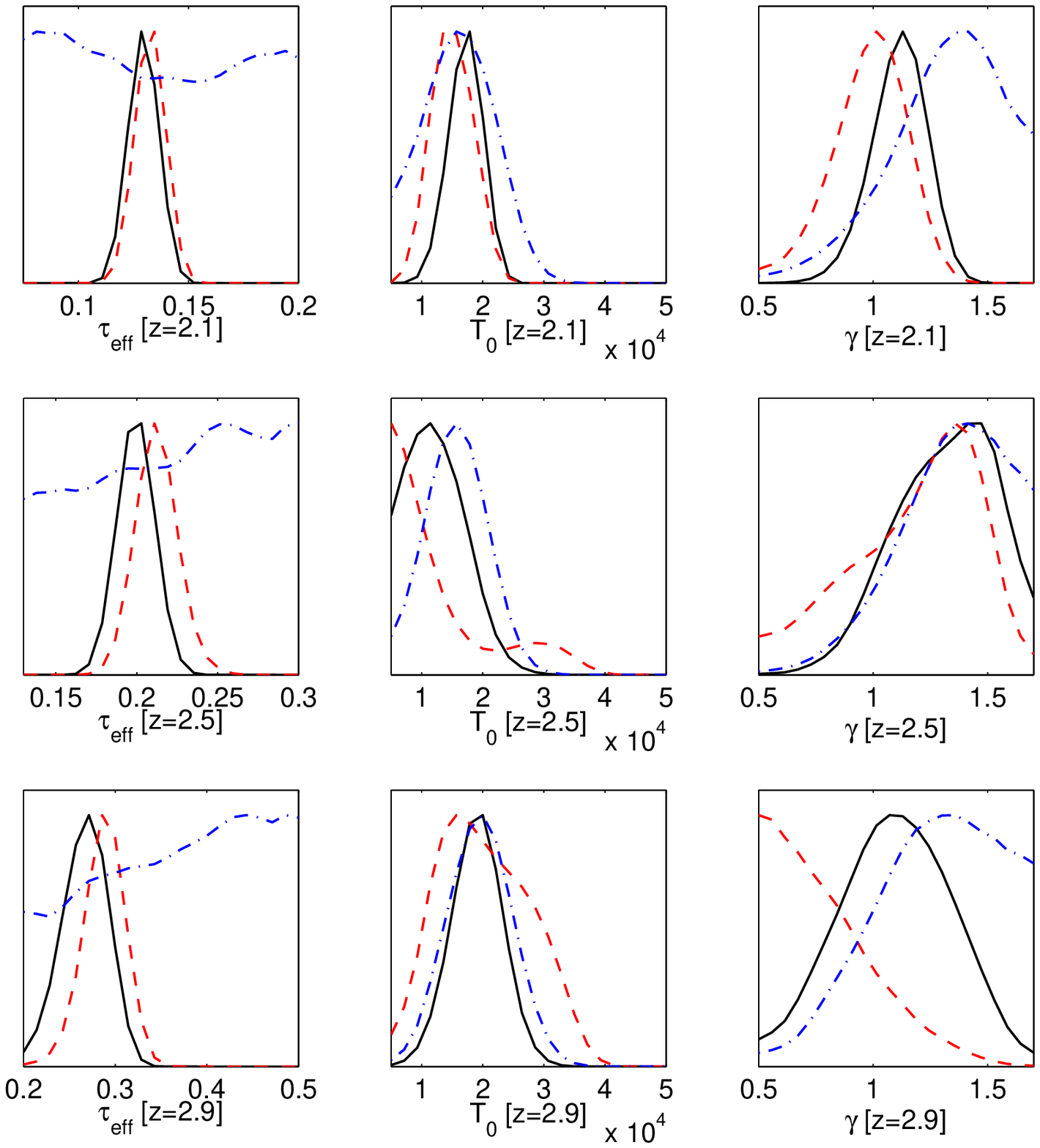} 
  \includegraphics[width=0.66\columnwidth]{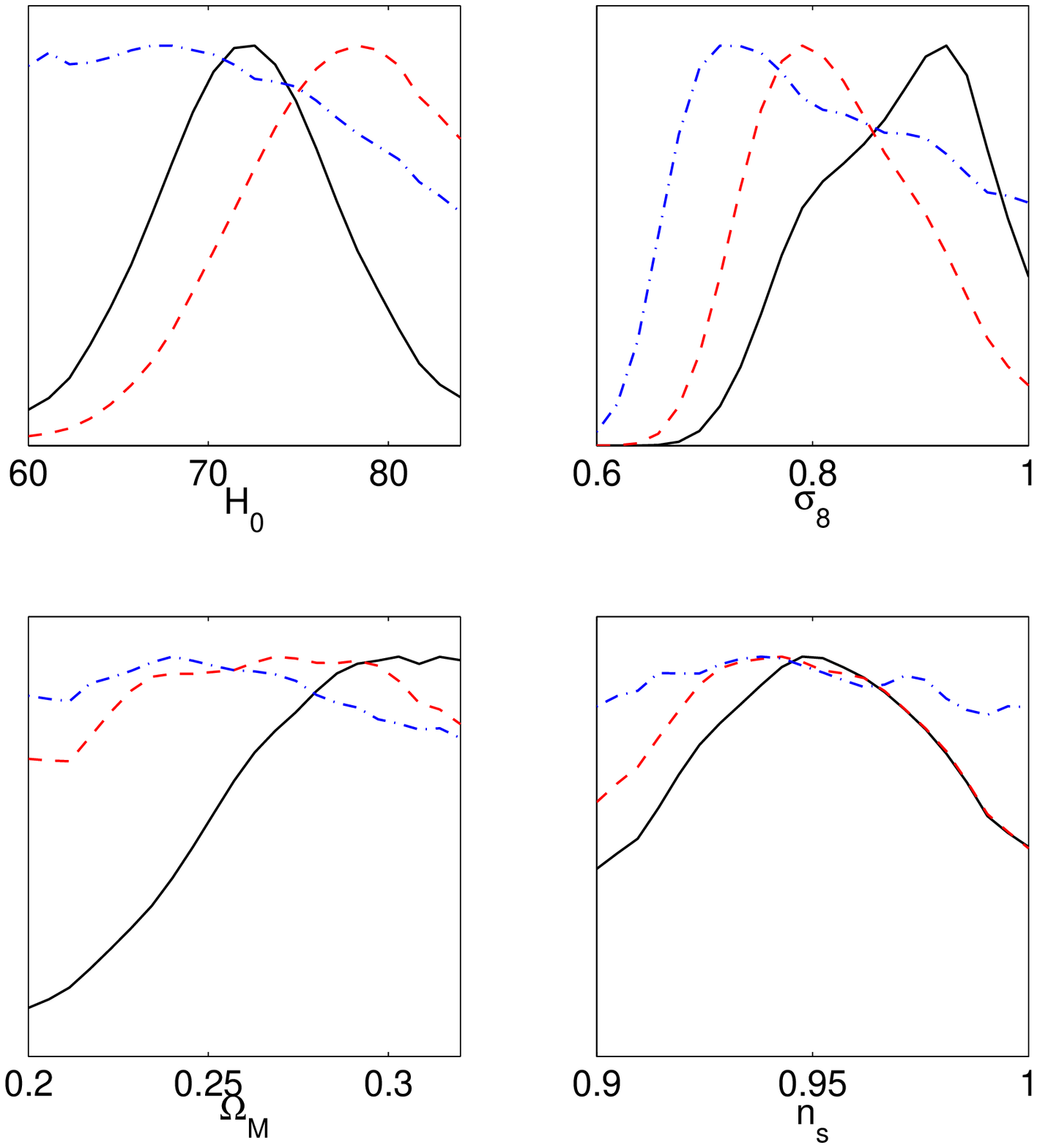} 
  \caption{One dimensional marginalised constraints, corresponding to
    the two dimensional constraints shown in
    Figure~\ref{fig:t0_gamma}. Note the constraints are shown as
    relative probabilities in each panel. The IGM temperature inferred
    from the wavelet amplitude PDF (dot dashed, blue) is in broad
    agreement with the flux PDF value (dashed red). The cosmological
    parameters are unconstrained, as is $\tau_{\rm eff}$ for the
    wavelet amplitude PDF. The joint constraints are also shown
    (solid, black).}
  \label{fig:t0_gamma_cosmo1d}
\end{figure}

Our one dimensional parameter constraints are shown in
Figure~\ref{fig:t0_gamma_cosmo1d} in which we can compare the results
from the wavelet and flux PDF side by side. It is clear that the
cosmological parameters are unconstrained by the wavelet PDF owing to
their weak effect, with perhaps only a lower bound on $\sigma_8$ being
found.  The flux PDF -- a one point statistic of the unfiltered
spectra -- puts the stronger constraint on the values of $\tau_{\rm
eff}$; the values of $\tau_{\rm eff}$ we find for these redshift bins
agree at the 1-2$\sigma$ level with those determined by
\citet{Faucher08}.  This is in fact our main motivation for performing
a `joint analysis' of the flux PDF and wavelet PDF together (combining
their likelihoods with equal weight), in order to self-consistently
add the $\tau_{\rm eff}$ constraint derived from the flux PDF onto the
parameter space consistent with the observed wavelet PDF. The final
results for the joint analysis are summarised in Table~\ref{tab:bins}
and the lower three panels of Figure~\ref{fig:t0_gamma}. The joint
constraints favour an IGM temperature-density relation which is close
to isothermal, with temperatures at mean density which lie in the
range $10\,000$--$20\,000 \rm\,K$.  However, the 1$\sigma$
uncertainties are too large to infer any significant redshift
evolution in these quantities.

\subsection{Comparison with previous constraints}

Finally, we compare our results to previous constraints in the
literature and consider the implications for the thermal history of
the IGM.  In Figure~\ref{fig:modelcomparison_t0} we present a
comparison of our results to models for the redshift evolution of the
temperature at mean density, $T_0(z)$, from the literature.  We show
our joint constraints on $T_0$ (black error bars) and from the wavelet
PDF alone (red error bars) together with the models of \HeII
reionisation from~\cite{mcquinn2009} (upper panel) and blazar
heating from~\cite{puchwein2011} (lower panel). Specifically, we
have compared the available data with the `D1' and `S4b' models
from~\cite{mcquinn2009}; the latter model implements a harder quasar
UV spectral index of $0.6$ compared to their fiducial $1.6$.  Our
constraints are generally consistent with the \cite{mcquinn2009} model
of \HeII reionisation, with the softer UV spectral index model
preferred.

{The wavelet only constraints tend to favour the weak blazar heating
  model of \cite{puchwein2011}, whose median temperature at mean
  density is shown.  In contrast, \cite{puchwein2011} concluded that
  their intermediate blazar heating was preferred based on a
  comparison of the temperature of their models calculated at the
  `optimal overdensity' (which rises to $\Delta\sim6$ at $z=2$) probed
  by the~\cite{becker2010} curvature constraints.  The differences
  between the blazar heating models are more pronounced at mean
  density; the temperature-density relations predicted by
  \cite{puchwein2011} are similar for all models at $\Delta\geq
  2$--$3$, which partially accounts for why \cite{puchwein2011}
  conclude the intermediate model is favoured.  We note, however, the
  wavelet PDF is sensitive to gas temperatures over a range of
  densities (including the mean density) as it is a {\it distribution}
  rather than a single number (i.e. the curvature statistic used by
  \citealt{becker2010}).  A precise measurement of the
  temperature-density relation could in principle rule out the blazar
  heating model if $\gamma>1$, as the volumetric heating rate used in
  the blazar heating models produce a strongly inverted
  temperature-density relation by $z=2$.

We also compare our constraints on $T_{0}$ with the measurements of
\citet{lidz2010}, also using the wavelet PDF technique, and the
curvature measurements of~\cite{becker2010} (plotted assuming
$\gamma=1.3$, which is within $\sim 1.5 \sigma$ of our joint
constraints in all redshift bins).  Note that although they used a
different measurement technique and data, \cite{becker2010} use the
same set of hydrodynamical simulations as us.  There is good agreement
between the measurements at $z<3$, although there does appear to be
some tension between the~\citet{becker2010} and~\citet{lidz2010}
results in the redshift range $z=3-3.5$.  As pointed out by
\citet{becker2010}, however, differences in the effective optical
depth assumed in the two studies may play a role here; as we have
demonstrated the wavelet PDF is rather sensitive to $\tau_{\rm eff}$.
The \citet{becker2010} constraints also have significantly smaller
error bars compared to the wavelet PDF measurements.  Note, however,
the \citet{becker2010} measurements do not attempt to simultaneously
measure both $T_{0}$ and $\gamma$.  As noted previously, they instead
measure the IGM temperature at the characteristic density,
$\bar{\Delta}(z)$, probed by the Lyman-$\alpha$ forest.  Their
constraints do not marginalise over the uncertain values of $\gamma$
and $\tau_{\rm eff}$, and so a direct comparison of their
uncertainties to our results is less straightforward.

In Figure~\ref{fig:gammacomparison_gamma} we attempt an analogous
comparison of our constraints (thick black error bars) with the
available models for $\gamma(z)$ as well as previous measurements.
Our results are shown with the analytical reionisation models
of~\cite{hui1997} (with $T_{\rm reion}=25\,000$ K at $z=6$) and the
extended \HeII reionisation models `L1' and `L1b' from the radiative
transfer simulations of~\cite{mcquinn2009}, which implement
$\gamma=1.3$ and 1.0 at $z=6$, respectively. We compare our results
with the limits from~\citet{lidz2010}, and measurements
from~\citet{ricotti2000}, \citet{schaye2000} and
\citet{2001ApJ...562...52M}.  Overall, it is clear that $\gamma$
remains relatively poorly constrained, although the data appear to
prefer a temperature-density relation which is shallower than the
$\gamma\sim 1.6$ expected if \HeII reionisation occurred at $z \gg 3$.
Overall, we find our results are in agreement with previous studies of
the IGM temperature that suggest there may be additional heating in
the IGM at $z<4$, most likely due to the reionisation of \HeII by
quasars (see also recent studies of the \HeII Lyman-$\alpha$ forest,
e.g. \citealt{Shull10,Worseck11,Syphers11}). We conclude improved
measurements of the slope of the temperature-density relation will be
required for testing blazar heating models in detail.

\begin{table*}
\centering
\begin{tabular}{lllll}
&&  Wavelet PDF&Flux PDF& Joint analysis\\
  \hline
$\left<z\right>=2.1$&   $T_0\,[10^3\,{\rm K}]$  & $ 16 \pm 5 $       & $ 15 \pm 3 $        & $ 17 \pm 2 $ \\
&   $\gamma $             & $ > 0.86 $       & $ 0.99 \pm 0.14 $        & $ 1.11 \pm 0.11 $ \\
&   $\tau_{\rm eff}$       & $ (0.14 \pm 0.04)  $       & $ 0.133 \pm 0.004 $         & $ 0.130 \pm 0.004 $ \\ 
  \hline
$\left<z\right>=2.5$&   $T_0\,[10^3\,{\rm K}]$  & $ 16 \pm 4 $       & $ 14 \pm 9 $        & $ 13 \pm 4 $ \\
&   $\gamma $             & $ > 0.92 $       & $ > 0.69 $        & $ > 0.95 $ \\
&   $\tau_{\rm eff}$       & $ (0.22 \pm 0.05) $       & $ 0.212 \pm 0.011 $         & $ 0.200 \pm 0.009 $ \\ 
  \hline
$\left<z\right>=2.9$&   $T_0\,[10^3\,{\rm K}]$  & $ 20 \pm 5 $       & $ 21 \pm 7 $        & $ 19 \pm 4 $ \\
&   $\gamma $             & $ > 0.80 $       & $ < 1.24 $        & $ 1.1 \pm 0.2 $ \\
&   $\tau_{\rm eff}$       & $ (0.36 \pm 0.09) $       & $ 0.290 \pm 0.019 $         & $ 0.27 \pm 0.02 $ \\ 
  \hline
\end{tabular}
\caption{1$\sigma$ constraints on the IGM thermal parameters for the
  wavelet PDF, flux PDF and joint analysis. We find broad consistency
  between the methods, with the flux PDF favouring a slightly lower
  value of $\gamma$. Entries in parentheses are prior dominated (no
  detection) and the limits quoted are 95\% confidence.}
\label{tab:bins}
\end{table*}

\section{Conclusions}

We have analyzed eighteen metal cleaned Lyman-$\alpha$ forest spectra
in the redshift range 1.7--3.2 using both the flux PDF and wavelet
PDF. The results have been interpreted using a suite of hydrodynamical
simulations to place constraints on the thermal state of the
intergalactic medium while marginalizing over the uncertainty in the
cosmological parameters.  Our wavelet analysis is similar to the
analysis performed~\cite{theuns2002}, following most closely the
technique employed by~\cite{lidz2010},  but using independently
reduced spectra and a different approach to simulating the IGM.  An
analysis of the constraints on the IGM thermal state obtained from the
flux PDF using the same data set furthermore enables us to explore any
systematic differences between the two methodologies.  The main
results of our study are as follows:

\begin{itemize}

\item{The constraints on the IGM thermal state at $z=1.7$--$3.2$
  derived from the wavelet PDF and flux PDF analysis are formally
  consistent with each other within the rather large uncertainties.
  However, we find there is some mild tension between the two
  measurements, with the flux PDF measurements generally preferring a
  lower value for the slope of the temperature-density relation at all
  redshifts.}\\

\item{We have checked that the impact of a continuum which has been
  placed 3 per cent too low on the wavelet and flux PDF is small
  compared to the effect of varying other free parameters such as
  $T_{0}$, $\gamma$ and $\tau_{\rm eff}$.  The flux PDF is indeed more
  sensitive to changes in the continuum placement, but the effect
  remains small and is minimal within the flux range of $F=[0.1,0.8]$
  we fit in our analysis.  We conclude it is unlikely that the
  continuum placement is fully responsible for the systematic offset
  found in the wavelet and flux PDF constraints.}\\

 \item{We have explicitly confirmed that varying cosmological
   parameters within a narrow range has little impact on the wavelet
   amplitude PDF, with the strongest effect being that of $\sigma_8$
   in our lowest redshift bin. We also confirm that there is a
   significant degeneracy between the parameters $T_{0}$, $\gamma$ and
   $\tau_{\rm eff}$ inferred from the wavelet amplitude PDF. }\\

\item{The flux PDF puts a much stronger constraint on $\tau_{\rm eff}$
  compared to the wavelet PDF. We therefore perform a joint analysis of
  the flux PDF and wavelet PDF in order to add the $\tau_{\rm eff}$
  constraint derived from the flux PDF.  We find the joint constraints
  on the IGM temperature at mean density, $T_{0}$, obtained at
  $z=[2.1,2.5,2.9]$ are in good agreement with other recent
  measurements.  The constraints are consistent with the models
  of~\cite{mcquinn2009}, in which an extended \HeII reionisation epoch
  completes around $z=3$, driven by quasars with an EUV index of
  $\alpha\simeq 1.6$.  We have also performed a rudimentary comparison
  with the recently proposed blazar heating models
  of~\cite{puchwein2011}, and find their weak blazar heating matches
  our $T_{0}$ constraints most closely.}\\

\item{We find the slope of the temperature-density relation obtained
  from the joint analysis is consistent with $\gamma\sim 1.1$--$1.3$,
  although the uncertainties on this measurement remain large and
  remain consistent with an inverted ($\gamma<1$) temperature-density
  relation within $1$--$1.5\sigma$.  A more precise measurement of the
  temperature-density relation will be necessary for stringently
  testing competing IGM heating models, such as the volumetric heating
  rate from blazar heating which produces a strongly inverted
  temperature-density relation by $z=2$ (\citealt{2011arXiv1106.5504C}).}

\end{itemize}
\begin{figure}
    \centering
    \includegraphics[width=0.75\columnwidth, angle=-90]{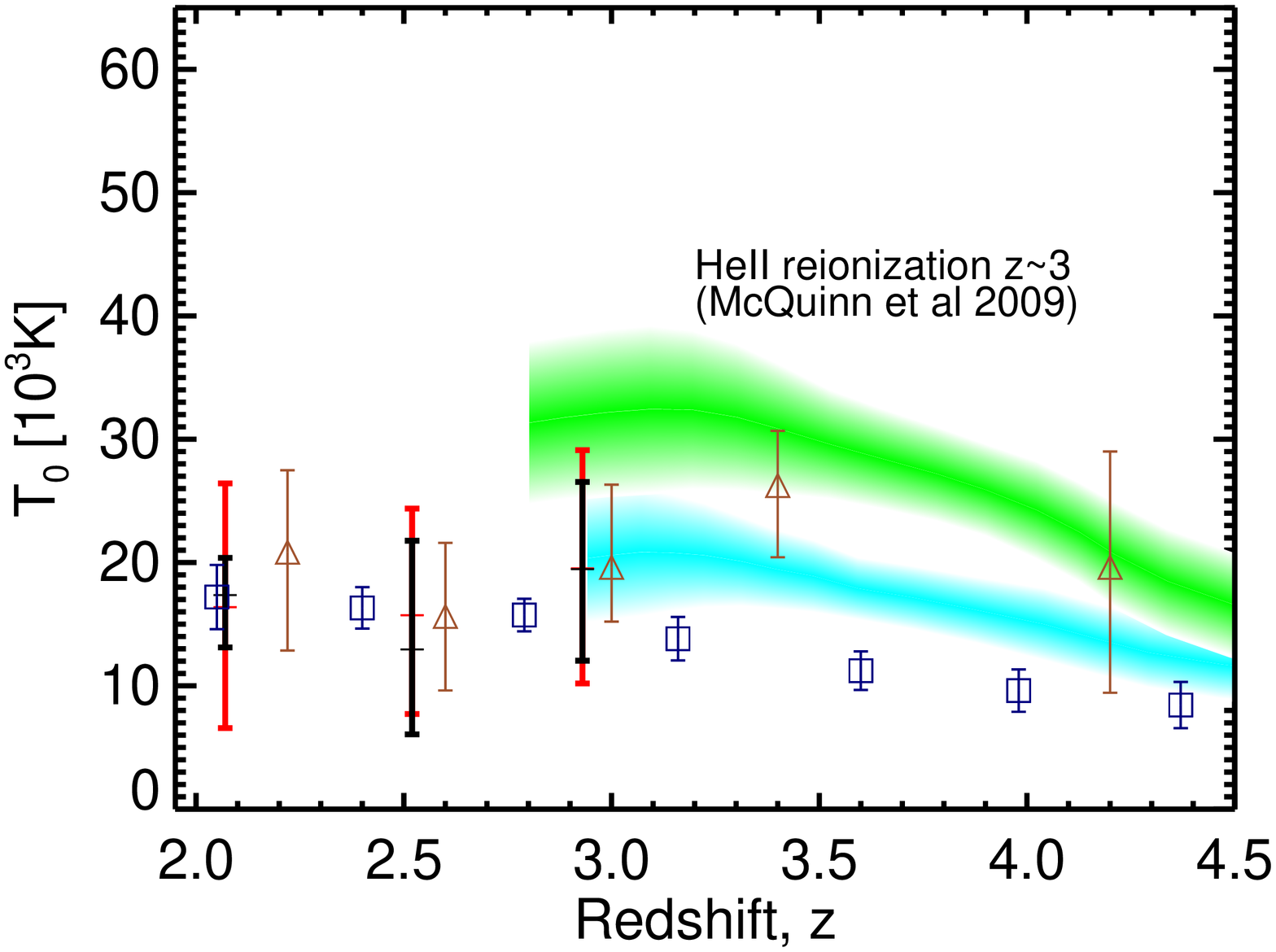}
    \includegraphics[width=0.75\columnwidth, angle=-90]{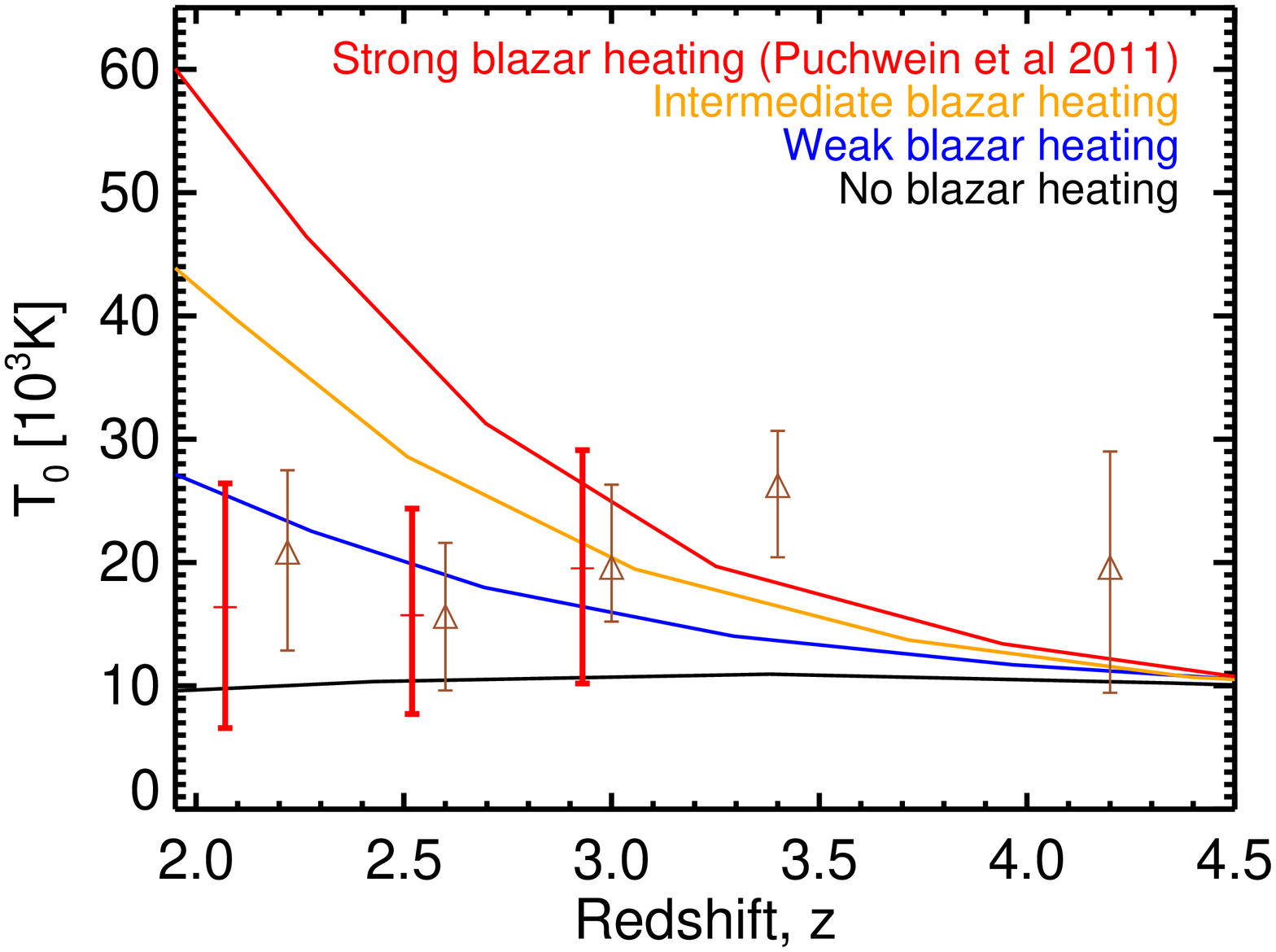}
    \caption{A comparison of our constraints with literature models
      for $T(z)$. Our constraints are the thick data points: Wavelet
      PDF $2\sigma$ (outer, red), joint analysis $2\sigma$ (inner,
      black), \citet{lidz2010} wavelet analysis $2\sigma$ (triangles),
      and the~\citet{becker2010} curvature analysis $2\sigma$
      (squares) which assumes $\gamma=1.3$. The shaded regions (upper
      panel) are two \HeII reionisation models
      from~\citet{mcquinn2009} (`S4b' upper, `D1' lower with quasar UV
      spectral indices of $0.6$ and $1.6$, respectively) while the
      lines (lower panel) show  the temperature at mean
      density of the blazar heating models
      from~\citet{puchwein2011}.}
    \label{fig:modelcomparison_t0}
\end{figure}

\begin{figure}
    \centering
    \includegraphics[width=0.75\columnwidth, angle=-270]{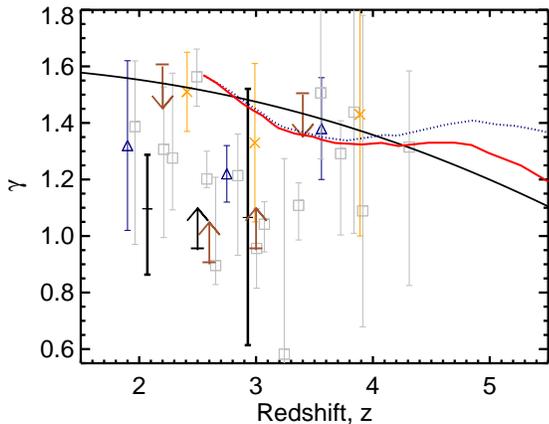}
    \caption{A comparison of our constraints on the slope of the
      temperature-density relation with various models for $\gamma(z)$
      and other observational constraints from the literature.  Our
      $2\sigma$ constraints and limit from the joint analysis (thick,
      black error bars and arrow), can be compared with the 2$\sigma$
      limits from~\citet{lidz2010} (brown arrows), the 1$\sigma$
      constraints from~\citet{ricotti2000} (triangles, navy), the
      1$\sigma$ constraints from~\citet{schaye2000} (squares, grey)
      and the 1$\sigma$ constraints from~\citet{2001ApJ...562...52M}
      (crosses, orange). The extended solid line shows a model with
      sudden \HeII reionisation heating the IGM to $25\,000$ K at
      $z=6$ (solid), while the broken lines are the `L1' and `L1b'
      models of \HeII reionisation (solid and dotted)
      from~\citet{mcquinn2009}, which implement $\gamma=1.3$ and
      1.0 at $z=6$, respectively.}
    \label{fig:gammacomparison_gamma}
\end{figure}

Overall, our results are consistent with previous observations that
  indicate there may be additional sources of heating in the IGM at
  $z<4$. This heating could be due to a number of effects such as an
  extended epoch of \HeII reionisation which has yet to complete at
  $z=3$~\citep{Shull10,Worseck11,Syphers11}, heating of the low
  density IGM by blazars~\citep{puchwein2011} or feedback either in
  the form of galactic winds and/or AGN
  feedback~\citep{torna,boothschaye}.  We note, however, the latter
  will provide only a partial explanation due to the small volume
  filling factor of the shock-heated gas
  (\citealt{2002ApJ...578L...5T}).

We believe that analyses of the flux PDF
(\citealt{viel2009,Calura2012}), the wavelet PDF (\citealt{lidz2010})
and other techniques such as the curvature (\citealt{becker2010}) and
absorption line widths (\citealt{schaye2000}) provide generally
consistent constraints on the IGM thermal state at $z<3$ when applied
to high resolution spectra.  This is encouraging given the wide range
of different methodologies used in the existing literature.  However,
significant uncertainties on measurements of the IGM thermal state
remain.  Ideally, one should also aim at reaching full consistency
between different data sets (high and low resolution QSO spectra) as
well as different methods (see e.g. \citealt{viel2009}).  In the near
future, large scale surveys of the Lyman-$\alpha$ forest at moderate
spectral resolution, such as the SDSS Baryon Oscillation Spectroscopic
Survey (BOSS) (\citealt{Slosar2011}), will provide valuable new
insights into the physical state of the IGM with a high degree of
statistical precision.  Fully understanding the potential systematic
uncertainties associated with these measurements are therefore vital
for further unravelling the thermal history of the IGM following
reionisation.

\section*{Acknowledgments}
We thank Jason Dick and Christoph Pfrommer for useful discussions and
the referee, Tom Theuns, for constructive comments which helped
improve this paper.  MV is supported by INFN PD51, ASI/AAE contract,
PRIN INAF, PRIN MIUR and the FP7 ERC Starting Grant
``cosmoIGM''. Numerical computations were performed using the COSMOS
Supercomputer in Cambridge (UK), which is sponsored by SGI, Intel,
HEFCE and the Darwin Supercomputer of the University of Cambridge High
Performance Computing Service (http://www.hpc.cam.ac.uk/), provided by
Dell Inc. using Strategic Research Infrastructure Funding from the
Higher Education Funding Council for England. Part of the analysis has
been performed at CINECA with a Key project on the intergalactic
medium obtained through a CINECA/INAF grant.  JSB acknowledges the
support of an ARC Australian postdoctoral fellowship (DP0984947).

\bibliographystyle{mn2e}
\bibliography{main}

\appendix

\section{Additional material}

\subsection{Dependence of the wavelet amplitude PDF on cosmological parameters}\label{app:cosmo}

In this appendix we explicitly show the wavelet amplitude PDF depends
almost negligibly on the cosmological parameters in our
analysis. Figure~\ref{fig:pdf_cosmo} shows the effect of varying
$H_0$, $\sigma_8$, $\Omega_{\rm m}$, and $n_{\rm s}$ on the wavelet
PDF.

\begin{figure*}
  \centering
  \includegraphics[width=0.65\columnwidth, angle=270]{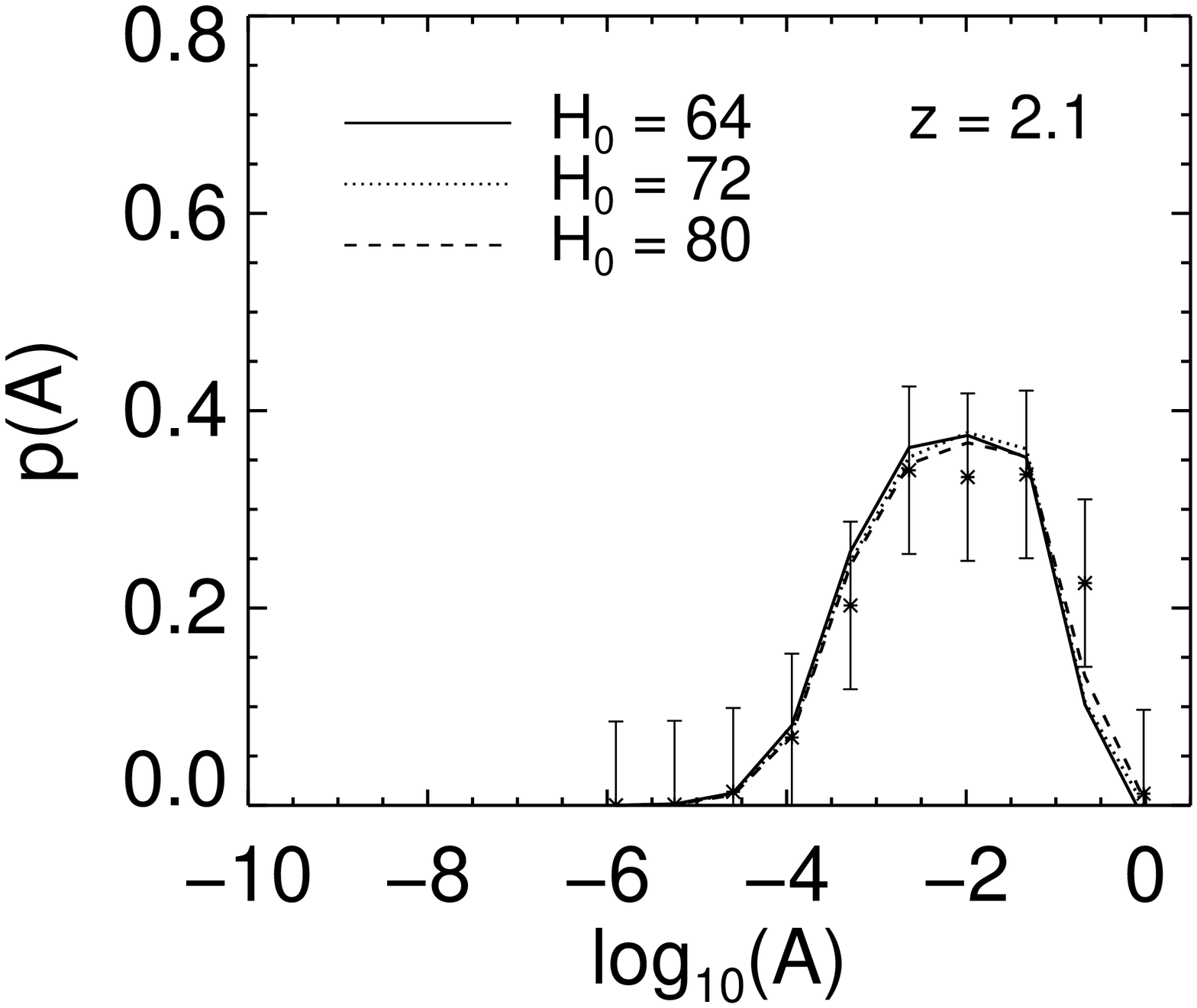}
  \includegraphics[width=0.65\columnwidth, angle=270]{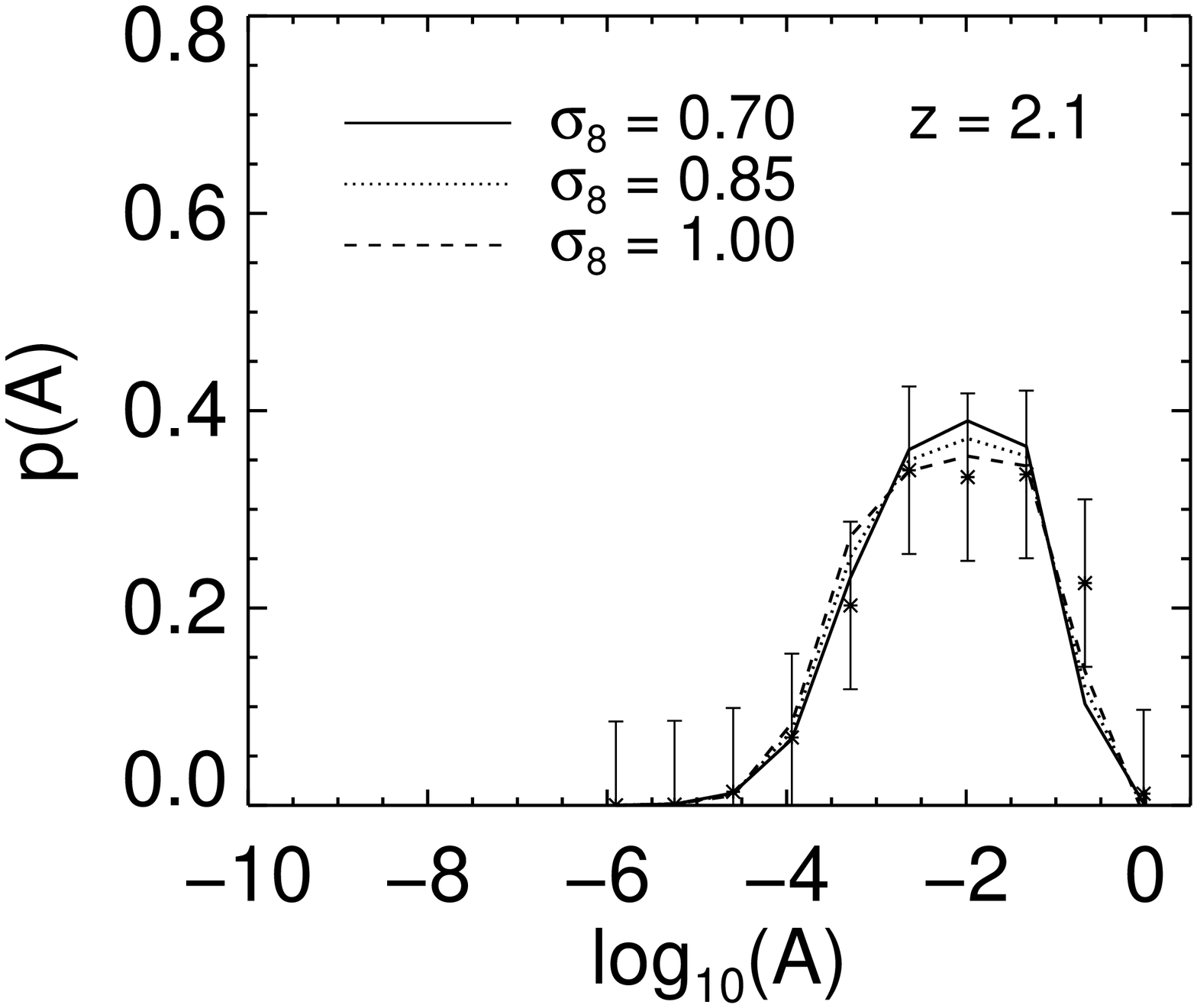}
  \includegraphics[width=0.65\columnwidth, angle=270]{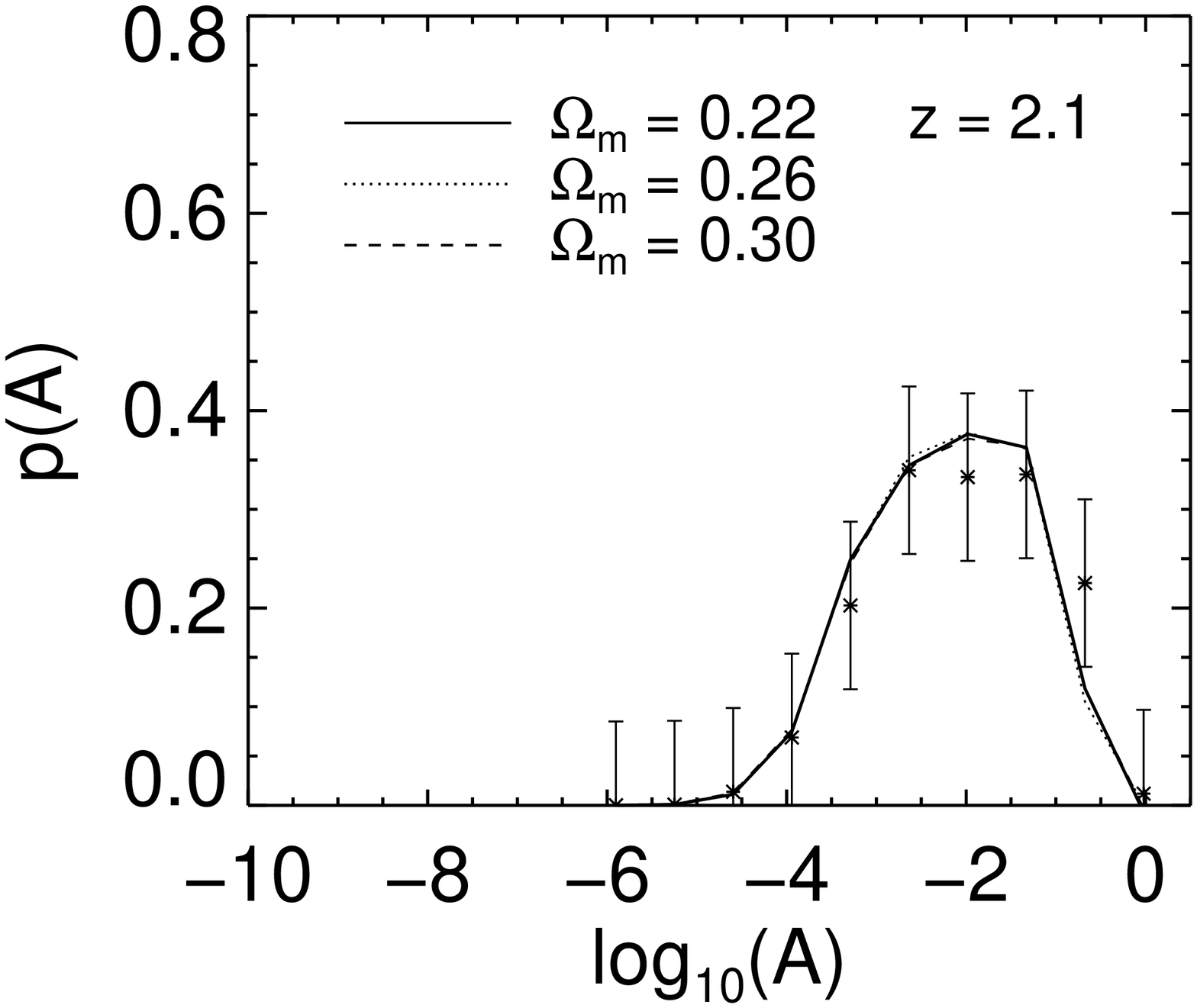}
  \includegraphics[width=0.65\columnwidth, angle=270]{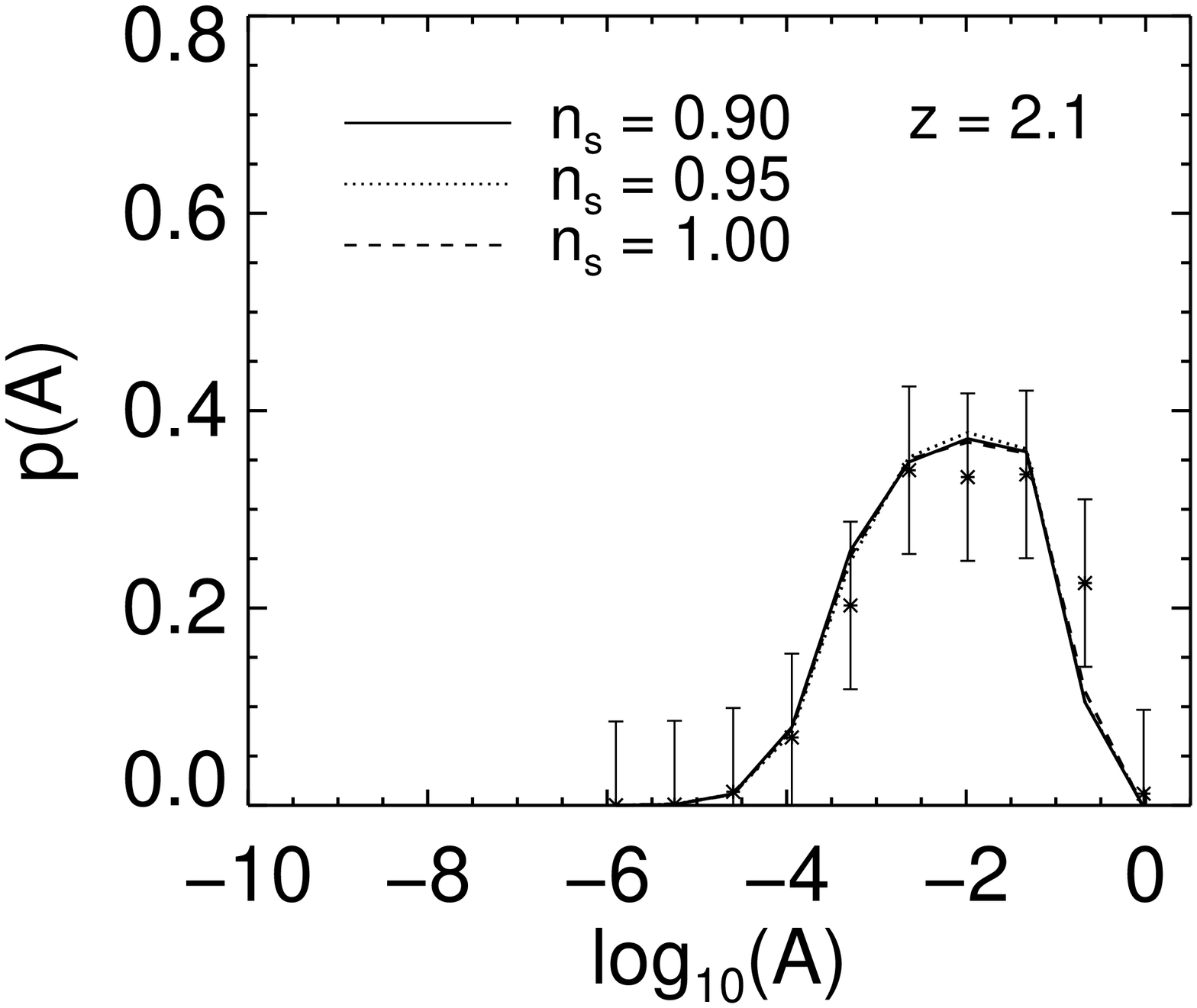}
  \caption{The wavelet amplitude PDF obtained from simulated spectra
    varying a single cosmological parameter about the fiducial
    model. Top left: Varying $H_0$. Top right: Varying
    $\sigma_8$. Bottom left: Varying $\Omega_{\rm m}$. Bottom right:
    Varying $n_{\rm s}$.  The wavelet amplitude PDF has a very weak
    dependence on the cosmological parameters, but with a mild
    dependence on $\sigma_8$ in our lowest redshift bin $z=2.1$.}
  \label{fig:pdf_cosmo}
\end{figure*}

\subsection{Simulation parameters and convergence tests}\label{app:simparams}
In Table~\ref{tab:simulations} we list the parameters of our
simulations, and catalogue how they have been used in the
interpolation scheme described in Section~\ref{sec:interpolation}.
Using simulations R1, R2, R3 and C15 we have tested the convergence of
the wavelet amplitude PDF with gas particle mass and box
size. Figure~\ref{fig:resolution} demonstrates the stability of the
wavelet amplitude PDF under this change.

\begin{table*}
\centering
\caption{Parameters of the simulations used for describing the thermal
  state of the IGM: $L$ is the comoving box-length; $N$ the number of
  gas and dark matter particles in the simulation; $M_{\rm gas}$ is
  the mass of each gas particle in the simulation box; $\zeta$ and
  $\xi$ are the scaling parameters used to modify the photo-heating
  rates. $T_0$ and $\gamma$ are not a priori parameters of the
  simulations, but are determined by fitting the temperature-density
  relation at each redshift. The table is divided into five sections:
  our reference model (D15); simulations for interpolating over
  $\gamma$; simulations for interpolating over $T_0$; the validation
  set; resolution and box-size checks.}
\label{tab:simulations}
\begin{tabular}{lccccccccccc}
  \hline
  \noalign{\smallskip}
  Model & $L$ & $N$ & $M_{\rm gas}$ &
  $\zeta$ & $\xi$ & $T_0/10^3{\rm K}$ & $T_0/10^3{\rm K}$ & $T_0/10^3{\rm K}$ &
  $\gamma$ & $\gamma$ & $\gamma$ \\
  &$[h^{-1}$ Mpc$]$ & & $[10^4 h^{-1}M_{\varodot}]$ &
  & & $[z=2.17]$ & $[z=2.55]$ & $[z=2.98]$ &
  $[z=2.17]$ & $[z=2.55]$ & $[z=2.98]$ \\

  \noalign{\smallskip}
  \hline
  \noalign{\smallskip}
  D15        & 10 & $2 \times 512^3$ & $9.2 $ & 2.20 &  0.00
  & 16.0 & 17.1 & 18.2 & 1.57 & 1.56 & 1.55 \\ 
  \hline
  D07        & 10 & $2 \times 512^3$ & $9.2 $ & 2.20 & -1.60
  & 16.0 & 16.8 & 17.9 & 0.76 & 0.73 & 0.71 \\ 
  D10        & 10 & $2 \times 512^3$ & $9.2 $ & 2.20 & -1.00
  & 16.0 & 17.0 & 18.1 & 1.07 & 1.05 & 1.03 \\ 
  D13        & 10 & $2 \times 512^3$ & $9.2 $ & 2.20 & -0.45
  & 16.0 & 17.0 & 18.1 & 1.35 & 1.33 & 1.32 \\ 

  \hline
  A15        & 10 & $2 \times 512^3$ & $9.2 $ & 0.3  &  0.00
  & 4.6  & 4.8  & 5.1 & 1.55 & 1.54 & 1.52 \\ 
  B15        & 10 & $2 \times 512^3$ & $9.2 $ & 0.8  &  0.00
  & 8.5  & 9.1  & 9.6 & 1.56 & 1.55 & 1.54 \\ 
  C15        & 10 & $2 \times 512^3$ & $9.2 $ & 1.45 &  0.00
  & 12.4 & 13.2 & 14.0 & 1.57 & 1.56 & 1.54 \\ 
  E15        & 10 & $2 \times 512^3$ & $9.2 $ & 3.10 &  0.00
  & 19.6 & 21.0 & 22.5 & 1.57 & 1.56 & 1.55 \\ 
  F15        & 10 & $2 \times 512^3$ & $9.2 $ & 4.20 &  0.00
  & 23.6 & 25.3 & 27.0 & 1.57 & 1.56 & 1.55 \\ 
  G15        & 10 & $2 \times 512^3$ & $9.2 $ & 5.30 &  0.00
  & 27.1 & 29.0 & 31.0 & 1.57 & 1.56 & 1.55 \\ 
  \hline
  C10        & 10 & $2 \times 512^3$ & $9.2 $ & 1.45 & -1.00
  & 12.3 & 13.1 & 13.7 & 1.06 & 1.04 & 1.02 \\ 
  E10        & 10 & $2 \times 512^3$ & $9.2 $ & 3.10 & -1.00
  & 19.7 & 21.0 & 22.2 & 1.07 & 1.06 & 1.04 \\
  \hline
  R1         & 10 & $2 \times 256^3$ & $7.4 $ & 1.45 &  0.00
  & 12.5 & 13.2 & 14.0 & 1.56 & 1.54 & 1.53 \\ 
  R2         & 10 & $2 \times 128^3$ & $5.9 $ & 1.45 &  0.00
  & 12.8 & 13.5 & 14.3 & 1.54 & 1.53 & 1.51 \\ 
  R3         & 20 & $2 \times 256^3$ & $5.9 $ & 1.45 &  0.00
  & 12.8 & 13.6 & 14.3 & 1.54 & 1.53 & 1.51 \\ 

  R4         & 40 & $2 \times 512^3$ & $5.9 $ & 1.45 &  0.00
  & 12.8 & 13.6 & 14.5 & 1.55 & 1.52 & 1.52 \\ 

  \hline
\end{tabular}
\end{table*}

\begin{figure*}
  \centering
  \includegraphics[width=0.43\columnwidth, angle=270]{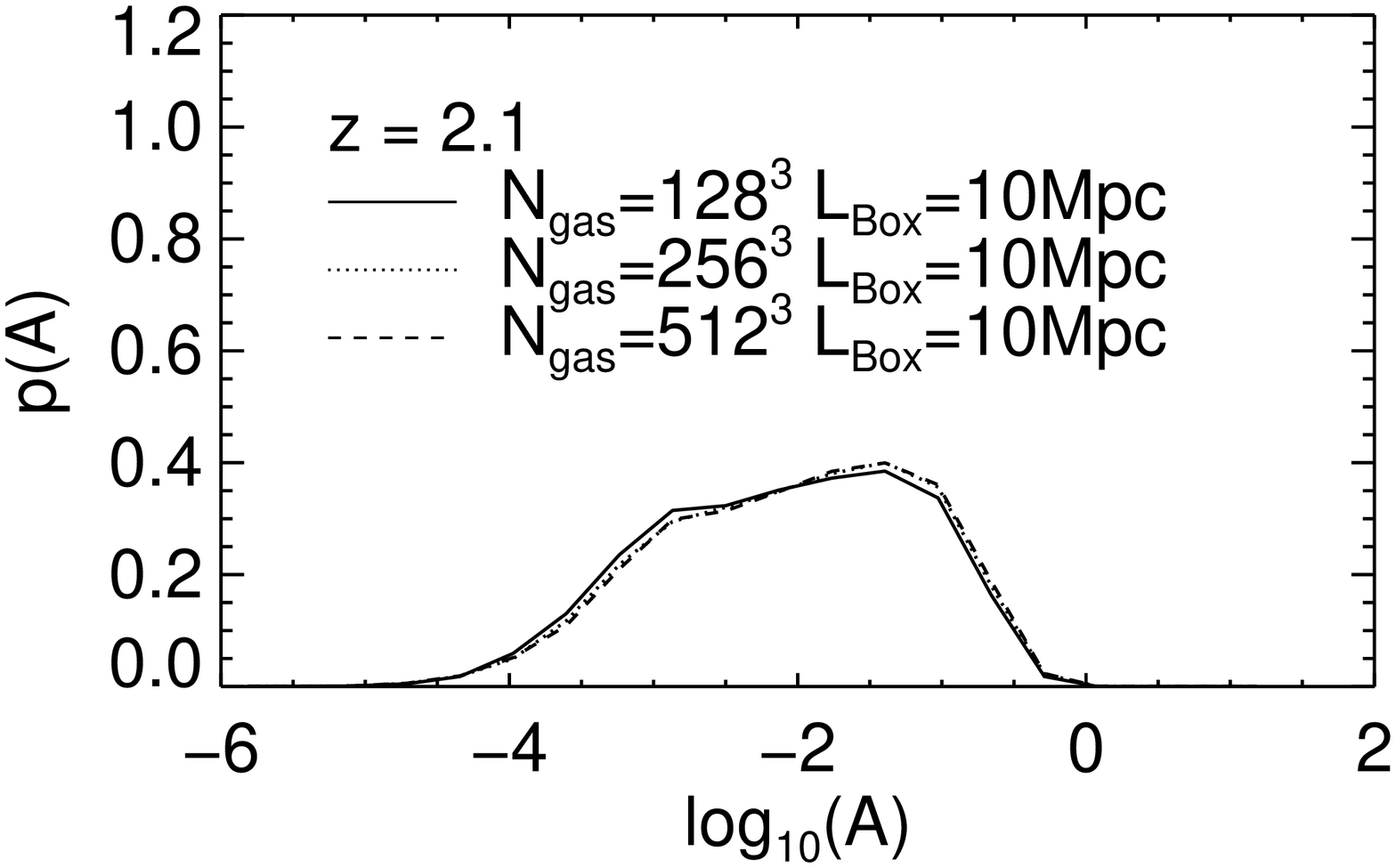}
  \includegraphics[width=0.43\columnwidth, angle=270]{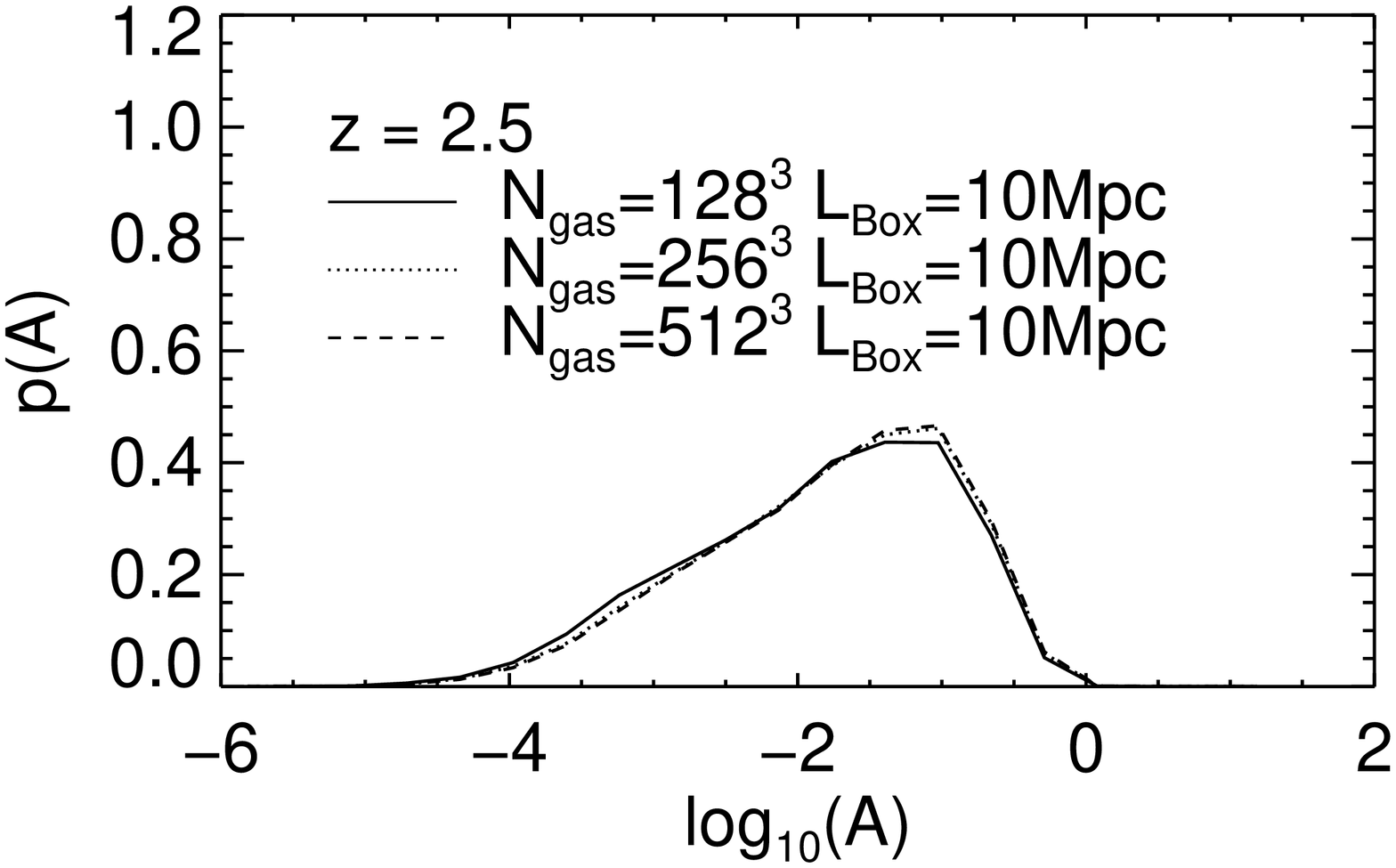}
  \includegraphics[width=0.43\columnwidth, angle=270]{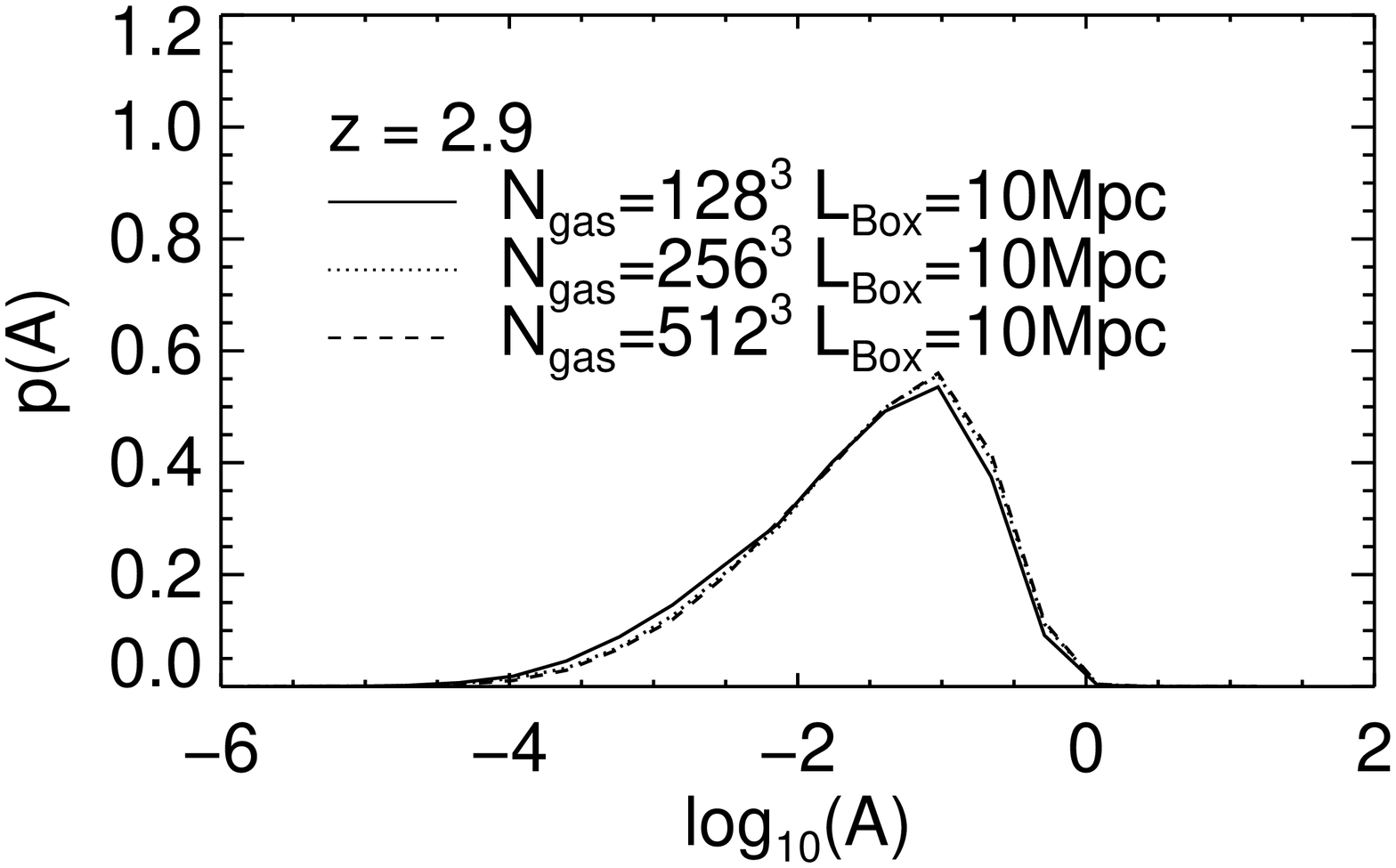}
  \includegraphics[width=0.43\columnwidth, angle=270]{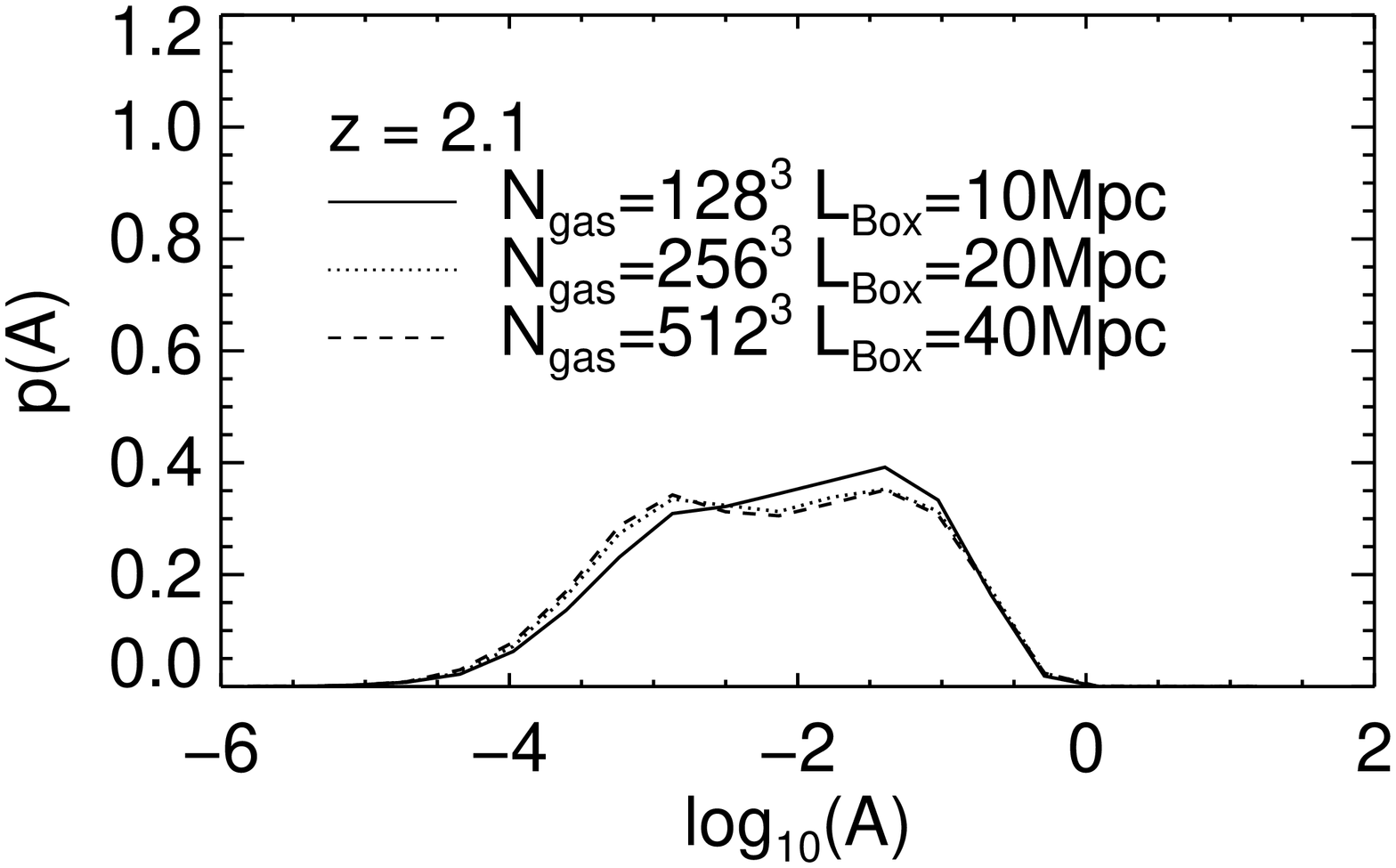}
  \includegraphics[width=0.43\columnwidth, angle=270]{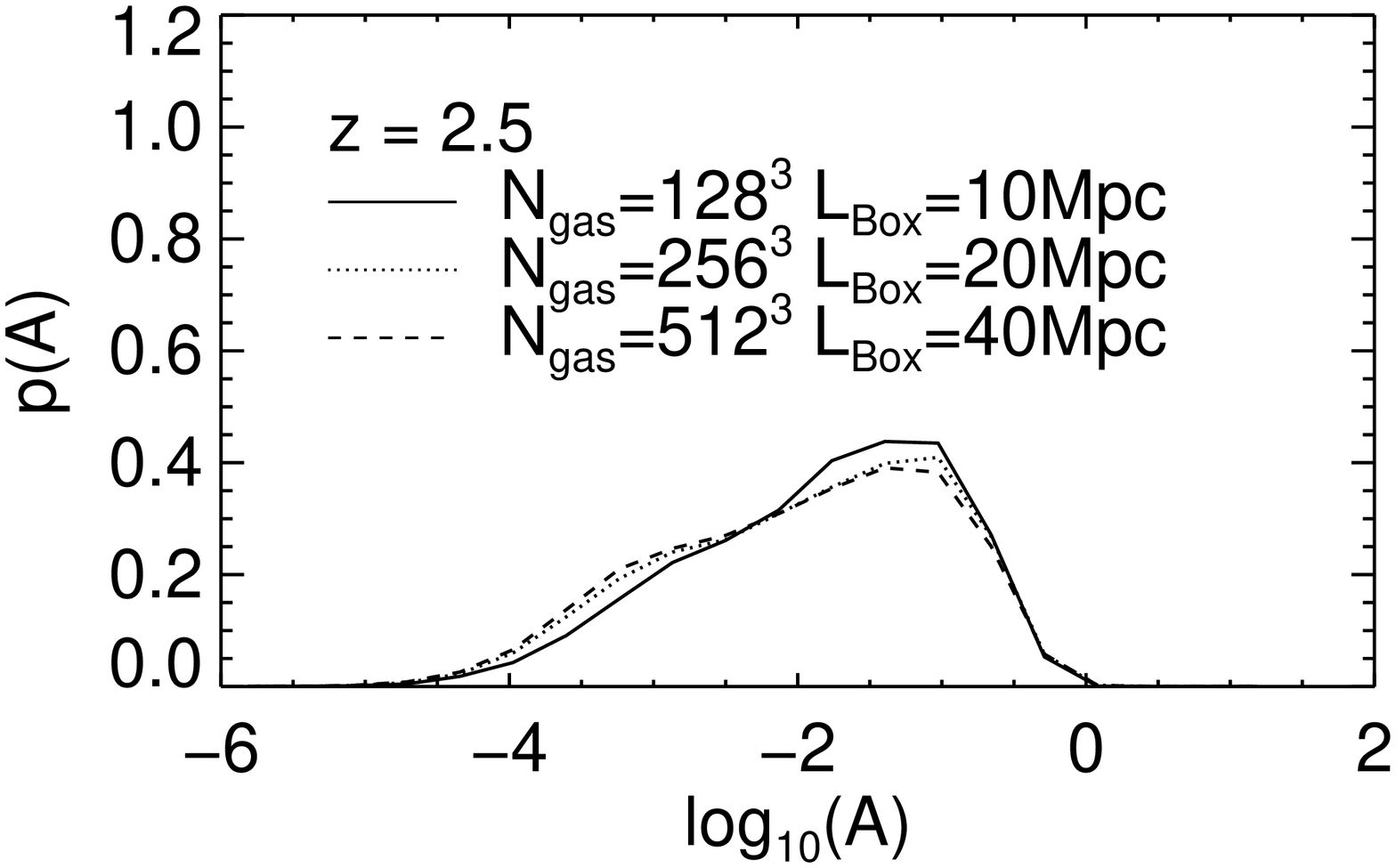}
  \includegraphics[width=0.43\columnwidth, angle=270]{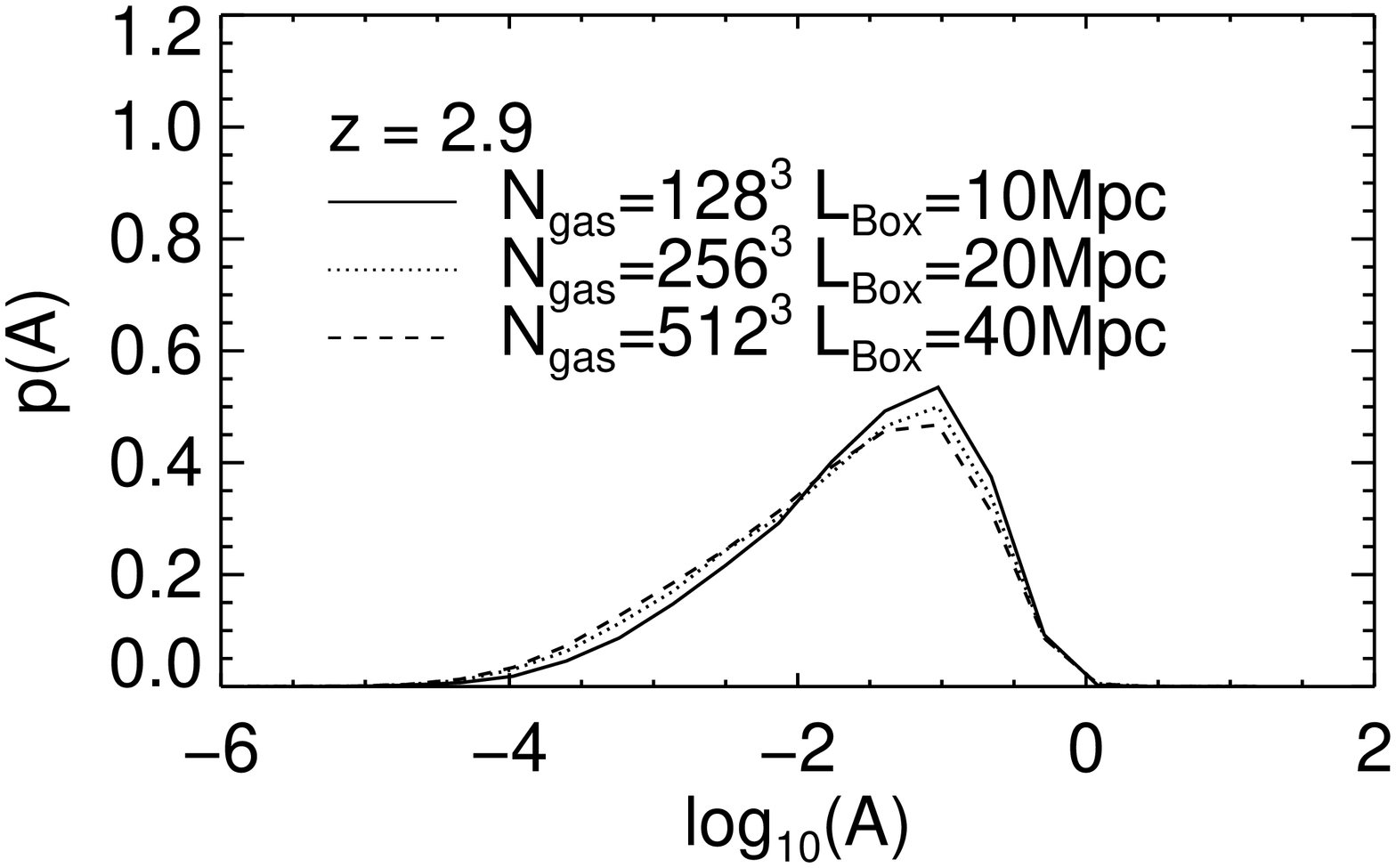}
  \caption{Convergence of the wavelet amplitude PDF with increasing
    gas particle mass at fixed box size (upper row) and increasing box
    size at fixed gas particle mass (lower row). Our conclusions from
    these tests are described in Section~\ref{sec:sims}.}
  \label{fig:resolution}
\end{figure*}

\subsection{Continuum test}\label{app:continuum}

Figure~\ref{fig:continuum_effect} demonstrates the effect of lowering
the continuum on the simulated spectra by 3 per cent on the wavelet
amplitude and flux PDFs.  A brief discussion of the results from this
test is provided in Section~\ref{sec:constraints}.

\begin{figure*}
  \centering
  \includegraphics[width=0.60\columnwidth, angle=270]{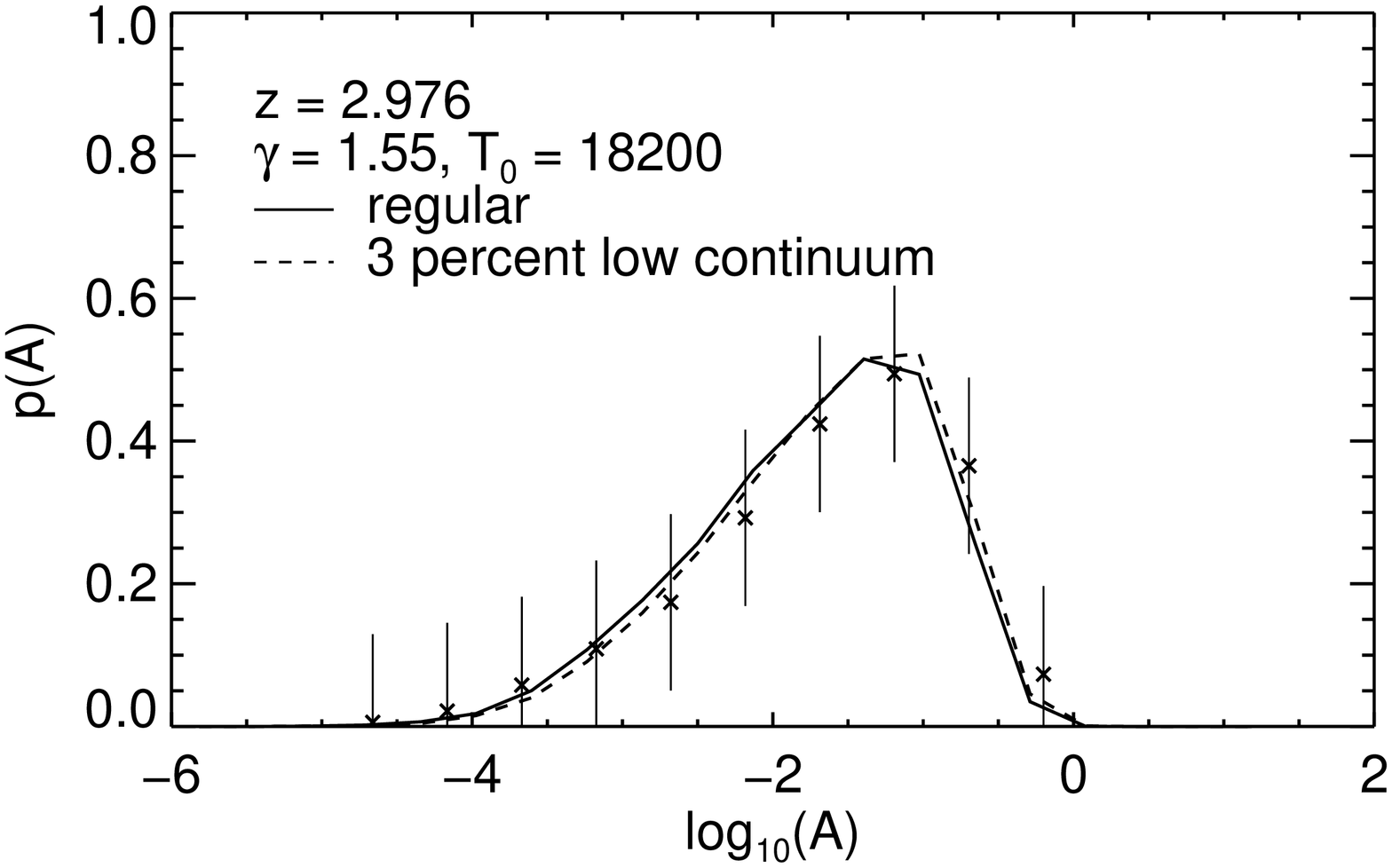}
  \includegraphics[width=0.60\columnwidth, angle=270]{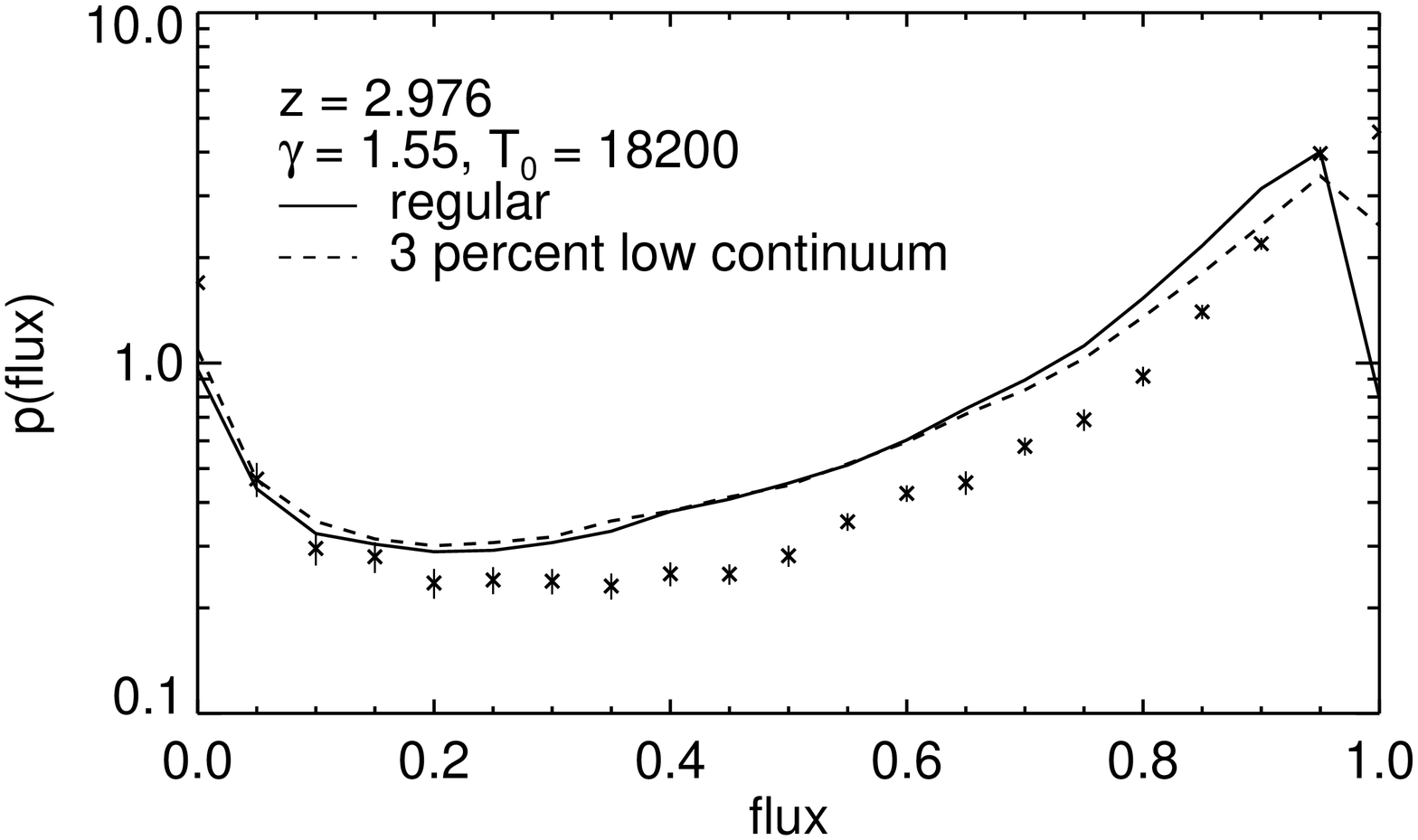}
  \includegraphics[width=0.60\columnwidth, angle=270]{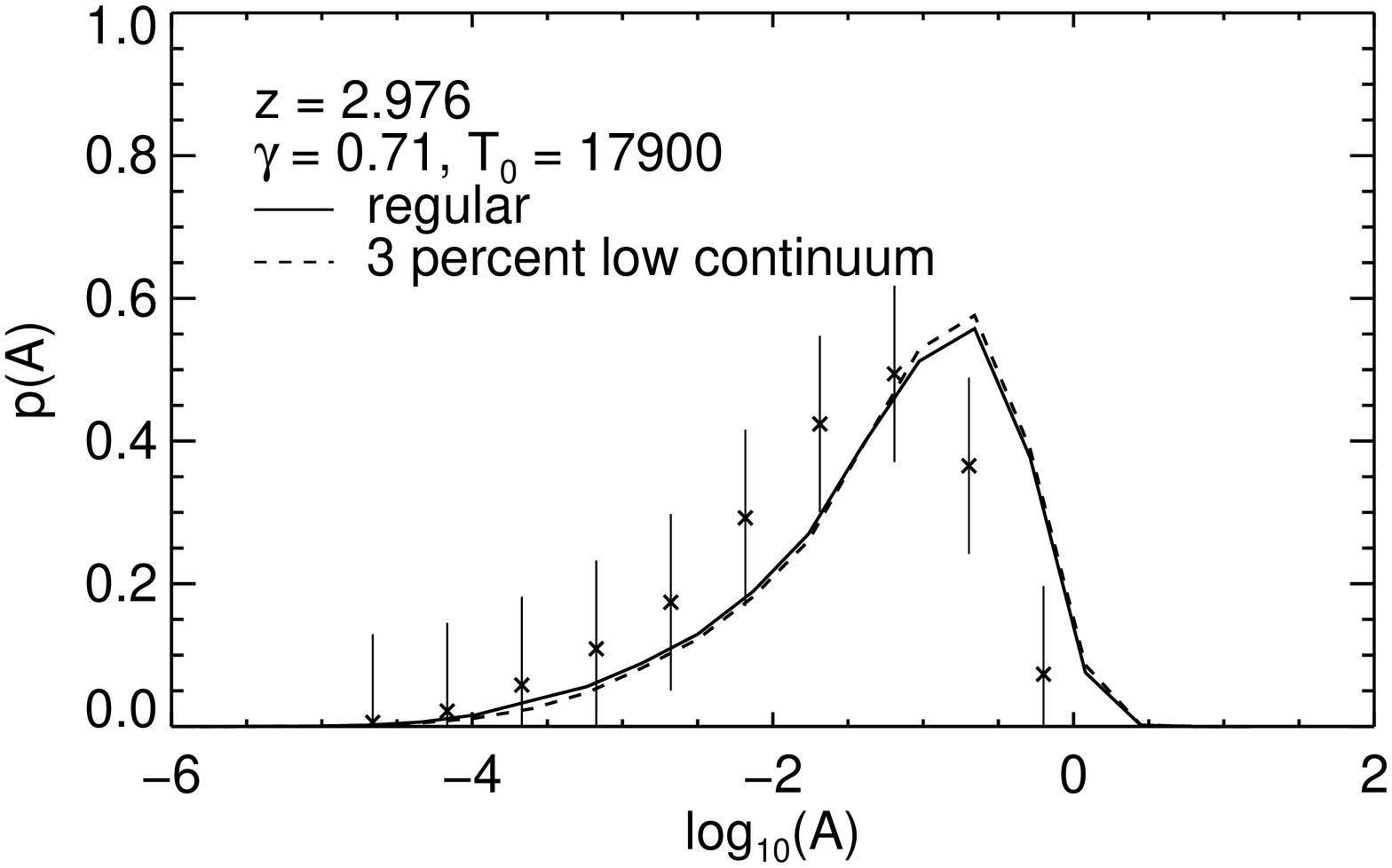}
  \includegraphics[width=0.60\columnwidth, angle=270]{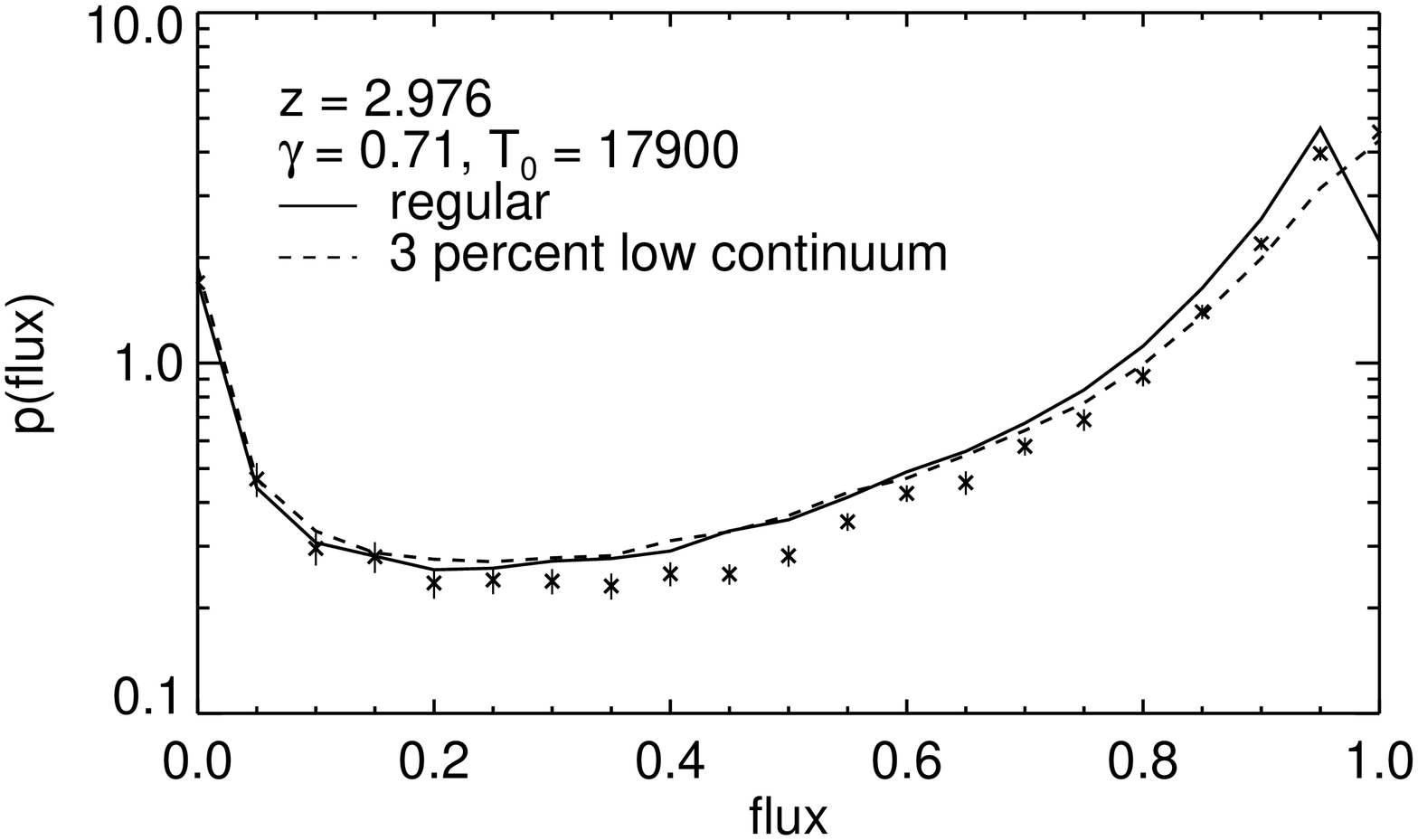}
  \caption{The effect of the continuum level on the wavelet amplitude
    and flux PDFs at redshift $z=2.9$. In the left panels we show the
    wavelet PDFs for the simulations D15 (above) and D07 (below); in
    the right panels we show the flux PDFs for the same
    simulations. The dotted lines are the results for the native
    continuum level in the simulations (indicated with \emph{regular})
    whereas the dashed lines are for a continuum level which has been
    lowered by $3$ per cent. For comparison, we also show with a solid
    line the data with the error bars used in the two analyses.}
  \label{fig:continuum_effect}
\end{figure*}

\end{document}